\documentstyle[12pt]{article}
\newcommand{\svec}[1]{ \stackrel{\rightarrow}{#1} }
\newcommand{\dvec}[1]{ \stackrel{\leftrightarrow}{#1} }
\newcommand{\lvec}[1]{ \stackrel{\leftarrow}{#1} }
\newcommand{\swav}[1]{ \stackrel{\sim}{#1} }
\newcommand{\tdot}[1]{ \stackrel{\cdot}{#1} }
\newcommand{\uhat}{ \hat U }
\newcommand{\ehat}{ \hat U_{\epsilon} }
\newcommand{\mhat}[1]{ \hat U_{\epsilon_{#1}} }
\newcommand{\define}{ \stackrel{\triangle}{=} }

\def\be{\begin{equation}}
\def\ee{\end{equation}}
\def\ba{\begin{array}}
\def\ea{\end{array}}

\begin{document}
\title{\bf Renormalizable Quantum Gauge Theory of Gravity }
\author{{Ning Wu}
\thanks{email address: wuning@heli.ihep.ac.cn}
\\
\\
{\small Institute of High Energy Physics, P.O.Box 918-1,
Beijing 100039, P.R.China}}
\maketitle
\vskip 0.8in

~~

\vskip 0.4in

\begin{abstract}
The quantum gravity is formulated based on gauge principle. 
The model discussed in this paper has local gravitational 
gauge symmetry and gravitational field is represented 
by gauge potential. A preliminary study on gravitational
gauge group is presented. Path integral quantization of the
theory is discussed in the paper. A strict proof on the
renormalizability of the theory is also given. In leading order
approximation, the gravitational gauge field theory gives out
classical Newton's theory of gravity. It can also give out
an Einstein-like field equation with cosmological term. 
The prediction for cosmological constant given by
this model is well consistent with experimental results. 
For classical tests, it gives out the same theoretical 
predictions as those of general relativity. 
Combining cosmological principle with the field equation
of gravitational gauge field, we can also set up
a cosmological model which is consistent with recent
observations. 
\\

\end{abstract}

\newpage

\Roman{section}

\section{Introduction}
\setcounter{equation}{0} 

In 1686, Isaac Newton published his book {\it MATHEMATICAL
 PRINCIPLES OF NATURAL PHILOSOPHY}. In this
book, through studying the motion of planet in solar system, he
found that gravity obeys the inverse square law and the magnitude
of gravity is proportional to the mass of the object\cite{01}. The
Newton's classical theory of gravity is kept unchanged until 1916. At
that year, Einstein proposed General
Relativity\cite{02,03}. In this great work, he founded a
relativistic theory on gravity, which is based on principle of
general relativity and equivalence principle. Newton's classical
theory for gravity appears as
a classical limit of  general relativity.\\

One of the biggest revolution in human kind in the last century is
the foundation of quantum theory. The quantum hypothesis was first
introduced into physics by Max Plank in 1900. 
In 1916, Albert Einstein points out that quantum effects must
lead to modifications in the theory of general relativity\cite{5}.
Soon after the foundation of  quantum mechanics,
physicists try to found a theory that could describe the
quantum behavior of the full gravitational field. In the 70
years attempts, physicists have found two theories based
on quantum mechanics that attempt to unify general
relativity and quantum mechanics, one is canonical quantum
gravity and another is superstring theory. But for quantum
field theory, there are different kinds of mathematical
infinities that naturally occur in quantum descriptions of
fields. These infinities should be removed by the technique
of perturbative renormalization. However, the perturbative
renormalization does not work for the quantization of Einstein's
theory of gravity, especially in canonical quantum gravity. In
superstring theory, in order to make perturbative renormalization
to work, physicists have to introduce six extra dimensions. But
up to now, none of the extra dimensions have been observed.
To found a consistent theory that can unify general relativity and
quantum mechanics is a long dream for physicists. \\

The "relativity revolution" and the "quantum revolution" are among
the greatest successes of twentieth century physics, yet two
theories appears to be fundamentally incompatible. General
relativity remains a purely classical theory which describes the
geometry of space and time as smooth and continuous, on the
contrary, quantum mechanics divides everything into discrete
quanta. The underlying theoretical incompatibility between two
theories arises from the way  that they treat the geometry of
space and time. This situation makes some physicists still wonder
whether quantum theory is a truly fundamental theory of Nature, or
just a convenient description of some aspects of the microscopic
world. Some physicists even consider the twentieth century as the
century of the incomplete revolution. To set up a consistent
quantum theory of gravity is considered to be the last challenge
of quantum theory\cite{50,51}. In other words, combining general
relativity with quantum mechanics is considered to be the last
hurdle to be overcome in the "quantum revolution".
\\

In 1921, H.Weyl introduced the concept of gauge transformation
into physics\cite{d01,d05}, which is one of the most important
concepts in modern physics, though his original theory is not
successful. Later, V.Fock, H.Weyl and W.Pauli found that quantum
electrodynamics is a gauge invariant theory\cite{d02,d03,d04}. In
1954, Yang and Mills proposed non-Abel gauge field theory\cite{1}.
This theory was soon applied to elementary particle physics.
Unified electroweak theory\cite{2,3,4} and quantum chromodynamics
are all based on gauge field theory. The predictions of unified
electroweak theory have been confirmed in a large number of
experiments, and the intermediate gauge bosons $W^{\pm}$ and $Z^0$
which are predicted by unified electroweak model are also found in
experiments. The $U(1)$ part of the unified electroweak model,
quantum electrodynamics, now become one of the most accurate and
best-tested theories of modern physics. All these achievements of
gauge field theories suggest that gauge field theory is a
fundamental theory that describes fundamental interactions. Now,
it is generally  believed that four kinds of fundamental
interactions in Nature are all gauge interactions and they can be
described by gauge field theory. From theoretical point of view,
the principle of local gauge invariance plays a fundamental role
in particle's interaction theory.  \\

Gauge treatment of gravity was suggested immediately after the
gauge theory birth itself\cite{b1,b2,b3}. In the traditional gauge
treatment of gravity, Lorentz group is localized, and the
gravitational field is not represented by gauge
potential\cite{b4,b5,b6}. It is represented by metric field. 
The theory has beautiful mathematical forms, but up to now, its
renormalizability is not proved. In other words,
it is conventionally considered to be non-renormalizable. 
There is also some other attempts to use Yang-Mills theory
to reformulate quantum gravity\cite{c1,c2,c3,c4,cc5}. In these new
approaches, the importance of gauge fields is emphasized. Some
physicists also try to use gauge potential to represent
gravitational field, some suggest that we should pay more
attention on translation group.
\\

I will not talk too much on the history of quantum gravity and
the incompatibilities between quantum mechanics and general
relativity here. Materials on these subject can be widely found
in literatures. Now we want to ask that, except for traditional
canonical quantum gravity and superstring theory, whether exists
another approach to set up a fundamental theory, in which general
relativity and quantum mechanics are compatible. \\

In literatures \cite{wu01,wu02,wu03},
a new formulation of quantum gauge theory of gravity is
proposed By N.Wu. In this new attempt, the quantum gravity 
is based on gauge principle, which is a renormalizable quantum 
theory. In literature \cite{wu02}, another model, which
uses different selections of $\eta_2$ and $J(C)$, is discussed.
In this paper, we will give a 
systematic formulation of the model which is discussed
in literature \cite{wu02}. The path integral quantization
of the theory and the proof the renormalizability of the 
theory are discussed. Because the selections of the Lagragian
are different, the proof on the renormalizability of the 
theory is different from that of the model in 
literature \cite{wu01}.  \\

As we have mentioned above, gauge field theory provides a
fundamental tool to study fundamental interactions. In this paper,
we will use this tool to study quantum gravity. We will use a
completely new language to express the quantum theory of gravity.
In order to do this, we first need to introduce some
transcendental foundations of this new theory, which is the most
important thing to formulate the whole theory. Then we will
discuss a new kind of non-Abel gauge group, which will be the
fundamental symmetry of quantum gravity. For the  sake of
simplicity, we call this group gravitational gauge group.  After
that, we will construct a Lagrangian which has local gravitational
gauge symmetry. In this Lagrangian, gravitational field appears as
the gauge field of the gravitational gauge symmetry. Then we will
discuss the gravitational interactions between scalar field (or
Dirac field or vector field ) and gravitational field. Just as
what Albert Einstein had ever said in 1916 that quantum effects
must lead to modifications in the theory of general relativity,
there are indeed quantum modifications in this new quantum gauge
theory of gravity. In other words, the local gravitational gauge
symmetry requires some additional interaction terms other than
those given by general relativity. This new quantum  theory of
gravity can even give out an exact relationship between
gravitational fields and space-time metric in generally
relativity. The classical limit of this new quantum theory will
give out classical Newton's theory of gravity and general
relativity. In other words, the leading order approximation of the
new theory gives out classical Newton's theory of gravity. 
It can also give out the Einstein's field equation with
cosmological contant. A rough estimation shows that the theoretical
expectation of cosmological contant is of the order of
$10^{-52} m^{-2}$, which is well consistent with experimental
results.  For classical tests, it gives out the same theoretical
predictions as those of general relativity.
Then we will discuss quantization of
gravitational gauge field. Something most important is that this
new quantum theory of gravity is a renormalizable theory. A formal
strict proof on the renormalizability of this new quantum theory
of gravity is given in this paper. I hope that the
effort made in this paper will be beneficial to our understanding
on the quantum aspects of gravitational field. The relationship
between this new quantum theory of gravity and traditional
canonical quantum gravity or superstring theory is not study now,
and I hope that this work will be done in the near future. Because
the new quantum theory of gravity is logically independent of
traditional quantum gravity, we need not discuss traditional
quantum gravity first. Anyone who is familiar with traditional
non-Abel gauge field theory can understand the whole paper. In
other words, readers who never study anything on traditional
quantum gravity can understand this new quantum theory of gravity.
\\

\section{The Transcendental Foundations}
\setcounter{equation}{0}

It is known that action principle is one of the most important
fundamental principle in quantum field theory. Action principle
says that any quantum system is described by an action. The action
of the system contains all interaction information, contains all
information of the fundamental dynamics. The least of the action
gives out the classical equation of motion of a field. Action
principle is widely used in quantum field theory. We will accept
it as one of the most fundamental principles in this new quantum
theory of gravity. The rationality of action principle will not be
discussed here, but it is well know that the rationality of the
action principle has already been tested by a huge amount of
experiments. However, this principle is not a special principle
for quantum gravity, it is a ubiquitous principle in quantum field
theory. Quantum gravity discussed in this paper is a kind of
quantum field theory, it's naturally to accept action principle as
one of its fundamental principles. \\

We need a special fundamental principle to introduce quantum
gravitational field, which should be the foundation of all kinds
of fundamental interactions in Nature. This special principle is
gauge principle. In order to introduce this important principle,
let's first study some fundamental laws in some fundamental
interactions other than gravitational interactions. We know that,
except for gravitational interactions, there are strong
interactions, electromagnetic interactions and weak interactions,
which are described by quantum chromodynamics,  quantum
electrodynamics and unified electroweak theory respectively. Let's
study these three fundamental interactions one by one. \\

Quantum electrodynamics (QED) is one of the most successful theory
in physics which has been tested by most accurate experiments. Let's
study some logic in QED. It is know that QED theory has $U(1)$
gauge symmetry. According to Noether's theorem, there is a
conserved charge corresponding to the global $U(1)$ gauge
transformations. This conserved charge is just the ordinary
electric charge. On the other hand, in order to keep local $U(1)$
gauge symmetry of the system, we had to introduce a $U(1)$ gauge
field, which transmits electromagnetic interactions. This $U(1)$
gauge field is just the well-know electromagnetic field. The
electromagnetic interactions between charged particles and the
dynamics of electromagnetic field are completely determined by the
requirement of local $U(1)$ gauge symmetry. The source of this
electromagnetic field is just the conserved charge which is given
by Noether's theorem. After quantization of the field, this
conserved charge becomes the generator of the quantum $U(1)$ gauge
transformations. The quantum $U(1)$ gauge transformation has only
one generator, it has no generator other than the quantum electric
charge.
\\

Quantum chromodynamics (QCD) is a prospective fundamental theory
for strong interactions. QCD theory has $SU(3)$ gauge symmetry.
The global $SU(3)$ gauge symmetry of the system gives out
conserved charges of the theory, which are usually called color
charges. The local $SU(3)$ gauge symmetry of the system requires
introduction of a set of $SU(3)$ non-Abel gauge fields, and the
dynamics of non-Abel gauge fields and the strong interactions
between color charged particles and gauge fields are completely
determined by the requirement of local $SU(3)$ gauge symmetry of
the system. These $SU(3)$ non-Abel gauge fields are usually call
gluon fields. The sources of gluon fields are color charges. After
quantization, these color charges become generators of quantum
$SU(3)$ gauge transformation. Something which is different from
$U(1)$ Abel gauge symmetry is that gauge fields themselves carry
color charges. \\

Unified electroweak model is the fundamental theory for
electroweak interactions. Unified electroweak model is usually
called the standard model. It has $SU(2)_L \times U(1)_Y$
symmetry. The global $SU(2)_L \times U(1)_Y$ gauge symmetry of the
system also gives out conserved charges of the system, The
requirement of local $SU(2)_L \times U(1)_Y$ gauge symmetry needs
introducing a set of $SU(2)$ non-Abel gauge fields and one $U(1)$
Abel gauge field. These gauge fields transmit weak interactions
and electromagnetic interactions, which correspond to intermediate
gauge bosons $W^{\pm}$, $Z^0$ and photon. The sources of these
gauge fields are just the conserved Noether charges. After
quantization, these conserved charges become generators of
quantum $SU(2)_L \times U(1)_Y$ gauge transformation. \\

QED, QCD and the standard model are three fundamental theories
of three kinds of fundamental interactions. Now we want to
summarize some fundamental laws of Nature on interactions.
Let's first ruminate over above discussions. Then we will find
that our formulations on three different fundamental interaction
theories are almost completely the same, that is the global gauge
symmetry of the system gives out conserved Noether charges,
in order to keep local gauge symmetry of the system, we have to
introduce gauge field or a set of gauge fields, these gauge
fields transmit interactions, and the source of these gauge fields are
the conserved charges and these conserved Noether charges
become generators of quantum gauge transformations after
quantization. These will be the main content of gauge principle.
\\

Before we formulate gauge principle formally, we need to study
something more on symmetry. It is know that not all symmetries can
be localized, and not all symmetries can be regarded as gauge
symmetries and have corresponding gauge fields. For example, time
reversal symmetry, space reflection symmetry, $\cdots$ are those
kinds of symmetries. We can not find any gauge fields or
interactions which correspond to these symmetries. It suggests
that symmetries can be divided into two different classes in
nature. Gauge symmetry is a special kind of symmetry which has the
following properties: 1) it can be localized; 2) it has some
conserved charges related to it; 3) it has a kind of interactions
related to it; 4) it is usually a continuous symmetry. This
symmetry can completely determine the dynamical behavior of a kind
of fundamental interactions. For the sake of simplicity, we call
this kind of symmetry dynamical symmetry or gauge symmetry. Any
kind of fundamental interactions has a gauge  symmetry
corresponding to it. In QED, the $U(1)$ symmetry is a gauge
symmetry, in QCD, the color $SU(3)$ symmetry is a gauge symmetry
and in the standard model, the $SU(2)_L \times U(1)_Y$ symmetry is
also a gauge symmetry. The gravitational gauge symmetry which we
will discuss below is also a kind of gauge symmetry. The time
reversal symmetry and space reflection symmetry are not gauge
symmetries. Those global symmetries which can not be localized are
not gauge symmetries either. Gauge symmetry is a fundamental
concept for gauge principle. \\

Gauge principle can be formulated as follows: Any kind of
fundamental interactions has a gauge symmetry corresponding to it;
the gauge symmetry completely determines the forms of
interactions. In principle, the gauge principle has the following
four different contents:
\begin{enumerate}

\item {\bf Conservation Law:} the global gauge symmetry  gives out
conserved current and conserved charge;

\item {\bf Interactions:} the requirement of the local gauge
symmetry requires introduction of gauge field or a set of gauge
fields; the interactions between gauge fields and matter fields
are completely determined by the requirement of local gauge
symmetry; these gauge fields transmit the corresponding
interactions;

\item {\bf Source:} qualitative speaking,
the conserved charge given by global gauge  symmetry is the source
of gauge field; for non-Abel gauge field, gauge field is also the
source of itself;

\item {\bf Quantum Transformation:} the conserved charges given
by global gauge symmetry become generators of quantum gauge
transformation after quantization, and for this kind of of
interactions, the quantum transformation can not have generators
other than quantum conserved charges given by global gauge
symmetry.

\end{enumerate}
It is known that conservation law is the objective origin of gauge
symmetry, so gauge symmetry is the exterior exhibition of the
interior conservation law. The conservation law is the law that
exists in fundamental interactions, so fundamental interactions
are the logic precondition and foundation of the conservation law.
Gauge principle tells us how to study conservation law and
fundamental interactions through symmetry. Gauge principle is one
of the most important transcendental fundamental principles for
all kinds of fundamental interactions in Nature; it reveals the
common nature of all kinds of fundamental interactions in Nature.
It is also the transcendental foundation of the quantum gravity
which is formulated in this paper. It will help us to select the
gauge symmetry for quantum gravitational theory and help us to
determine the Lagrangian of the system. In a meaning, we can say
that without gauge principle, we can not set up this new
renormalizable quantum gauge theory of gravity.
\\

Another transcendental principle that widely used in quantum field
theory is the microscopic causality principle. The central idea of the
causality principle is that any changes in the objective world have
their causation. Quantum field theory is a relativistic theory. It is know
that, in the special theory of relativity, the limit
spread speed is the speed
of light. It means that, in a definite reference system, the limit
spread speed
of the causation of some changes is the speed of light. Therefore,
the special theory of relativity exclude the possibility of the existence
of any kinds of non-local interactions in a fundamental theory. Quantum
field theory inherits this basic idea and calls it the microscopic causality
principle. There are several expressions of the microscopic causality
principle in quantum field theory. One expression say that two events
which happen at the same time but in different space position are two
independent events. The mathematical formulation for microscopic
causality principle is that
\be
[ O_1(\svec{x},t)~~,~~O_2(\svec{y},t) ] =0,
\label{2.1}
\ee
when $\svec{x} \not= \svec{y}$. In the above relation,
$ O_1(\svec{x},t)$ and $ O_2(\svec{y},t)$ are two different
arbitrary local bosonic operators. Another important expression
of the microscopic causality principle is that, in the Lagrangian
of a fundamental theory, all operators appear in the same point
of space-time. Gravitational interactions are a kind of physical
interactions, the fundamental theory of gravity should also obey
microscopic causality principle. This requirement is realized
in the construction of the Lagrangian for gravity. We will require
that all field operators in the Lagrangian should be at the same
point of space-time. \\

Because quantum field theory is a kind of relativistic theory, it
should obey some fundamental principles of the special theory of
relativity, such as principle of special relativity and principle
of invariance of light speed. These two principles conventionally
exhibit themselves through Lorentz invariance. So, in constructing
the Lagrangian of the quantum theory of gravity, we require that
it should have Lorentz invariance. This is also a transcendental
requirement for the quantum theory of gravity. But what we treat
here that is different from that of general relativity is that we
do not localize Lorentz transformation. Because gauge principle
forbids us to localize Lorentz transformation, asks us only to
localize gravitational gauge transformation. We will discuss this
topic in details later. However, it is important to remember that
global Lorentz invariance of the Lagrangian is a fundamental
requirement. The requirement of global Lorentz invariance can also
be treated as a transcendental principle of the quantum theory
of gravity.  \\

It is well-known that two transcendental principles of general
relativity are principle of general relativity and principle of
equivalence. It should be stated that, in the new gauge theory of
gravity, the principle of general relativity appears in  another
way, that is, it realized itself through local gravitational gauge
symmetry. From mathematical point of view, the local gravitational
gauge invariance is just the general covariance in general
relativity. In the new quantum theory of gravity, principle of
equivalence plays no role. In other words, we will not accept
principle of equivalence as a transcendental principle of the new
quantum theory of gravity, for gauge principle is enough for us to
construct quantum theory of gravity. We will discuss something
more about principle of equivalence later. \\

\section{Gravitational Gauge Group}
\setcounter{equation}{0}

Before we start our mathematical formulation of gravitational
gauge theory, we have to determine  which group is the
gravitational gauge group, which is the starting point of the
whole theory. It is know that, in the traditional quantum gauge
theory of gravity, Lorentz group is localized. We will not follow
this way, for it contradicts with gauge principle. Now, we use
gauge principle to determine which group is the exact group for
gravitational gauge theory. \\

Some of the most important properties of gravity can be seen
from Newton's classical gravity. In this classical theory of
gravity, gravitational force between two point objects is given by:
\be
f = G \frac{m_1 m_2}{r^2}
\label{3.1}
\ee
with $m_1$ and $m_2$ masses of two objects, $r$ the distance
between two objects. So, gravity is proportional to the masses of
both objects, in other words, mass is the source of gravitational
field. In general relativity, Einstein's gravitational equation is
the equation which gives out the relation between energy-momentum
tensor and space-time curvature, which is essentially the relation
between energy-momentum tensor and gravitational field. In the
Einstein's gravitational equation, energy-momentum is treated as
the source of gravity. This opinion is inherited in the new quantum
theory gauge of gravity. In other words, the starting point of
the new quantum gauge theory of gravity is that the energy-momentum is
the source of gravitational field. According to rule 3 and rule 1
of gauge principle, we know that, energy-momentum is the conserved
charges of the corresponding global symmetry, which is just the
symmetry for gravity. According to quantum field theory,
energy-momentum is the conserved charge of global space-time
translation, the corresponding conserved current is
energy-momentum tensor. Therefore, the global space-time
translation is the global gravitational gauge transformation.
According to rule 4, we know that, after quantization, the
energy-momentum operator becomes the generator of gravitational
gauge transformation. It also states that, except for energy-momentum
operator, there is no other generator for gravitational
gauge transformation, such as, angular momentum operator $M_{\mu
\nu}$ can not be the generator of gravitational gauge
transformation. This is the reason why we do not localize Lorentz
transformation in this new quantum gauge theory of gravity, for
the generator of Lorentz transformation is not energy-momentum
operator. According to rule 2 of gauge principle, the
gravitational interactions will be completely determined by the
requirement of the local gravitational gauge symmetry. These are
the basic ideas of the new quantum gauge theory of gravity, and
they are completely deductions of gauge principle. \\

We know that the generator of Lorentz group is angular momentum
operator $M_{\mu \nu}$. If we localize Lorentz group, according to
gauge principle, angular momentum will become source of a new
filed, which transmits direct spin interactions. This kind of
interactions does not belong to traditional Newton-Einstein
gravity. It is a new kind of interactions. Up to now, we do not
know that whether this kind of interactions exists in Nature or
not. Besides, spin-spin interaction is a kind of
non-renormalizable interaction. In other words, a quantum theory
which contains spin-spin interaction is a non-renormalizable
quantum theory. For these reasons, we will not localize Lorentz
group in this paper. We only localize translation group in this
paper. We will find that go along this way, we can set up a
consistent quantum gauge theory of gravity which is
renormalizable. In other words, only localizing space-time
translation group is enough for us to set up a
consistent quantum gravity. \\

From above discussions, we know that, from mathematical point of
view,  gravitational gauge transformation is the inverse
transformation of  space-time translation, and gravitational gauge
group is space-time translation group. Suppose that there is an
arbitrary function $\phi(x)$ of space-time coordinates $x^{\mu}$.
The global space-time translation is:
\be
x^{\mu} \to x'^{\mu} = x^{\mu} + \epsilon^{\mu}.
\label{3.2}
\ee
The corresponding transformation for function $\phi(x)$ is
\be
\phi(x) \to \phi'(x')=\phi(x) = \phi(x' - \epsilon).
\label{3.3}
\ee
According to Taylor series expansion, we have:
\be
\phi(x - \epsilon) = \left(1 + \sum_{n=1}^{\infty} \frac{(-1)^n}{n!}
\epsilon ^{\mu_1} \cdots \epsilon^{\mu^n}
\partial_{\mu_1} \cdots \partial_{\mu_n}  \right) \phi(x),
\label{3.4}
\ee
where
\be
\partial_{\mu_i}=\frac{\partial}{\partial x^{\mu_i}}.
\label{3.5}
\ee
\\

Let's define a special exponential operation here. Define
\be
E^{a^{\mu} \cdot b_{\mu}} \define 1 +
\sum_{n=1}^{\infty} \frac{1}{n!}
a^{\mu_1} \cdots a^{\mu_n} \cdot
b_{\mu_1} \cdots b_{\mu_n}.
\label{3.6}
\ee
This definition is quite different from that of ordinary
exponential function. In general cases, operators $a^{\mu}$ and
$b_{\mu}$ do not commutate each other, so
\be
E^{a^{\mu} \cdot b_{\mu}} \not=
E^{b_{\mu} \cdot a^{\mu}},
\label{3.7}
\ee
\be
E^{a^{\mu} \cdot b_{\mu}} \not=
e^{a^{\mu} \cdot b_{\mu}},
\label{3.8}
\ee
where $ e^{a^{\mu} \cdot b_{\mu}}$ is the ordinary exponential
function whose definition is
\be
e^{a^{\mu} \cdot b_{\mu}} \equiv 1 +
\sum_{n=1}^{\infty} \frac{1}{n!}
(a^{\mu_1} \cdot b_{\mu_1}) \cdots
(a_{\mu_n} \cdot b_{\mu_n}).
\label{3.9}
\ee
If operators $a^{\mu}$ and $b_{\mu}$ commutate each other, we will
have
\be
E^{a^{\mu} \cdot b_{\mu}}  =
E^{b_{\mu} \cdot a^{\mu}},
\label{3.10}
\ee
\be
E^{a^{\mu} \cdot b_{\mu}} =
e^{a^{\mu} \cdot b_{\mu}}.
\label{3.11}
\ee
The translation operator $\ehat$ can be defined by
\be
\ehat \equiv 1 + \sum_{n=1}^{\infty} \frac{(-1)^n}{n!}
\epsilon^{\mu_1} \cdots \epsilon^{\mu_n}
\partial_{\mu_1} \cdots \partial_{\mu_n}.
\label{3.12}
\ee
Then we have
\be
\phi(x - \epsilon) = ( \uhat_{ \epsilon} \phi(x)).
\label{3.13}
\ee
In order to have a good form which is similar to ordinary gauge
transformation operators, the form of $\ehat$ can also be
written as
\be
\ehat =  E^{- i \epsilon^{\mu} \cdot \hat{P}_{\mu}},
\label{3.14}
\ee
where
\be
\hat{P}_{\mu} = -i \frac{\partial}{\partial x^{\mu}}.
\label{3.15}
\ee
$\hat{P}_{\mu}$ is just the energy-momentum operator in
space-time coordinate space. In the definition of $\ehat$ of
eq.(\ref{3.14}), $\epsilon^{\mu}$ can be independent of of space-time
coordinate or a function of space-time coordinate, in a ward, it
can be any functions of space time coordinate $x$.  \\

Some operation properties of translation operator $\ehat$ are
summarized below.
\begin{enumerate}

\item Operator $\ehat$ translate the space-time point of a field
from $x$ to $x - \epsilon$,
\be
\phi(x- \epsilon) = (\ehat \phi(x)),
\label{3.16}
\ee
where $\epsilon^{\mu}$ can be any function of space-time
coordinate. This relation can also be regarded as the definition
of the translation operator $\ehat$.

\item If $\epsilon$ is a function of space-time coordinate, that is
$\partial_{\mu} \epsilon^{\nu} \not= 0$, then
\be
\ehat =  E^{- i \epsilon^{\mu} \cdot \hat{P}_{\mu}}
\not= E^{- i \hat{P}_{\mu} \cdot \epsilon^{\mu}  },
\label{3.17}
\ee
and
\be
\ehat =  E^{- i \epsilon^{\mu} \cdot \hat{P}_{\mu}}
\not= e^{- i \epsilon^{\mu} \cdot \hat{P}_{\mu}}.
\label{3.18}
\ee
If $\epsilon$ is a constant, that is
$\partial_{\mu} \epsilon^{\nu} = 0$, then
\be
\ehat =  E^{- i \epsilon^{\mu} \cdot \hat{P}_{\mu}}
= E^{- i \hat{P}_{\mu} \cdot \epsilon^{\mu}  },
\label{3.19}
\ee
and
\be
\ehat =  E^{- i \epsilon^{\mu} \cdot \hat{P}_{\mu}}
= e^{- i \epsilon^{\mu} \cdot \hat{P}_{\mu}}.
\label{3.20}
\ee

\item Suppose that $\phi_1(x)$ and $\phi_2(x)$ are two arbitrary
functions of space-time coordinate, then we have
\be
\left( \ehat (\phi_1(x) \cdot \phi_2(x))\right)=
(\ehat \phi_1(x)) \cdot (\ehat \phi_2(x))
\label{3.21}
\ee

\item Suppose that $A^{\mu}$ and $B_{\mu}$ are two arbitrary
operators in Hilbert space, $\lambda$ is an arbitrary ordinary
c-number which is commutate with operators $A^{\mu}$ and
$B_{\mu}$, then we have
\be
\frac{\rm d}{{\rm d} \lambda}
E^{\lambda A^{\mu} \cdot B_{\mu}} =
A^{\mu} \cdot E^{\lambda A^{\mu} \cdot B_{\mu}} \cdot B_{\mu}.
\label{3.22}
\ee

\item Suppose that $\epsilon$ is an arbitrary function of space-time
coordinate, then
\be
(\partial_{\mu} \ehat) =
-i (\partial_{\mu} \epsilon^{\nu} ) \ehat \hat{P}_{\nu}.
\label{3.23}
\ee

\item Suppose that $A^{\mu}$ and $B_{\mu}$ are two arbitrary
operators in Hilbert space, then
\be
tr( E^{ A^{\mu} \cdot B_{\mu}} E^{ - B_{\mu} \cdot A^{\mu}} )=
tr {\bf I},
\label{3.24}
\ee
where $tr$ is the trace operation and ${\bf I}$ is the unit operator
in the Hilbert space.

\item Suppose that $A^{\mu}$, $B_{\mu}$ and $C^{\mu}$ are three
operators in Hilbert space, but operators $A^{\mu}$ and $C^{\nu}$
commutate each other:
\be
[ A^{\mu}~~,~~C^{\nu} ] = 0,
\label{3.25}
\ee
then
\be
tr( E^{ A^{\mu} \cdot B_{\mu}} E^{ B_{\nu} \cdot C^{\nu}} )=
tr( E^{ (A^{\mu}+ C^{\mu}) \cdot B_{\mu}}).
\label{3.26}
\ee

\item Suppose that $A^{\mu}$, $B_{\mu}$ and $C^{\mu}$ are three
operators in Hilbert space, they satisfy
\be
\begin{array}{rcl}
\lbrack A^{\mu}~~,~~C^{\nu} \rbrack & = &  0 , \\
\lbrack B_{\mu}~~,~~C^{\nu} \rbrack & = & 0,
\end{array}
\label{3.27}
\ee
then
\be
E^{ A^{\mu} \cdot B_{\mu}} E^{ C^{\nu} \cdot B_{\nu}}  =
E^{ (A^{\mu}+ C^{\mu}) \cdot B_{\mu}}.
\label{3.28}
\ee

\item Suppose that $A^{\mu}$, $B_{\mu}$ and $C^{\mu}$ are three
operators in Hilbert space, they satisfy
\be
\begin{array}{rcl}
\lbrack A^{\mu}~~,~~C^{\nu} \rbrack & = & 0, \\
\lbrack  \lbrack B_{\mu}~~,~~C^{\nu}
\rbrack ~~,~~ A^{\rho} \rbrack &=& 0,  \\
\lbrack  \lbrack B_{\mu}~~,~~C^{\nu}
\rbrack ~~,~~ C^{\rho} \rbrack  &=& 0,
\end{array}
\label{3.29}
\ee
then,
\be
E^{ A^{\mu} \cdot B_{\mu}} E^{ C^{\nu} \cdot B_{\nu}}  =
E^{ (A^{\mu}+ C^{\mu}) \cdot B_{\mu}}
+ [E^{ A^{\mu} \cdot B_{\mu}} ~~,~~C^{\sigma}]
E^{ C^{\nu} \cdot B_{\nu}} B_{\sigma}.
\label{3.30}
\ee

\item Suppose that $\mhat{1}$ and $\mhat{2}$ are two arbitrary
translation operators, define
\be
\mhat{3}=\mhat{2} \cdot \mhat{1},
\label{3.31}
\ee
then,
\be
\epsilon_3^{\mu}(x) = \epsilon_2^{\mu}(x) +
\epsilon_1^{\mu}(x- \epsilon_2(x) ).
\label{3.32}
\ee
This property means that the product to two translation operator
satisfy closure property, which is one of the conditions that any group
must satisfy.

\item Suppose that $\ehat$ is a non-singular translation operator,
then
\be
\ehat^{-1} = E^{ i \epsilon^{\mu}(f(x)) \cdot \hat{P}_{\mu}},
\label{3.33}
\ee
where $f(x)$ is defined by the following relations:
\be
f(x- \epsilon (x)) = x.
\label{3.34}
\ee
$\ehat^{-1}$ is the inverse operator of $\ehat$, so
\be
\ehat^{-1} \ehat = \ehat \ehat^{-1} = {\rm\bf 1},
\label{3.35}
\ee
where {\bf 1} is the unit element of the gravitational gauge group.

\item The product operation of translation also satisfies associative
law. Suppose that $\mhat{1}$ , $\mhat{2}$ and $\mhat{3}$ are three
arbitrary translation operators, then
\be
\mhat{3} \cdot ( \mhat{2} \cdot \mhat{1} )
 = ( \mhat{3} \cdot \mhat{2} ) \cdot \mhat{1} .
\label{3.36}
\ee

\item Suppose that $\ehat$ is an arbitrary translation operator and
$\phi(x)$ is an arbitrary function of space-time coordinate, then
\be
\ehat \phi(x) \ehat^{-1} = f(x- \epsilon(x)).
\label{3.37}
\ee
This relation is quite useful in following discussions.

\item Suppose that $\ehat$ is an arbitrary translation operator.
Define
\be
\Lambda^{\alpha}_{~~\beta} =
\frac{\partial x^{\alpha}}{\partial ( x - \epsilon (x) )^{\beta}},
\label{3.38}
\ee
\be
\Lambda_{\alpha}^{~~\beta} =
\frac{\partial ( x - \epsilon (x))^{\beta}}{\partial x^{\alpha}}.
\label{3.39}
\ee
They satisfy
\be
\Lambda_{\alpha}^{~~ \mu} \Lambda^{\alpha}_{~~ \nu}
= \delta^{\mu}_{\nu},
\label{3.40}
\ee
\be
\Lambda_{\mu}^{ ~~\alpha} \Lambda^{\nu}_{~~ \alpha}
= \delta_{\mu}^{\nu}.
\label{3.41}
\ee
Then we have following relations:
\be
\ehat \hat{P}_{\alpha} \ehat ^{-1}
= \Lambda^{\beta}_{~~\alpha}  \hat{P}_{\beta},
\label{3.42}
\ee
\be
\ehat {\rm d}x^{\alpha} \ehat ^{-1}
= \Lambda_{\beta}^{~~\alpha}  {\rm d}x^{\beta}.
\label{3.43}
\ee
These give out the the transformation laws of  $\hat{P}_{\alpha}$
and d$x^{\alpha}$ under local gravitational gauge transformations.

\end{enumerate}

Gravitational gauge group (GGG) is a transformation group which
consists of all non-singular translation operators $\ehat$. We can
easily see that gravitational gauge group is indeed a  group, for
\begin{enumerate}

\item the product of two arbitrary non-singular translation operators
is also a non-singular translation operator, which is also an element
of the gravitational gauge group. So, the product of the group satisfies
closure property which is expressed in eq(\ref{3.31});

\item the product of the gravitational gauge group also satisfies the
associative law which is expressed in eq(\ref{3.36});

\item the gravitational gauge group has
its unit element {\bf 1}, it satisfies
\be
{\rm\bf 1} \cdot \ehat = \ehat \cdot {\rm\bf 1} = \ehat;
\label{3.44}
\ee

\item every non-singular element $\ehat$ has its inverse element
which is given by eqs(\ref{3.33}) and (\ref{3.35}).

\end{enumerate}
According to gauge principle, the gravitational gauge group is the symmetry
of gravitational interactions. The global invariance of gravitational gauge
transformation will give out conserved charges which is just the ordinary
energy-momentum; the requirement of local gravitational gauge invariance
needs introducing gravitational gauge field, and gravitational
interactions are completely determined by the local gravitational gauge
invariance. \\

The generators of gravitational gauge group is just the energy-momentum
operators $\hat{P}_{\alpha}$.  This is required by gauge principle. It can
also be seen from the form of infinitesimal transformations. Suppose that
$\epsilon$ is an infinitesimal quantity, then we have
\be
\ehat \simeq 1 - i \epsilon^{\alpha} \hat{P}_{\alpha}.
\label{3.45}
\ee
Therefore,
\be
i \frac{\partial \ehat}{\partial \epsilon^{\alpha}} \mid _{\epsilon = 0}
\label{3.46}
\ee
gives out  generators $\hat{P}_{\alpha}$ of gravitational gauge group.
It is known  that generators of gravitational gauge
group commute each other
\be
\lbrack \hat{P}_{\alpha} ~~,~~ \hat{P}_{\beta} \rbrack = 0.
\label{3.47}
\ee
However, the commutation property of generators does not mean that
gravitational gauge group is an Abel group, because two general elements
of gravitational gauge group do not commute:
\be
\lbrack \mhat{1} ~~,~~ \mhat{2} \rbrack \not= 0.
\label{3.48}
\ee
Gravitational gauge group is a kind of non-Abel gauge group. The
non-Able nature of  gravitational gauge group will cause
self-interactions of gravitational gauge field. \\

In order to avoid confusion, we need to pay some attention to some
differences between two concepts: space-time translation group and
gravitational gauge group. Generally speaking, space-time
translation is a kind of coordinates transformation, that is, the
objects or fields in space-time are kept fixed while the
space-time coordinates that describe the motion of objective
matter undergo transformation. But gravitational gauge
transformation is a kind of system transformation rather than a
kind of coordinates transformation. In system transformation, the
space-time coordinate system is kept unchanged while objects or
fields undergo transformation. From mathematical point of view,
space-time translation and gravitational gauge transformation are
essentially the same, and the space-time translation is the
inverse transformation of the gravitational gauge transformation;
but from physical point of view, space-time translation and
gravitational gauge transformation are quite different, especially
when we discuss gravitational gauge transformation of
gravitational gauge field. For gravitational gauge field, its
gravitational gauge transformation is not the inverse
transformation of its space-time translation. In a meaning,
space-time translation is a kind of mathematical transformation,
which contains little dynamical information of interactions; while
gravitational gauge transformation is a kind of physical
transformation, which contains all dynamical information of
interactions and is convenient for us to study physical
interactions. Through gravitational gauge symmetry, we can
determine the whole gravitational interactions among various kinds
of fields. This is the reason why we do not call gravitational
gauge transformation space-time translation. This is important for
all of our discussions on gravitational
gauge transformations of various kinds of fields. \\

Suppose that $\phi(x)$ is an arbitrary scalar field. Its gravitational
gauge transformation is
\be
\phi(x) \to \phi'(x) = ( \ehat \phi(x)).
\label{3.49}
\ee
Similar to ordinary $SU(N)$ non-able gauge field theory, there are
two kinds of scalars. For example, in chiral perturbative theory, the
ordinary $\pi$ mesons are scalar fields, but they are vector fields in
isospin space. Similar case exists in gravitational gauge field theory.
A Lorentz scalar can be a scalar or a vector or a tensor in the space
of gravitational gauge group. If $\phi (x)$ is a scalar in the space of
gravitational gauge group, we just simply denote it as $\phi (x)$ in
gauge group space. If it is a vector in the space of gravitational
gauge group, it can be expanded in the gravitational gauge group
space in the following way:
\be
\phi(x)  =  \phi^{\alpha}(x) \cdot \hat{P}_{\alpha}.
\label{3.50}
\ee
The transformation of component field is
\be
\phi^{\alpha}(x)  \to \phi^{\prime  \alpha}(x) =
\Lambda^{\alpha}_{~~\beta} \ehat \phi^{\beta}(x) \ehat^{-1} .
\label{3.51}
\ee
The important thing that we must remember is that, the $\alpha$
index is not a Lorentz index, it is just a group index. For
gravitation gauge group, it is quite special that a group index looks
like a Lorentz index. We must be carefully on this important thing.
This will cause some fundamental changes on quantum gravity.
Lorentz scalar $\phi(x)$ can also be a tensor in gauge group
space. suppose that it is a $n$th order tensor in gauge group
space, then it can be expanded as
\be
\phi(x)  =  \phi^{\alpha_1 \cdots \alpha_n}(x)
\cdot \hat{P}_{\alpha_1}  \cdots \hat{P}_{\alpha_n}.
\label{3.52}
\ee
The transformation of component field is
\be
\phi^{\alpha_1 \cdots \alpha_n}(x)  \to
\phi ^{\prime \alpha_1 \cdots \alpha_n}(x) =
\Lambda^{\alpha_1}_{~~\beta_1} \cdots \Lambda^{\alpha_n}_{~~\beta_n}
\ehat \phi^{\beta_1 \cdots \beta_n}(x) \ehat^{-1} .
\label{3.53}
\ee
\\

If $\phi (x)$ is a spinor field, the above discussion is also valid.
That is, a spinor can also be a scalar or a vector or a tensor in
the space of gravitational gauge group. The gravitational
gauge transformations of the component fields are also given by
eqs.(\ref{3.49}-\ref{3.53}). 
There is no transformations in spinor space, which
is different from that of the Lorentz transformation of a spinor.  \\

Suppose that $A_{\mu} (x)$ is an arbitrary vector field. Here,
the index $\mu$ is a Lorentz index. Its gravitational gauge
transformation is:
\be
A_{\mu}(x) \to A'_{\mu}(x) = ( \ehat A_{\mu}(x)).
\label{3.54}
\ee
Please remember that there is no rotation in the space of Lorentz
index $\mu$, while in the general coordinates transformations of
general relativity, there is rotation in the space of Lorentz index
$\mu$. The reason is that gravitational gauge transformation is a
kind of system transformation, while in general relativity, the
general coordinates transformation is a kind of coordinates
transformation. If $A_{\mu}(x)$ is a scalar in the space of
gravitational gauge group, eq(\ref{3.54}) is all for its gauge
transformation. If $A_{\mu}(x)$ is a vector in the space of
gravitational gauge group, it can be expanded as:
\be
A_{\mu}(x)  =  A_{\mu}^{\alpha}(x) \cdot \hat{P}_{\alpha}.
\label{3.55}
\ee
The transformation of component field  is
\be
A_{\mu}^{\alpha}(x)  \to A_{ \mu}^{\prime\alpha}(x) =
\Lambda^{\alpha}_{~~\beta} \ehat A_{\mu}^{~~\beta}(x) \ehat^{-1} .
\label{3.56}
\ee
If $A_{\mu}(x)$ is a $n$th order tensor in the space of
gravitational gauge group, then
\be
A_{\mu}(x)  =  A_{\mu}^{\alpha_1 \cdots \alpha_n}(x)
\cdot \hat{P}_{\alpha_1}  \cdots \hat{P}_{\alpha_n}.
\label{3.57}
\ee
The transformation of component fields is
\be
A_{\mu}^{\alpha_1 \cdots \alpha_n}(x)  \to
A_{\mu}^{\prime\alpha_1 \cdots \alpha_n}(x) =
\Lambda^{\alpha_1}_{~~\beta_1} \cdots \Lambda^{\alpha_n}_{~~\beta_n}
\ehat A_{\mu}^{\beta_1 \cdots \beta_n}(x) \ehat^{-1} .
\label{3.58}
\ee
Therefore, under gravitational gauge transformations, the behavior
of a group index is quite different from that of a Lorentz index. However,
they have the same behavior in global Lorentz transformations. \\

Generally speaking, suppose that $T^{\mu_1 \cdots \mu_n}
_{\nu_1 \cdots \nu_m} (x)$ is an arbitrary tensor, its
gravitational gauge transformations are:
\be
T^{\mu_1 \cdots \mu_n}_{\nu_1 \cdots \nu_m} (x)
\to T'^{\mu_1 \cdots \mu_n}_{\nu_1 \cdots \nu_m} (x)
 = ( \ehat T^{\mu_1 \cdots \mu_n}_{\nu_1 \cdots \nu_m} (x)).
\label{3.59}
\ee
If it is a $p$th order tensor in group space, then
\be
T^{\mu_1 \cdots \mu_n}_{\nu_1 \cdots \nu_m} (x)
= T^{\mu_1 \cdots \mu_n ; \alpha_1 \cdots \alpha_p}
_{\nu_1 \cdots \nu_m} (x)
\cdot \hat{P}_{\alpha_1}  \cdots \hat{P}_{\alpha_p}.
\label{3.60}
\ee
The transformation of component fields is
\be
T^{\mu_1 \cdots \mu_n ; \alpha_1 \cdots \alpha_p}
_{\nu_1 \cdots \nu_m} (x)  \to
T'^{\mu_1 \cdots \mu_n ; \alpha_1 \cdots \alpha_p}
_{\nu_1 \cdots \nu_m} (x) =
\Lambda^{\alpha_1}_{~~\beta_1} \cdots \Lambda^{\alpha_p}_{~~\beta_p}
\ehat T^{\mu_1 \cdots \mu_n ; \beta_1 \cdots \beta_p}
_{\nu_1 \cdots \nu_m} (x)\ehat^{-1}.
\label{3.61}
\ee
\\

$\eta^{\mu \nu}$ is a second order Lorentz tensor, but it is a scalar in
group space. It is the metric of the coordinate space. A Lorentz
index can be raised or descended by this metric tensor.
In a special coordinate system, it is selected to be:
\be
\begin{array}{rcl}
\eta^{0~0} &=& -1,  \\
\eta^{1~1} &=&  1,  \\
\eta^{2~2} &=&  1,  \\
\eta^{3~3} &=&  1,
\end{array}
\label{3.73}
\ee
and other components of $\eta^{\mu \nu}$ vanish. $\eta^{\mu \nu}$
is the traditional Minkowski metric.
\\

\section{Physics Picture of Gravity}
\setcounter{equation}{0}
 
As we have mentioned above, quantum gauge theory of gravity
is logically independent of traditional quantum gravity. It
is know that, there are at least two pictures of gravity.
In one picture, gravity is treated as space-time geometry.
In this picture, space-time is curved and there is no
physical gravitational interactions, for all effects of
gravity are represented by space-time metric. In another
picture, gravity is treated as a kind of fundamental
interactions. In this picture, space-time is always flat
and space-time metric is always selected to be the
Minkowski metric. For the sake of simplicity, we call
the first picture gemeotry picture of gravity and the
second picture physics picture of gravity. \\

The concepts of "physics picture of gravity" and "geometrical
picture of gravity" are key important to understand the
present theory. In order to understand these important
things, I use quantum mechanics as an example.
In quantum mechanics, there are many pictures, such Schrodinger
picture, Heisenberg picture, $\cdots$ etc. In Schrodinger
picture, operators of physical quantities are fixed and do not
change with time, but wave functions are evolve with time.
On the contrary, in Heisenberg picture, wave functions are
 fixed and do not change with time, but operators evolve
with time. If we want to know whether wave functions is  changed
with time or not, you must first determine in which picture
you study wave functions. If you do not know in which picture
you study wave functions, you will not know whether
wave functions should be changed with time or not.
Now, similar case happens in quantum gauge theory of
gravity. If you want to know whether space-time
is curved or not, you must first determine in which
picture gravity is studied. In physics picture of
gravity, space-time is flat, but in geometry picture
of gravity, space-time is curved. Quantum gauge theory
of gravity is formulated in the physics picture of gravity,
classical Newton's theory of gravity is also
formulated in the physics picture of gravity,
and general relativity is formulated in the geometry
picture of gravity. Please do not discuss any
problem simultaneous in two pictures, which is dangerous.\\

Quantum gauge theory of gravity is foumulated in the
physics picture of gravity. So, in quantum gauge theory
of gravity, space-time is always flat and gravity is
treated as a kind of fundatmental interactions. In
order to avoid confusing, we do
not introduce any comcept of curved space-time
and we do not use any language of geometry
in this paper. It is suggest that anyone read this
paper do not try to find any geometrical meaning
of any physical quantities, do not use the language
of geometry to understand anything of this paper
and forget everything about the concept of fibre bundles,
connections, curved space-time metric, $\cdots$ etc,
for the present theory is not formulated in the
geometry picture of gravity. After we go into the
geometry picture of gravity and set up the geometrical
picture of quantum gauge theory of gravity, we
can use geometry language and study the geometry
meaning of the present theory. But at present, we
will not use the language of geometry. 
\\

There are mainly the following three reasons to
introduce the physics picture of gravity and formulate
quantun gauge theory of gravity in the physics
picture of gravity:
\begin{enumerate}

\item   It has a clear interaction picture, so we
        can use perturbation theory to calculate
        the amplitudes of physical process.
 
\item   We can use traditional gauge field theory to
        study quantum behavior of gravitational interactions,
        and four different kinds of fundamental interactions
        in Nature can be formulated in the same manner.
 
\item   The perturbatively renormalizability of the theory can be
        easily proved in the physics picture of gravity.\\
 
\end{enumerate}

Gravitational gauge transformation is different from space-time
translation. In gravitational gauge transformation, space-time
is fixed, space-time coordinates are not changed, only fields
and objects undergo some translation. In a meaning, gravitational
gauge transformation is a kind of physical transformation on
objects and fields. The traditional space-time translation
is a kind of transformation in which objects and fields are
kept unchanged while space-time coordinates undergo some
translation. In a meaning, space-time translation is a kind
of geometrical transformation on space-time. Because quantum
gauge theory of gravity is set up in the physics picture
of gravity, we have to use gravitational gauge transformation
in our discussion, for physics picture needs physical
transformation. I do not discuss translation transformation
of space-time and gauge translations in physics picture of
gravity.
In a meaning, space-time translation
is a kind of geometrical transformation,
which contains  geometrical information of
space-time structure and is convenient for us to study
space-time geometry; while
gravitational gauge transformation is a kind of physical
transformation, which contains all dynamical information of
gravitational interactions and is convenient
for us to study physical interactions.
Though from mathematical point of view, for global transformations,
space-time translation is the inverse transformation of
the gravitational gauge transformation. But from physical
point of view and for local transformation, they are not
the same. In the new theory, I do not gauge translation
group, but gauge gravitational gauge group, for translation
group is different from gravitational gauge group.
Translation group is the symmetry of space-time, but
 gravitational gauge group is the symmetry group
of physical fields and objects. They have essential difference
from physical point of view. However, in geometry picture
of gravity, the space-time transformation is used. 
\\

\section{Pure Gravitational Gauge Fields}
\setcounter{equation}{0}

Before we study gravitational field, we must determine which field
represents gravitational field. In the traditional gravitational
gauge theory, gravitational field is represented by space-time
metric tensor. If there is gravitational field in space-time, the
space-time metric will not be equivalent to Minkowski metric, and
space-time will become curved. In other words, in the traditional
gravitational gauge theory, quantum gravity is formulated in
curved space-time. In this paper, we will not follow this way. The
underlying point of view of this new quantum gauge theory of
gravity is that it is formulated in the framework of traditional 
quantum field theory, gravity is treated as a kind of physical
interactions in flat space-time and the 
gravitational field is represented by gauge
potential.  In other words, if we put gravity into the structure
of space-time, the space-time will become curved and there will be
no physical gravity in space-time, because all gravitational
effects are put into space-time metric and gravity is geometrized.
But if we study physical gravitational interactions, it is better
to rescue gravity from space-time metric and treat gravity as a
kind of 
physical interactions. In this case, space-time is flat and there is
physical gravity in Minkowski space-time. For this reason, we will
not introduce the concept of curved space-time to study quantum
gravity in first twelve chapters of 
this paper. So, in the first twelve chapters of this paper, 
the space-time is always flat,
gravitational field is represented by gauge potential and
gravitational interactions are always treated as physical
interactions. 
In fact, what gravitational field is represented by gauge 
potential is required by gauge principle.
\\

Now, let's begin to construct the Lagrangian of 
gravitational gauge theory. For the sake of simplicity, 
let's suppose that  $\phi (x)$ is a Lorentz scalar and 
gauge group scalar. According to above discussions, 
its gravitational gauge transformation is:
\be
\phi (x) \to \phi '(x) = (\ehat \phi (x)).
\label{4.1}
\ee
Because
\be
(\partial_{\mu} \ehat ) \not= 0,
\label{4.2}
\ee
partial differential of $\phi (x)$ does not transform covariantly
under gravitational gauge transformation:
\be
\partial_{\mu} \phi (x) \to \partial_{\mu} \phi '(x)
\not= ( \ehat \partial_{\mu} \phi (x) ).
\label{4.3}
\ee
In order to construct an action which is invariant under local
gravitational gauge transformation, gravitational gauge
covariant derivative is highly necessary. The gravitational gauge
covariant derivative is defined by
\be
D_{\mu} = \partial_{\mu} - i g C_{\mu} (x),
\label{4.4}
\ee
where $C_{\mu} (x)$ is the gravitational gauge field. It is a Lorentz
vector. Under gravitational gauge transformations, it transforms as
\be
C_{\mu}(x) \to  C'_{\mu}(x) =
\ehat (x) C_{\mu} (x) \ehat^{-1} (x)
+ \frac{i}{g} \ehat (x) (\partial_{\mu} \ehat^{-1} (x)).
\label{4.5}
\ee
Using the original definition of $\ehat$, we can strictly proved
that
\be
\lbrack \partial_{\mu} ~~,~~ \ehat \rbrack
= (\partial_{\mu} \ehat).
\label{4.6}
\ee
Therefor, we have
\be
\ehat \partial_{\mu} \ehat^{-1} =
\partial_{\mu}  + \ehat (\partial_{\mu} \ehat^{-1}),
\label{4.7}
\ee
\be
\ehat D_{\mu} \ehat^{-1} =
\partial_{\mu} - i g  C'_{\mu} (x).
\label{4.8}
\ee
So, under local gravitational gauge transformations,
\be
D_{\mu} \phi (x) \to
D'_{\mu} \phi '(x) =
(\ehat D_{\mu} \phi (x)) ,
\label{4.9}
\ee
\be
D_{\mu} (x) \to D'_{\mu} (x)
= \ehat D_{\mu} (x) \ehat^{-1}.
\label{4.10}
\ee
\\

Gravitational gauge field $C_{\mu} (x)$ is vector field, it is a
Lorentz vector. It is also a vector in gauge group space, so it can be
expanded as linear combinations of generators of gravitational gauge
group:
\be
C_{\mu} (x) = C_{\mu}^{\alpha}(x) \cdot \hat{P}_{\alpha}.
\label{4.11}
\ee
$C_{\mu}^{\alpha}$ are component fields of gravitational gauge
field. It looks like a second rank tensor. But according to our
previous discussion, it is not a tensor field, it is a vector
field. The index $\alpha$ is not a Lorentz index, it is just a gauge
group index. Gravitational gauge field $C_{\mu}^{\alpha}$ has only
one Lorentz index, so it is a kind of vector field. This is a
result of gauge principle. The gravitational gauge transformation  of
component field is:
\be
C_{\mu}^{\alpha}(x) \to C_{\mu}^{\prime \alpha}(x)=
\Lambda ^{\alpha}_{~~\beta} (\ehat C_{\mu}^{\beta}(x))
- \frac{1}{g} (\ehat  \partial_{\mu} \epsilon^{\alpha}(y)),
\label{4.12}
\ee
where $y$ is a function of space-time coordinates which satisfy:
\be
( \ehat y(x) )=x.
\label{4.13}
\ee
\\

Define matrix $G$ as
\be  \label{4.701}
G = (G_{\mu}^{\alpha}) = ( \delta_{\mu}^{\alpha} - g C_{\mu}^{\alpha}),
\ee
where $C_{\mu}^{\alpha}$ is the gravitational gauge field which
will be introduced below.  A simple form for matrix $G$ is
\be  \label{4.702}
G = I - gC,
\ee
where $I$ is a unit matrix and $C= (C_{\mu}^{\alpha})$. Therefore,
\be  \label{4.703}
G^{-1} = \frac{1}{I-gC}.
\ee
$G^{-1}$ is the inverse matrix of $G$, it satisfies
\be  \label{4.704}
(G^{-1})^{\mu}_{\beta} G^{\alpha}_{\mu}
= \delta_{\beta}^{\alpha},
\ee
\be  \label{4.705}
G^{\alpha}_{\mu} (G^{-1})^{\nu}_{\alpha}
= \delta^{\nu}_{\mu}.
\ee
Define
\be  \label{4.706}
g^{\alpha \beta} \define \eta^{\mu \nu}
G_{\mu}^{\alpha} G_{\nu}^{\beta}.
\ee
\be  \label{4.707}
g_{\alpha \beta} \define \eta_{\mu \nu}
(G^{-1})_{\alpha}^{\mu} (G^{-1})_{\beta}^{\nu}.
\ee
It can be easily proved that
\be  \label{4.708}
g_{\alpha \beta} g^{\beta \gamma} = \delta_{\alpha}^{\gamma},
\ee
\be  \label{4.708}
g^{\alpha \beta} g_{\beta \gamma} = \delta^{\alpha}_{\gamma}.
\ee
Under gravitatinal gauge transformations, they transform
as
\be  \label{4.709}
g_{\alpha \beta}(x)  \to g'_{\alpha \beta}(x)
= \Lambda_{\alpha}^{~\alpha_1} \Lambda_{\beta}^{~\beta_1}
(\ehat g_{\alpha_1 \beta_1}(x)),
\ee
\be  \label{4.710}
g^{\alpha \beta}(x)  \to g^{\prime \alpha \beta}(x)
= \Lambda^{\alpha}_{~\alpha_1} \Lambda^{\beta}_{~\beta_1}
(\ehat g^{\alpha_1 \beta_1}(x)),
\ee
\\

The strength of gravitational gauge field is defined by
\be
F_{\mu \nu} = \frac{1}{-i g}
\lbrack D_{\mu} ~~,~~ D_{\nu} \rbrack,
\label{4.14}
\ee
or
\be
F_{\mu \nu} = \partial_{\mu} C_{\nu}(x)
- \partial_{\nu} C_{\mu}(x)
- i g C_{\mu}(x) C_{\nu}(x)
+ i g  C_{\nu}(x) C_{\mu}(x).
\label{4.15}
\ee
$F_{\mu \nu}$ is a second order Lorentz tensor. It is a vector is
group space, so it can be expanded in group space,
\be
F_{\mu \nu} (x) = F_{\mu \nu}^{\alpha} (x) \cdot \hat{P}_{\alpha}.
\label{4.16}
\ee
The explicit form of component field strengths is
\be
F_{\mu \nu}^{\alpha} = \partial_{\mu} C_{\nu}^{\alpha}
- \partial_{\nu} C_{\mu}^{\alpha}
- g C_{\mu}^{\beta} \partial_{\beta} C_{\nu}^{\alpha}
+ g C_{\nu}^{\beta} \partial_{\beta} C_{\mu}^{\alpha}
\label{4.17}
\ee
The strength of gravitational gauge fields transforms covariantly
under gravitational gauge transformation:
\be
F_{\mu \nu} \to F'_{\mu \nu} =
\ehat F_{\mu \nu} \ehat^{-1}.
\label{4.18}
\ee
The gravitational gauge transformation of the 
component field strength  is
\be
F_{\mu \nu}^{\alpha} \to F_{\mu \nu}^{\prime \alpha} =
\Lambda^{\alpha}_{~~\beta} (\ehat F_{\mu \nu}^{\beta}).
\label{4.19}
\ee
\\

Similar to traditional gauge field theory, the kinematical term for
gravitational gauge field can be selected as
\be
{\cal L}_0 = - \frac{1}{4} \eta^{\mu \rho} \eta^{\nu \sigma}
g_{ \alpha \beta}
F_{\mu \nu}^{\alpha} F_{\rho \sigma}^{\beta}.
\label{4.20}
\ee
We can easily prove that this Lagrangian does not invariant under
gravitational gauge transformation, it transforms covariantly
\be
{\cal L}_0 \to {\cal L}'_0 = ( \ehat {\cal L}_0).
\label{4.21}
\ee
\\

In order to resume the gravitational gauge symmetry of the action,
we introduce an extremely important factor, whose form is
\be
J(C) = \sqrt{- {\rm det} g_{\alpha \beta} },
\label{4.22a}
\ee
where $g_{\alpha \beta}$ is given by eq.(\ref{4.707}).
The gravitational gauge transformations of $g_{\alpha\beta}$
is given by eq.(\ref{4.709}). Then $J(C)$ transforms as
\be
J(C) \to J'(C') = J \cdot (\ehat J(C)),
\ee
where $J$ is the Jacobian of the transformation,
\be
J = det \left(\frac{\partial (x - \epsilon)^{\mu}}
{\partial x^{\nu}} \right).
\label{4.2301}
\ee
The Lagrangian for gravitational gauge field is selected as
\be
{\cal L} = J(C) {\cal L}_0 =  \sqrt{- {\rm det} g_{\alpha \beta} }
\cdot {\cal L}_0, 
\label{4.24}
\ee
and the action for gravitational gauge field is
\be
S = \int {\rm d}^4 x {\cal L}.
\label{4.25}
\ee
It can be proved that this action has gravitational gauge symmetry.
In other words, it is invariant under gravitational gauge
transformation,
\be
S \to S' = S.
\label{4.26}
\ee
In order to prove the gravitational gauge symmetry of the action,
the following relation is important,
\be
\int {\rm d}^4 x J \left(\ehat f(x) \right) 
= \int {\rm d}^4 x f(x),
\label{4.27}
\ee
where $f(x)$ is an arbitrary function of space-time coordinate.
\\

According to gauge principle, the global gauge symmetry will give out
conserved charges. Now, let's discuss the conserved charges of
global gravitational gauge transformation. Suppose that
$\epsilon^{\alpha}$ is an infinitesimal constant 4-vector. Then,
in the first order approximation, we have
\be
\ehat = 1 - \epsilon^{\alpha} \partial_{\alpha} + o(\epsilon^2).
\label{4.29}
\ee
The first order variation of the gravitational gauge field is
\be
\delta C_{\mu}^{\alpha} (x) =
- \epsilon^{\nu} \partial_{\nu} C_{\mu}^{\alpha},
\label{4.30}
\ee
and the first order variation of action is:
\be
\delta S = \int {\rm d}^4 x ~ \epsilon^{\alpha} \partial_{\mu}
T_{i \alpha}^{\mu},
\label{4.31}
\ee
where $T_{i \alpha}^{\mu}$ is the inertial energy-momentum
tensor, whose definition is
\be
T_{i \alpha}^{\mu} \equiv J(C)
\left( - \frac{\partial {\cal L}_0}
{\partial \partial_{\mu} C_{\nu}^{\beta}}
\partial_{\alpha} C_{\nu}^{\beta} 
+ \delta^{\mu}_{\alpha} {\cal L}_0 \right).
\label{4.32}
\ee
It is a conserved current,
\be
\partial_{\mu} T_{i \alpha}^{\mu} = 0.
\label{4.33}
\ee
Except for the factor $J(C)$, the form of the inertial
energy-momentum tensor is almost completely the same as that
in the traditional quantum field theory. It means that gravitational
interactions will change energy-momentum of matter fields, which
is what we expected in general relativity.\\

The Euler-Lagrange equation for gravitational gauge field is
\be
\partial_{\mu} \frac{\partial \cal L}
{\partial \partial_{\mu} C_{\nu}^{\alpha}}
= \frac{\partial \cal L}{\partial C_{\nu}^{\alpha}}.
\label{4.34}
\ee
This form is completely the same as what we have ever seen in
quantum field theory. But if we insert eq.(\ref{4.24}) into it, 
we will get
\be
\partial_{\mu} \frac{\partial {\cal L }_0}
{\partial \partial_{\mu} C_{\nu}^{\alpha}}
= \frac{\partial {\cal L}_0}{\partial C_{\nu}^{\alpha}}
+g G_{\alpha}^{-1 \nu} {\cal L}_0 
- g G_{\alpha}^{-1 \nu} 
(  \partial_{\mu} C_{\nu}^{\alpha})
\frac{\partial {\cal L}_0}{\partial \partial_{\mu} C_{\nu}^{\alpha}}.
\label{4.35}
\ee
Eq.(\ref{4.17}) can be changed into
\be
F_{\mu \nu}^{\alpha} =
(D_{\mu} C_{\nu}^{\alpha})
- (D_{\nu} C_{\mu}^{\alpha}) ,
\label{4.36}
\ee
so the Lagrangian ${\cal L}_0$ depends on gravitational gauge fields
completely through its covariant derivative. Therefore,
\be
\frac{\partial {\cal L}_0}{\partial C_{\nu}^{\alpha}}
= - g \frac{\partial {\cal L}_0}{\partial D_{\nu} C_{\mu}^{\beta}}
\partial_{\alpha} C_{\mu}^{\beta}
+ \frac{g}{2} \eta^{\mu\rho} \eta^{\lambda\sigma}
g_{\alpha\beta} G^{-1\nu}_{\gamma}
F^{\gamma}_{\mu\lambda} F^{\beta}_{\rho\sigma}.
\label{4.37}
\ee
Using the above relations and
\be
\frac{\partial {\cal L}_0}{\partial \partial_{\mu} C_{\nu}^{\alpha}}
= - \eta^{\mu \lambda} \eta^{\nu \tau} \eta_{2 \alpha \beta}
F_{\lambda \tau}^{\beta}
- g \eta^{\nu \lambda} \eta^{\sigma \tau} \eta_{2 \alpha \beta}
F_{\lambda \tau}^{\beta} C_{\sigma}^{\mu},
\label{4.38}
\ee
the above equation of motion of gravitational gauge fields
are changed into
\be
\partial_{\mu} (\eta^{\mu \lambda} \eta^{\nu \tau} g_{\alpha \beta}
F_{\lambda \tau}^{\beta} )  = - g T_{g \alpha}^{\nu},
\label{4.41}
\ee
where
\be  \label{3.25}
\ba{rcl}
T_{g \alpha}^{\nu} & = &
 - \frac{\partial {\cal L}_0}{\partial D_{\nu} C_{\mu}^{\beta}}
\partial_{\alpha} C_{\mu}^{\beta}
+ G_{\alpha}^{-1\nu} {\cal L}_0
-  G^{-1 \lambda}_{\sigma}
(\partial_{\mu} C_{\lambda}^{\sigma})
\frac{\partial {\cal L}_0}
{\partial \partial_{\mu} C_{\nu}^{\alpha}} \\
&& \\
&&- \frac{1}{2} \eta^{\mu\rho} \eta^{\lambda\sigma}
g_{\alpha\beta} G^{-1\nu}_{\gamma}
F^{\gamma}_{\mu\lambda} F^{\beta}_{\rho\sigma}
+ \partial_{\mu} (\eta^{\nu \lambda}
\eta^{\sigma \tau}  g_{ \alpha\beta}
F_{\lambda \tau}^{\beta} C_{\sigma}^{\mu}).
\ea
\ee
$ T_{g \alpha}^{\nu}$ is also a conserved current, that is
\be
\partial_{\nu} T_{g \alpha}^{\nu} = 0,
\label{4.42}
\ee
because of the following identity
\be
\partial_{\nu} \partial_{\mu} (\eta^{\mu \lambda} \eta^{\nu \tau}
g_{\alpha \beta} F_{\lambda \tau}^{\beta} )  = 0.
\label{4.43}
\ee
$ T_{g \alpha}^{\nu}$ is called gravitational energy-momentum
tensor, which is the source of gravitational gauge field. Now we get
two different energy-momentum tensors, one is the inertial energy-momentum
tensor $ T_{i \alpha}^{\nu}$ and another is the gravitational
energy-momentum tensor $ T_{g \alpha}^{\nu}$. They are similar,
but they are different.The inertial energy-momentum tensor
$ T_{i \alpha}^{\nu}$ is given by conservation law which is associate
with global gravitational gauge symmetry, it gives out an
energy-momentum 4-vector:
\be
P_{i \alpha} = \int {\rm d}^3 \svec{x} T_{i  \alpha}^{0}.
\label{4.44}
\ee
It is a conserved charges,
\be
\frac{\rm d}{{\rm d} t}P_{i \alpha} =  0.
\label{4.45}
\ee
The time component of $P_{i \alpha}$, that is
$P_{i 0}$,  gives out the Hamiltonian $H$ of the system,
\be
H = - P_{i~0} =  \int {\rm d}^3 \svec{x} J(C)
(\pi_{\alpha}^{\mu} \tdot{C} _{\mu}^{\alpha} - {\cal L}_0).
\label{4.46}
\ee
According to our conventional belief, $H$ should be the inertial
energy of the system, therefore $P_{i \alpha}$ is the inertial
energy-momentum of the system. The gravitational energy-momentum
is given by equation of motion of gravitational gauge field, it is also
a conserved current. The space integration of the time component
of it gives out a conserved energy-momentum 4-vector,
\be
P_{g \alpha} = \int {\rm d}^3 \svec{x} T_{g  \alpha}^{0}.
\label{4.47}
\ee
It is also a conserved charge,
\be
\frac{\rm d}{{\rm d} t}P_{g \alpha} =  0.
\label{4.48}
\ee
The time component of it just gives out the gravitational energy
of the system. This can be easily seen. Set $\nu$  and $\alpha$
in eq.(\ref{4.41}) to $0$, we get
\be
\partial^i F_{i 0}^0   = - g T_{g 0}^{0}.
\label{4.49}
\ee
The field strength of gravitational field is defined by
\be
E^i = - F_{i 0}^0 .
\label{4.50}
\ee
The space integration of eq.(\ref{4.49}) gives out
\be
\oint {\rm d}\svec{\sigma} \cdot \svec{E}
=   g \int {\rm d}^3 \svec{x} T_{g  0}^0.
\label{4.51}
\ee
According to Newton's classical theory of gravity,
$\int {\rm d}^3 \svec{x} T_{g  0}^0$ in the right hand term is
just the gravitational mass of the system. Denote the gravitational
mass of the system as $M_g$, that is
\be
M_g
= - \int {\rm d}^3 \svec{x} T_{g  0}^0.
\label{4.52}
\ee
Then eq(\ref{4.51}) is changed into
\be
\oint {\rm d}\svec{\sigma} \cdot \svec{E}
= - g M_g.
\label{4.53}
\ee
This is just the classical Newton's law of universal gravitation.
It can be strictly proved that gravitational mass is different from
inertial mass. They are not equivalent. But their difference is at least
first order infinitesimal quantity if the gravitational field
$ g C_{\mu}^{\alpha}$, for this difference is
proportional to $g C_{\mu}^{\alpha}$. So, this difference is
too small to be detected in experiments. But in the environment
of strong gravitational field, the difference will become relatively
larger and will be easier to be detected. Much more highly precise
measurement of this difference is strongly needed to test this
prediction and to test the validity of the equivalence principle.
In the chapter of classical limit of quantum gauge theory of
gravity, we will return to discuss this problem again. \\

Now, let's discuss self-coupling of gravitational
field. The Lagrangian of gravitational gauge field is given by
eq(\ref{4.24}). Because
\be
J(C) = 1 + \sum_{m=1}^{\infty} \frac{1}{m!}
\left( \sum_{n=1}^{\infty} 
\frac{g^n}{n}{\rm tr} (C^n)
\right)^m
\label{4.54}
\ee
there are vertexes of  $n$ gravitational gauge fields in tree diagram
where $n$ can be arbitrary integer number that is greater than 3. This
property is important for renormalization of the theory. Because
the coupling constant of the gravitational gauge interactions has
negative mass dimension, any kind of regular vertex exists divergence.
In order to cancel these divergences, we need to introduce the
corresponding counterterms. Because of the existence of
the vertex of $n$ gravitational gauge fields in tree diagram in the
non-renormalized Lagrangian, we need not introduce any new
counterterm which does not exist in the non-renormalized
Lagrangian, what we need to do is to redefine gravitational coupling
constant $g$ and gravitational gauge field $C_{\mu}^{\alpha}$
in renormalization.
If there is no  $J(C)$ term in the original Lagrangian, then
we will have to introduce infinite counterterms in
renormalization, and therefore the theory is non-renormalizable.
Because of the existence of the factor $J(C)$, though
quantum gauge theory of gravity looks like a non-renormalizable
theory according to power counting law, it is indeed renormalizable.
In a word, the factor $J(C)$ is highly important for the
quantum gauge theory of gravity. \\

\section{Gravitational Interactions of Scalar Fields}
\setcounter{equation}{0}

Now, let's start to discuss gravitational interactions of matter
fields. First, we discuss gravitational interactions of scalar fields.
For the sake of simplicity, we first discuss real scalar field. Suppose
that $\phi(x)$ is a real scalar field. The traditional Lagrangian for the
real scalar field is
\be
- \frac{1}{2} \eta^{\mu \nu} \partial_{\mu} \phi(x)
\partial_{\nu} \phi(x) - \frac{m^2}{2} \phi^2 (x),
\label{5.1}
\ee
where $m$ is the mass of scalar field.
This is the Lagrangian for a free real scalar field. The Euler-Lagrangian
equation of motion of it is
\be
( \eta^{\mu \nu} \partial_{\mu} \partial_{\nu} - m^2 ) \phi(x) =0,
\label{5.2}
\ee
which is the famous Klein-Gordan equation. \\

Now, replace the ordinary partial derivative $\partial_{\mu}$
with gauge covariant derivative $D_{\mu}$, and add into the 
Lagrangian of pure gravitational gauge field, we get
\be
{\cal L}_0 = -\frac{1}{2} \eta^{\mu \nu} (D_{\mu} \phi)( D_{\nu} \phi)
-\frac{m^2}{2} \phi^2
- \frac{1}{4} \eta^{\mu \rho} \eta^{\nu \sigma} g_{\alpha \beta }
F_{\mu \nu}^{\alpha} F_{\rho \sigma}^{\beta}.
\label{5.3}
\ee
The full Lagrangian is selected to be
\be
{\cal L} = J(C) {\cal L}_0,
\label{5.4}
\ee
and the action $S$ is defined by
\be
S = \int {\rm d}^4 x ~ {\cal L}.
\label{5.5}
\ee
\\

Using our previous definitions of gauge covariant derivative $D_{\mu}$
and strength of gravitational gauge field $F_{\mu \nu}^{\alpha}$, we can
obtain an explicit form of Lagrangian ${\cal L}$,
\be
{\cal L} = {\cal L}_F  +  {\cal  L}_I,
\label{5.6}
\ee
with ${\cal L}_F$ the free Lagrangian and ${\cal L}_I$  the
interaction Lagrangian. Their explicit expressions are
\be
{\cal L}_F =
- \frac{1}{2} \eta^{\mu \nu} \partial_{\mu} \phi(x)
\partial_{\nu} \phi(x) - \frac{m^2}{2} \phi^2 (x)
- \frac{1}{4} \eta^{\mu \rho} \eta^{\nu \sigma} \eta_{\alpha \beta }
F_{0 \mu \nu}^{\alpha} F_{0 \rho \sigma}^{\beta},
\label{5.7}
\ee
\be
\begin{array}{rcl}
{\cal L}_I &=&
{\cal L}_F \cdot ( J(C) - 1 ) 
- \frac{1}{4} \eta^{\mu \rho} \eta^{\nu \sigma} 
(J(C) g_{\alpha \beta} - \eta_{\alpha \beta })
F_{0 \mu \nu}^{\alpha} F_{0 \rho \sigma}^{\beta}  \\
&&\\
&& + g J(C) \eta^{\mu \nu} C_{\mu}^{\alpha}
(\partial_{\alpha} \phi)(\partial_{\nu} \phi)
- \frac{g^2}{2} J(C) \eta^{\mu \nu} C_{\mu}^{\alpha}
C_{\nu}^{\beta} (\partial_{\alpha} \phi)(\partial_{\beta} \phi)  \\
&&\\
&&  + g  J(C) \eta^{\mu \rho} \eta^{\nu \sigma}
g_{\alpha \beta} (\partial_{\mu} C_{\nu }^{\alpha}
- \partial_{\nu} C_{\mu}^{\alpha})
C_{\rho}^{\delta} \partial_{\delta} C_{\sigma}^{\beta} \\
&&\\
&&  - \frac{1}{2} g^2 J(C) \eta^{\mu \rho} \eta^{\nu \sigma}
g_{ \alpha \beta}
(C_{\mu}^{\delta} \partial_{\delta} C_{\nu}^{\alpha}
- C_{\nu}^{\delta} \partial_{\delta} C_{\mu}^{\alpha} )
C_{\rho}^{\epsilon} \partial_{\epsilon} C_{\sigma}^{\beta},
\end{array}
\label{5.8}
\ee
where,
\be
F_{0 \mu \nu}^{\alpha} = \partial_{\mu} C_{\nu}^{\alpha}
- \partial_{\nu}  C_{\mu}^{\alpha}.
\label{5.9}
\ee
From eq.(\ref{5.8}), we can see that scalar field can directly couples
to any number of gravitational gauge fields. This is one of the most
important interaction properties of gravity. Other kinds of interactions,
such as strong interactions, weak interactions and electromagnetic
interactions do not have this kind of interaction properties. Because
the gravitational coupling constant has negative mass dimension,
renormalization of theory needs this kind of interaction properties.
In other words, if matter field can not directly couple to any
number of gravitational gauge fields, the theory will be
non-renormalizable. \\

The symmetries of the theory can be easily seen from eq.(\ref{5.3}).
First, let's discuss Lorentz symmetry. In eq.(\ref{5.3}), some indexes
are Lorentz indexes and some are group indexes. Lorentz indexes
and group indexes have different transformation law under
gravitational gauge transformation, but they have the same
transformation law under Lorentz transformation. Therefor,
it can be easily seen that both ${\cal L}_0$ and $J(C)$ are
Lorentz scalars, the Lagrangian ${\cal  L}$ and action $S$
are invariant under global Lorentz transformation. \\

Under gravitational gauge transformations, real scalar field
$\phi (x)$ transforms as
\be
\phi (x) \to \phi '(x) = ( \ehat \phi (x) ),
\label{5.10}
\ee
therefore,
\be
D_{\mu} \phi (x) \to D'_{\mu} \phi '(x) = ( \ehat D_{\mu} \phi (x) ).
\label{5.11}
\ee
It can be easily proved that ${\cal L}_0$ transforms covariantly
\be
{\cal L}_0 \to {\cal L}'_0  = (\ehat {\cal L}_0),
\label{5.12}
\ee
and the action eq.(\ref{5.5}) of the system is invariant,
\be
S \to S' =  S.
\label{5.13}
\ee
Please remember that eq.(\ref{4.27}) is an important relation to be used
in the proof of the gravitational gauge symmetry of the action. \\

Global gravitational gauge symmetry gives out conserved charges.
Suppose that $\ehat$ is an infinitesimal gravitational gauge
transformation, it will have the form of eq.(\ref{4.29}). The first order
variations of fields are
\be
\delta C_{\mu}^{\alpha} (x) =
- \epsilon^{\nu} (\partial_{\nu} C_{\mu}^{\alpha} (x)),
\label{5.14}
\ee
\be
\delta \phi (x) =
- \epsilon^{\nu} (\partial_{\nu} \phi (x)),
\label{5.15}
\ee
Using Euler-Lagrange equation of motions for scalar fields and
gravitational gauge fields, we can obtain that
\be
\delta S = \int {\rm d}^4 x
\epsilon^{\alpha} \partial_{\mu} T_{i \alpha}^{\mu},
\label{5.16}
\ee
where
\be
T_{i \alpha}^{\mu} \equiv J(C)
( - \frac{\partial {\cal L}_0}{\partial \partial_{\mu} \phi} \partial_{\alpha} \phi
- \frac{\partial {\cal L}_0}{\partial \partial_{\mu} C_{\nu}^{\beta}}
\partial_{\alpha} C_{\nu}^{\beta} + \delta^{\mu}_{\alpha} {\cal L}_0 ).
\label{5.17}
\ee
Because action is invariant under global gravitational gauge
transformation,
\be
\delta S = 0,
\label{5.18}
\ee
and $\epsilon^{\alpha}$ is an  arbitrary  infinitesimal
constant 4-vector,   we obtain,
\be
\partial_{\mu} T_{i \alpha}^{\mu} = 0.
\label{5.19}
\ee
This is the conservation equation for inertial energy-momentum
tensor. $T_{i \alpha}^{\mu}$ is the conserved current which
corresponds to the global gravitational gauge symmetry.
The space integration of the time component of inertial
energy-momentum tensor gives out the conserved charge,
which is just the inertial energy-momentum of the system.
The time component of the conserved charge is the Hamilton of
the system, which is
\be
H = - P_{i~0} =  \int {\rm d}^3 \svec{x} J(C)
( \pi_{\phi} \tdot{\phi} +
\pi_{\alpha}^{\mu} \tdot{C} _{\mu}^{\alpha} - {\cal L}_0),
\label{5.20}
\ee
where $\pi_{\phi}$ and $\pi_{\alpha}^{\mu}$ are  canonical conjugate
momenta of the real scalar field and gravittational field
\be
\pi_{\phi} =  \frac{\partial {\cal L}_0 }{\partial \tdot{\phi} } ,
\label{5.2001}
\ee
\be
\pi_{\alpha}^{\mu} =
\frac{\partial {\cal L}_0 }{\partial \tdot{C_{\mu}^{\alpha}} } .
\label{5.2002}
\ee
The inertial space momentum of the system is given by
\be
P^i  = P_{i ~i} =  \int {\rm d}^3 \svec{x} J(C)
( - \pi_{\phi} \partial_i {\phi} -
\pi_{\alpha}^{\mu} \partial_i {C} _{\mu}^{\alpha} ).
\label{5.21}
\ee
According to gauge principle, after quantization, they will become
generators of quantum gravitational gauge transformation. \\

Using the difinion (\ref{4.706}), we can change the 
Lagrangian given by eq.(\ref{5.3}) into
\be
{\cal L}_0 = -\frac{1}{2} g^{\alpha \beta} 
( \partial_{\alpha} \phi)
(\partial_{\beta} \phi ) -  \frac{m^2}{2} \phi^2
- \frac{1}{4} \eta^{\mu \rho} \eta^{\nu \sigma} g_{\alpha \beta }
F_{\mu \nu}^{\alpha} F_{\rho \sigma}^{\beta}.
\label{5.23}
\ee
$g^{\alpha \beta}$ is the metric tensor of curved group space-time.
We can easily see that, when there is no gravitational field 
in space-time, that is,
\be
C_{\mu}^{\alpha} = 0,
\label{5.25}
\ee
the group space-time will be flat
\be
g^{ \alpha \beta} = \eta^{\alpha \beta}.
\label{5.26}
\ee
This is what we expected in general relativity. We do not talk to much
on this problem here, for we will discuss this problem again
in details in the chapter on Einstein-like field equation
with cosmological term.  \\

Euler-Lagrange equations of motion can be easily deduced from
action principle. Keep gravitational gauge field $C_{\mu}^{\alpha}$
fixed and let real scalar field vary infinitesimally, then the first order
infinitesimal variation of action is
\be
\delta S =
\int {\rm d}^4 x  J(C)
\left( \frac{\partial {\cal L}_0}{\partial \phi}
- \partial_{\mu} \frac{\partial {\cal L}_0}{\partial \partial_{\mu} \phi}
- g G^{-1 \nu}_{\alpha} (\partial_{\mu} C_{\nu}^{\alpha} )
\frac{\partial {\cal L}_0}
{\partial \partial_{\mu} \phi} \right) \delta \phi.
\label{5.27}
\ee
Because $\delta \phi$ is an arbitrary variation of scalar field, according
to action principle, we get
\be
\frac{\partial {\cal L}_0}{\partial \phi}
- \partial_{\mu} \frac{\partial {\cal L}_0}{\partial \partial_{\mu} \phi}
- g G^{-1 \nu}_{\alpha} (\partial_{\mu} C_{\nu}^{\alpha} )
\frac{\partial {\cal L}_0}{\partial \partial_{\mu} \phi} = 0.
\label{5.28}
\ee
 Because of the existence of the factor $J(C)$,
the equation of motion for scalar field is quite different from the
traditional form in quantum field theory. But the difference is
a second order infinitesimal quantity if we suppose that both gravitational
coupling constant and gravitational gauge field are first order
infinitesimal quantities. Because
\be
\frac{\partial {\cal L}_0}{\partial \partial_{\alpha} \phi}
= - g^{\alpha \beta} \partial_{\beta} \phi,
\label{5.29}
\ee
\be
\frac{\partial {\cal L}_0}{\partial \phi}
= - m^2 \phi,
\label{5.30}
\ee
the explicit form of the equation of motion of scalar field is
\be
g^{\alpha \beta} \partial_{\alpha} \partial_{\beta} \phi
- m^2 \phi + (\partial_{\alpha} g^{\alpha \beta})
\partial_{\beta} \phi
+ g g^{\alpha \beta} (\partial_{\beta} \phi )
G^{-1 \nu}_{\gamma} (\partial_{\alpha} C_{\nu}^{\gamma} )
= 0.
\label{5.31}
\ee
The equation of motion for gravitational gauge field is:
\be
\partial_{\mu} (\eta^{\mu \lambda} \eta^{\nu \tau} 
g_{\alpha \beta} F_{\lambda \tau}^{\beta} )  
= - g T_{g \alpha}^{\nu},
\label{5.32}
\ee
where $ T_{g \alpha}^{\nu}$ is the gravitational energy-momentum
tensor, whose definition is:
\be
\begin{array}{rcl}
T_{g \alpha}^{\nu}&=&
 - \frac{\partial {\cal L}_0}{\partial D_{\nu} C_{\mu}^{\beta}}
\partial_{\alpha} C_{\mu}^{\beta}
- \frac{\partial {\cal L}_0}{\partial D_{\nu} \phi}
\partial_{\alpha} \phi
+ G_{\alpha}^{-1 \nu} {\cal L}_0  
- G^{-1 \lambda}_{\beta} (\partial_{\mu} C_{\lambda}^{\beta} )
\frac{\partial {\cal L}_0}
{\partial \partial_{\mu} C_{\nu}^{\alpha}}
\\
&&\\
&& 
- \frac{1}{2} \eta^{\mu\rho} \eta^{\lambda\sigma}
g_{\alpha\beta} G^{-1\nu}_{\gamma}
F^{\gamma}_{\mu\lambda} F^{\beta}_{\rho\sigma}
 + \partial_{\mu} (\eta^{\nu \lambda}
\eta^{\sigma \tau} g_{ \alpha \beta}
F_{\lambda \tau}^{\beta} C_{\sigma}^{\mu}),
\end{array}
\label{5.33}
\ee
We can see again that, for matter field, its inertial energy-momentum
tensor is also different from the gravitational energy-momentum
tensor, this difference completely originate from the influences
of gravitational gauge field. Compare eq.(\ref{5.33}) with eq.(\ref{5.17}),
and set gravitational gauge field to zero, that is
\be
D_{\mu} \phi = \partial_{\mu} \phi ,
\label{5.34}
\ee
\be
J(C) = 1,
\label{5.35}
\ee
then we find that two energy-momentum tensors are completely
the same:
\be
T_{i \alpha}^{\mu } = T_{g \alpha}^{\mu }.
\label{5.36}
\ee
It means that the equivalence principle only strictly hold in a space-time
where there is no gravitational field. In the environment of strong
gravitational field, such as in black hole, the equivalence principle
will be strongly violated. \\

Define
\be
{\bf L} = \int {\rm d}^3 \svec{x} {\cal L}
= \int {\rm d}^3 \svec{x} J(C) {\cal L}_0.
\label{5.37}
\ee
Then, we can easily prove that
\be
\frac{\delta {\bf L} }{\delta \phi} =
J(C) \left( \frac{\partial {\cal L}_0 }{\partial \phi}
- \partial_i \frac{\partial {\cal L}_0}{\partial \partial_i \phi}
-g G^{-1 \mu}_{\alpha} (\partial_i C_{\mu}^{\alpha} ) 
\frac{\partial {\cal L}_0}{\partial \partial_i \phi} \right),
\label{5.38}
\ee
\be
\frac{\delta {\bf L} }{\delta \tdot{\phi} } =
J(C) \frac{\partial {\cal L}_0 }{\partial \tdot{\phi} } ,
\label{5.39}
\ee
\be
\frac{\delta {\bf L} }{\delta C_{\nu}^{\alpha}} =
J(C) \left( \frac{\partial {\cal L}_0 }{\partial C_{\nu}^{\alpha}}
- \partial_i \frac{\partial
{\cal L}_0}{\partial \partial_i C_{\nu}^{\alpha}}
+ g G_{\alpha}^{-1 \nu} {\cal L}_0
-g G^{-1 \mu}_{\beta} (\partial_i C_{\mu}^{\beta} )
\frac{\partial {\cal L}_0}{\partial \partial_i C_{\nu}^{\alpha} } \right),
\label{5.40}
\ee
\be
\frac{\delta {\bf L} }{\delta \tdot{ C_{\nu}^{\alpha}} } =
J(C) \frac{\partial {\cal L}_0 }{\partial \tdot{ C_{\nu}^{\alpha}} } .
\label{5.41}
\ee
Then, Hamilton's action principle gives out the following equations
of motion:
 \be
\frac{\delta {\bf L} }{\delta \phi}
- \frac{\rm d}{{\rm d}t}
\frac{\delta {\bf L} }{\delta \tdot{\phi}} = 0 ,
\label{5.42}
\ee
\be
\frac{\delta {\bf L} }{\delta C_{\nu}^{\alpha} }
- \frac{\rm d}{{\rm d}t}
\frac{\delta {\bf L} }{\delta \tdot{ C_{\nu}^{\alpha} }} = 0 .
\label{5.43}
\ee
These two equations of motion are essentially the same as the
Euler-Lagrange equations of motion which we have obtained before.
But these two equations have more beautiful forms. \\

The Hamiltonian of the system is given by a Legendre transformation,
\be
\begin{array}{rcl}
H  &= & \int {\rm d}^3 \svec{x}
(\frac{\delta {\bf L} }{\delta \tdot{ \phi} }  \tdot{\phi}
+ \frac{\delta {\bf L} }{\delta \tdot{ C_{\mu}^{\alpha}} }
\tdot{ C_{\mu}^{\alpha}} ) - {\bf L}  \\
&=&  \int {\rm d}^3 \svec{x} J(C)
( \pi_{\phi} \tdot{\phi} +
\pi_{\alpha}^{\mu} \tdot{C} _{\mu}^{\alpha} - {\cal L}_0),
\end{array}
\label{5.44}
\ee
where $\pi_{\phi}$ and $\pi_{\alpha}^{\mu}$ are
canonical conjugate momenta whose definitions are
given by (\ref{5.2001}) and (\ref{5.2002}).
It can be easily seen that the Hamiltonian given by Legendre
transformation is completely the same as that given by
inertial energy-momentum tensor. After Legendre transformation,
$\phi$, $C_{\mu}^{\alpha}$, $J(C) \pi_{\phi}$ and
$J(C) \pi_{\alpha}^{\mu}$ are canonical independent
variables. Let these variables vary infinitesimally, we can get
\be
\frac{\delta H }{\delta \phi}
= - \frac{\delta {\bf L} }{\delta \phi} ,
\label{5.47}
\ee
\be
\frac{\delta H }{\delta ( J(C) \pi_{\phi})}
= \tdot{\phi} ,
\label{5.48}
\ee
\be
\frac{\delta H }{\delta C_{\nu}^{\alpha}}
= - \frac{\delta {\bf L} }{\delta C_{\nu}^{\alpha}} ,
\label{5.49}
\ee
\be
\frac{\delta H }{\delta ( J(C) \pi_{\alpha}^{\nu})}
= \tdot{ C_{\nu}^{\alpha}}.
\label{5.50}
\ee
Then, Hamilton's equations of motion read:
\be
\frac{\rm d}{{\rm d}t} \phi =
\frac{\delta H }{\delta ( J(C) \pi_{\phi})} ,
\label{5.51}
\ee
\be
\frac{\rm d}{{\rm d}t} (J(C) \pi_{\phi} ) =
- \frac{\delta H }{\delta \phi} ,
\label{5.52}
\ee
\be
\frac{\rm d}{{\rm d}t} C_{\nu}^{\alpha} =
\frac{\delta H }{\delta ( J(C) \pi_{\alpha}^{\nu})} ,
\label{5.53}
\ee
\be
\frac{\rm d}{{\rm d}t} (J(C) \pi_{\alpha}^{\nu} ) =
- \frac{\delta H }{\delta C_{\nu}^{\alpha}} .
\label{5.54}
\ee
The  forms of the Hamilton's equations of motion are completely
the same as those appears in usual quantum field theory and usual
classical analytical mechanics. Therefor, the introduction of the
factor $J(C)$ does not affect the forms of Lagrange equations
of motion and Hamilton's equations of motion. \\

The Poisson brackets of two general functional of canonical arguments
can be defined by
\be
\begin{array}{rcl}
\lbrace  A ~~,~~ B \rbrace  &=& \int {\rm d}^3 \svec{x}
( \frac{\delta A}{\delta \phi} \frac{\delta B}{\delta (J(C) \pi_{\phi})}
- \frac{\delta A}{\delta (J(C) \pi_{\phi})}
\frac{\delta B}{\delta \phi}  \\
&&\\
&&+\frac{\delta A}{\delta C_{\nu}^{\alpha}}
\frac{\delta B}{\delta (J(C) \pi_{\alpha}^{\nu})}
- \frac{\delta A}{\delta (J(C) \pi_{\alpha}^{\nu})}
\frac{\delta B}{\delta C_{\nu}^{\alpha}} ).
\end{array}
\label{5.55}
\ee
According to this definition, we have
\be
\lbrace  \phi(\svec{x},t) ~~,
~~ (J(C) \pi_{\phi})(\svec{y},t) \rbrace
= \delta^3(\svec{x} - \svec{y}),
\label{5.56}
\ee
\be
\lbrace  C_{\nu}^{\alpha}(\svec{x},t) ~~,
~~ (J(C) \pi_{\beta}^{\mu})(\svec{y},t) \rbrace
= \delta_{\nu}^{\mu} \delta^{\alpha}_{\beta}
\delta^3(\svec{x} - \svec{y}).
\label{5.57}
\ee
These two relations can be used as the starting point of
canonical quantization of quantum gravity.\\

Using Poisson brackets, the Hamilton's equations of motion can
be expressed in other forms,
\be
\frac{\rm d}{{\rm d}t} \phi (\svec{x},t) =
\lbrace  \phi(\svec{x},t) ~~,~~ H \rbrace  ,
\label{5.58}
\ee
\be
\frac{\rm d}{{\rm d}t} (J(C) \pi_{\phi}) (\svec{x},t) =
\lbrace  (J(C) \pi_{\phi}) (\svec{x},t) ~~,~~ H \rbrace  ,
\label{5.59}
\ee
\be
\frac{\rm d}{{\rm d}t} C_{\nu}^{\alpha} (\svec{x},t) =
\lbrace C_{\nu}^{\alpha}(\svec{x},t) ~~,~~ H \rbrace  ,
\label{5.60}
\ee
\be
\frac{\rm d}{{\rm d}t} (J(C) \pi_{\alpha}^{\nu}) (\svec{x},t) =
\lbrace  (J(C) \pi_{\alpha}^{\nu} ) (\svec{x},t) ~~,~~ H \rbrace  .
\label{5.61}
\ee
Therefore, if $A$ is an arbitrary functional of the canonical arguments
$\phi$, $C_{\mu}^{\alpha}$, $J(C) \pi_{\phi}$ and
$J(C) \pi_{\alpha}^{\mu}$, then we have
\be
\tdot{A} =
\lbrace  A ~~,~~ H \rbrace  .
\label{5.62}
\ee
After quantization, this equation will become the Heisenberg equation.
\\

If $\phi(x)$ is a complex scalar field, its traditional Lagrangian is
\be
-  \eta^{\mu \nu} \partial_{\mu} \phi(x)
\partial_{\nu} \phi^{*}(x) - m^2 \phi(x) \phi^{*}(x).
\label{5.63}
\ee
Replace ordinary partial derivative with gauge covariant derivative,
and add into the Lagrangian for pure gravitational gauge field,
we get,
\be
{\cal L}_0 = - \eta^{\mu \nu}( D_{\mu} \phi)  (D_{\nu} \phi)^*
- m^2\phi \phi^{*}
- \frac{1}{4} \eta^{\mu \rho} \eta^{\nu \sigma} g_{\alpha \beta }
F_{\mu \nu}^{\alpha} F_{\rho \sigma}^{\beta}.
\label{5.64}
\ee
Repeating all above discussions, we can get the whole theory
for gravitational interactions of complex scalar fields. We will not
repeat this discussion here.\\

\section{Gravitational Interactions of Dirac Field}
\setcounter{equation}{0}

In the usual quantum field theory, the Lagrangian for Dirac field is
\be
- \bar{\psi} (\gamma^{\mu} \partial_{\mu} + m) \psi.
\label{6.1}
\ee
Replace ordinary partial derivative with gauge covariant derivative,
and add into the Lagrangian of pure gravitational gauge field,
we get,
\be
{\cal L}_0 =
- \bar{\psi} (\gamma^{\mu} D_{\mu} + m) \psi
- \frac{1}{4} \eta^{\mu \rho} \eta^{\nu \sigma} g_{\alpha \beta }
F_{\mu \nu}^{\alpha} F_{\rho \sigma}^{\beta}.
\label{6.2}
\ee
The full Lagrangian of the system is
\be
{\cal L} = J(C) {\cal L}_0 ,
\label{6.3}
\ee
and the corresponding action is
\be
S =  \int {\rm d}^4 x {\cal L}
=\int {\rm d}^4 x  ~~ J(C) {\cal L}_0 .
\label{6.4}
\ee
This Lagrangian can be separated into two parts,
\be
{\cal L} = {\cal L}_F + {\cal L}_I ,
\label{6.5}
\ee
with ${\cal L}_F$ the free Lagrangian and ${\cal L}_I$ the interaction
Lagrangian. Their explicit forms are
\be
{\cal L}_F =
- \bar{\psi} (\gamma^{\mu} \partial_{\mu} + m) \psi
- \frac{1}{4} \eta^{\mu \rho} \eta^{\nu \sigma} \eta_{\alpha \beta }
F_{0 \mu \nu}^{\alpha} F_{0 \rho \sigma}^{\beta},
\label{6.6}
\ee
\be
\begin{array}{rcl}
{\cal L}_I &=&
{\cal L}_F \cdot (J(C) - 1 )
- \frac{1}{4} \eta^{\mu \rho} \eta^{\nu \sigma} 
( J(C) g_{\alpha \beta} - \eta_{\alpha \beta })
F_{0 \mu \nu}^{\alpha} F_{0 \rho \sigma}^{\beta}  \\
&& \\
&& + g J(C) \bar{\psi} \gamma^{\mu} (\partial_{\alpha}
\psi) C_{\mu}^{\alpha}  
  + g J(C) \eta^{\mu \rho} \eta^{\nu \sigma}
g_{\alpha \beta} (\partial_{\mu} C_{\nu }^{\alpha}
- \partial_{\nu} C_{\mu}^{\alpha})
C_{\rho}^{\delta} \partial_{\delta} C_{\sigma}^{\beta} \\
&&\\
&&  - \frac{1}{2} g^2 J(C) \eta^{\mu \rho} \eta^{\nu \sigma}
g_{\alpha \beta}
(C_{\mu}^{\delta} \partial_{\delta} C_{\nu}^{\alpha}
- C_{\nu}^{\delta} \partial_{\delta} C_{\mu}^{\alpha} )
C_{\rho}^{\epsilon} \partial_{\epsilon} C_{\sigma}^{\beta}.
\end{array}
\label{6.7}
\ee
From ${\cal L}_I$, we can see that Dirac field can directly
couple to any number of gravitational gauge fields, the mass
term of Dirac field also take part in gravitational interactions. All these
interactions are completely determined by the requirement of
gravitational gauge symmetry. The Lagrangian function before
renormalization almost contains all kind of divergent vertex,
which is important in the renormalization of the theory.
Besides, from eq.(\ref{6.7}), we can directly write out Feynman
rules of the corresponding interaction vertexes.
\\

Because the traditional Lagrangian function eq.(\ref{6.1}) is invariant
under global Lorentz transformation, which is already proved in the
traditional quantum field theory, and the covariant derivative
has the same behavior as partial derivative under global Lorentz
transformation, the first two terms of Lagrangian ${\cal L}$ are
global Lorentz invariant. We have already prove that the Lagrangian
function for pure gravitational gauge field is invariant under
global Lorentz transformation. Therefor, ${\cal L}$ has global
Lorentz symmetry.  \\

The gravitational gauge transformation of Dirac field is
\be
\psi(x) \to \psi ' (x) = (\ehat \psi (x) ).
\label{6.8}
\ee
$\bar{\psi}$ transforms similarly,
\be
\bar{\psi}(x) \to \bar{\psi} ' (x) = (\ehat \bar{\psi} (x) ).
\label{6.9}
\ee
Dirac $\gamma$-matrices is not a physical field, so it keeps unchanged
under gravitational gauge transformation,
\be
\gamma^{\mu} \to \gamma^{\mu}.
\label{6.10}
\ee
It can be proved that, under gravitational gauge transformation,
${\cal L}_0$ transforms as
\be
{\cal L}_0 \to {\cal L}'_0 = (\ehat {\cal L}_0 ).
\label{6.11}
\ee
So,
\be
{\cal L} \to {\cal L}'  =  J (\ehat {\cal L}_0 ),
\label{6.12}
\ee
where $J$ is the Jacobi of the corresponding space-time
translation. Then using eq.(\ref{4.27}), we can prove that the action $S$
has gravitational gauge symmetry. \\

Suppose that $\ehat$ is an infinitesimal global transformation, then
the first order infinitesimal variations of Dirac field are
\be
\delta \psi = - \epsilon^{\nu} \partial_{\nu} \psi,
\label{6.13}
\ee
\be
\delta \bar{\psi}  = - \epsilon^{\nu} \partial_{\nu} \bar{\psi}.
\label{6.14}
\ee
The first order variation of action is
\be
\delta S = \int {\rm d}^4 x
\epsilon^{\alpha} \partial_{\mu} T_{i \alpha}^{\mu},
\label{6.15}
\ee
where $ T_{i \alpha}^{\mu}$ is the inertial energy-momentum
tensor whose definition is,
\be
T_{i \alpha}^{\mu} \equiv J(C)
\left( - \frac{\partial {\cal L}_0}{\partial \partial_{\mu} \psi}
\partial_{\alpha} \psi
- \frac{\partial {\cal L}_0}{\partial \partial_{\mu} C_{\nu}^{\beta}}
\partial_{\alpha} C_{\nu}^{\beta}
+ \delta^{\mu}_{\alpha} {\cal L}_0 \right).
\label{6.16}
\ee
The global gravitational gauge symmetry of action gives out conservation
equation of the inertial energy-momentum tensor,
\be
\partial_{\mu} T_{i \alpha}^{\mu} = 0.
\label{6.17}
\ee
The inertial energy-momentum tensor is the conserved current
which expected by gauge principle. The space integration of its
time component gives out the conserved energy-momentum of the
system,
\be
H = - P_{i~0} =  \int {\rm d}^3 \svec{x} J(C)
( \pi_{\psi} \tdot{\psi} +
\pi_{\alpha}^{\mu} \tdot{C} _{\mu}^{\alpha} - {\cal L}_0),
\label{6.18}
\ee
\be
P^i  = P_{i~i} =  \int {\rm d}^3 \svec{x} J(C)
( - \pi_{\psi} \partial_i {\psi} -
\pi_{\alpha}^{\mu} \partial_i {C} _{\mu}^{\alpha} ),
\label{6.19}
\ee
where
\be
\pi_{\psi} = \frac{\partial {\cal L}_0}{\partial \tdot{\psi}}.
\label{6.20}
\ee

The equation of motion for Dirac field is
\be
(\gamma^{\mu} D_{\mu} + m) \psi = 0.
\label{6.21}
\ee
From this expression, we can see that the factor $J(C)$ does not
affect the equation of motion of Dirac field. This is caused by the
asymmetric form of the Lagrangian. If we use a symmetric form of
Lagrangian, the factor $J(C)$ will also affect the equation
of motion of Dirac field, which will be discussed later. \\

The equation of motion of gravitational gauge field can be 
easily deduced,
\be
\partial_{\mu} (\eta^{\mu \lambda} \eta^{\nu \tau} g_{\alpha \beta}
F_{\lambda \tau}^{\beta} )  = - g T_{g \alpha}^{\nu},
\label{6.22}
\ee
where $ T_{g \alpha}^{\nu}$ is the gravitational energy-momentum
tensor, whose definition is:
\be
\begin{array}{rcl}
T_{g \alpha}^{\nu}&=&
 - \frac{\partial {\cal L}_0}{\partial D_{\nu} C_{\mu}^{\beta}}
\partial_{\alpha} C_{\mu}^{\beta}
- \frac{\partial {\cal L}_0}{\partial D_{\nu} \psi}
\partial_{\alpha} \psi
+ G_{\alpha}^{-1 \nu} {\cal L}_0  
- G^{-1 \lambda}_{\beta} 
(\partial_{\mu} C_{\lambda}^{\beta} )
\frac{\partial {\cal L}_0}
{\partial \partial_{\mu} C_{\nu}^{\alpha}}  \\
&&\\
&& 
 + \partial_{\mu} (\eta^{\nu \lambda}
\eta^{\sigma \tau} g_{\alpha \beta}
F_{\lambda \tau}^{\beta} C_{\sigma}^{\mu})
- \frac{1}{2} \eta^{\mu\rho} \eta^{\lambda\sigma}
g_{\alpha\beta} G^{-1\nu}_{\gamma}
F^{\gamma}_{\mu\lambda} F^{\beta}_{\rho\sigma}.
\end{array}
\label{6.23}
\ee
We see again that the gravitational energy-momentum tensor is
different from the  inertial energy-momentum tensor. \\

In usual quantum field theory, the Lagrangian for Dirac field
has a more symmetric form, which is
\be
- \bar{\psi} (\gamma^{\mu} \dvec{\partial}_{\mu} + m) \psi,
\label{6.24}
\ee
where
\be
 \dvec{\partial}_{\mu} =
\frac{\partial_{\mu} - \lvec{\partial_{\mu}}}{2}.
\label{6.25}
\ee
The Euler-Lagrange equation of motion of eq.(\ref{6.24}) also
gives out the conventional Dirac equation. \\

Now replace ordinary space-time partial derivative with
covariant derivative, and add into the Lagrangian of pure
gravitational gauge field, we get,
\be
{\cal L}_0 =
- \bar{\psi} (\gamma^{\mu} \dvec{D}_{\mu} + m) \psi
- \frac{1}{4} \eta^{\mu \rho} \eta^{\nu \sigma} g_{\alpha \beta }
F_{\mu \nu}^{\alpha} F_{\rho \sigma}^{\beta},
\label{6.26}
\ee
where $\dvec{D}_{\mu}$ is defined by
\be
 \dvec{D}_{\mu} =
\frac{D_{\mu} - \lvec{D}_{\mu}}{2}.
\label{6.27}
\ee
Operator $\lvec{D}_{\mu}$ is understood in the following way
\be
f(x) \lvec{D}_{\mu} g(x) =
(D_{\mu} f(x)) g(x),
\label{6.28}
\ee
with $f(x)$ and $g(x)$ two arbitrary functions. The Lagrangian density
${\cal L}$ and action $S$ are also defined by 
eqs.(\ref{6.3}-\ref{6.4}). In this case,
the free Lagrangian ${\cal L}_F$ and interaction Lagrangian ${\cal L}_I$
are given by
\be
{\cal L}_F =
- \bar{\psi} (\gamma^{\mu} \dvec{\partial}_{\mu} + m) \psi
- \frac{1}{4} \eta^{\mu \rho} \eta^{\nu \sigma} 
\eta_{\alpha \beta }
F_{0 \mu \nu}^{\alpha} F_{0 \rho \sigma}^{\beta},
\label{6.29}
\ee
\be
\begin{array}{rcl}
{\cal L}_I &=&
{\cal L}_F \cdot (J(C) - 1 )
- \frac{1}{4} \eta^{\mu \rho} \eta^{\nu \sigma}
( J(C) g_{\alpha \beta} - \eta_{\alpha \beta })
F_{0 \mu \nu}^{\alpha} F_{0 \rho \sigma}^{\beta} \\
&&\\
&& + g J(C)( \bar{\psi} \gamma^{\mu} \dvec{\partial}_{\alpha}
\psi) C_{\mu}^{\alpha}  
  + g J(C) \eta^{\mu \rho} \eta^{\nu \sigma}
g_{\alpha \beta} (\partial_{\mu} C_{\nu }^{\alpha}
- \partial_{\nu} C_{\mu}^{\alpha})
C_{\rho}^{\delta} \partial_{\delta} C_{\sigma}^{\beta} \\
&&\\
&&  - \frac{1}{2} g^2 J(C) \eta^{\mu \rho} \eta^{\nu \sigma}
g_{\alpha \beta}
(C_{\mu}^{\delta} \partial_{\delta} C_{\nu}^{\alpha}
- C_{\nu}^{\delta} \partial_{\delta} C_{\mu}^{\alpha} )
C_{\rho}^{\epsilon} \partial_{\epsilon} C_{\sigma}^{\beta}.
\end{array}
\label{6.30}
\ee
\\

The Euler-Lagrange equation of motion for Dirac field is
\be
\frac{\partial {\cal L}_0}{\partial \bar{\psi}}
- \partial_{\mu} \frac{\partial {\cal L}_0}{\partial
\partial_{\mu} \bar{\psi}}
- g G^{-1 \nu}_{\alpha} ( \partial_{\mu} C_{\nu}^{\alpha})
\frac{\partial {\cal L}_0}{\partial
\partial_{\mu} \bar{\psi}} = 0.
\label{6.31}
\ee
Because
\be
\frac{\partial {\cal L}_0}{\partial \bar{\psi}}
= -\frac{1}{2} \gamma^{\mu} D_{\mu} \psi - m \psi ,
\label{6.32}
\ee
\be
\frac{\partial {\cal L}_0}{\partial \partial_{\mu} \bar{\psi}}
=   \frac{1}{2} \gamma^{\alpha} G_{\alpha}^{\mu} \psi ,
\label{6.33}
\ee
eq.(\ref{6.31}) will be changed into
\be
(\gamma^{\mu} D_{\mu} + m) \psi =
- \frac{1}{2}\gamma^{\mu}
(\partial_{\alpha} G_{\mu}^{\alpha}) \psi
-  \frac{1}{2} g \gamma^{\mu} \psi
G_{\beta}^{-1 \nu} (D_{\mu}  C_{\nu}^{\beta} ).
\label{6.34}
\ee
If gravitational gauge field vanishes, this equation of motion
will return to the traditional Dirac equation. \\

The inertial energy-momentum tensor now becomes
\be
T_{i \alpha}^{\mu} = J(C)
\left( - \frac{\partial {\cal L}_0}{\partial \partial_{\mu} \psi}
\partial_{\alpha} \psi
- (\partial_{\alpha} \bar{\psi})
\frac{\partial {\cal L}_0}{\partial \partial_{\mu} \bar{\psi}}
- \frac{\partial {\cal L}_0}{\partial \partial_{\mu} C_{\nu}^{\beta}}
\partial_{\alpha} C_{\nu}^{\beta} 
+ \delta^{\mu}_{\alpha} {\cal L}_0 \right),
\label{6.35}
\ee
and the gravitational energy-momentum tensor becomes
\be
\begin{array}{rcl}
T_{g \alpha}^{\nu}&=&
 - \frac{\partial {\cal L}_0}{\partial D_{\nu} C_{\mu}^{\beta}}
\partial_{\alpha} C_{\mu}^{\beta}
- \frac{\partial {\cal L}_0}{\partial D_{\nu} \psi}
\partial_{\alpha} \psi
- (\partial_{\alpha} \bar{\psi})
\frac{\partial {\cal L}_0}{\partial D_{\nu} \bar{\psi}}
+ G_{\alpha}^{-1 \nu} {\cal L}_0  \\
&&\\
&& -  G^{-1 \lambda}_{\beta} (\partial_{\mu} C_{\lambda}^{\beta} ) 
\frac{\partial {\cal L}_0}{\partial \partial_{\mu} C_{\nu}^{\alpha}}
 + \partial_{\mu} (\eta^{\nu \lambda}
\eta^{\sigma \tau} g_{\alpha \beta}
F_{\lambda \tau}^{\beta} C_{\sigma}^{\mu})  \\
&&\\
&& - \frac{1}{2} \eta^{\mu\rho} \eta^{\lambda\sigma}
g_{\alpha\beta} G^{-1\nu}_{\gamma}
F^{\gamma}_{\mu\lambda} F^{\beta}_{\rho\sigma}.
\end{array}
\label{6.36}
\ee
Both of them are conserved energy-momentum tensor.  But they
are not equivalent.
\\

\section{Gravitational Interactions of Vector Field}
\setcounter{equation}{0}

The traditional Lagrangian for vector field is
\be
- \frac{1}{4} \eta^{\mu \rho} \eta^{\nu \sigma}
A_{\mu \nu} A_{\rho \sigma}
- \frac{m^2}{2}  \eta^{\mu \nu} A_{\mu} A_{\nu},
\label{7.1}
\ee
where $A_{\mu \nu}$ is the strength of vector field which is
given by
\be
 \partial_{\mu} A_{\nu}
- \partial_{\nu} A_{\mu}.
\label{7.2}
\ee
The Lagrangian ${\cal L}_0$ that describes gravitational interactions
between vector field and gravitational fields is
\be
{\cal L}_0 =
- \frac{1}{4} \eta^{\mu \rho} \eta^{\nu \sigma}
A_{\mu \nu} A_{\rho \sigma}
- \frac{m^2}{2}  \eta^{\mu \nu} A_{\mu} A_{\nu}
- \frac{1}{4} \eta^{\mu \rho} \eta^{\nu \sigma} g_{\alpha \beta }
F_{\mu \nu}^{\alpha} F_{\rho \sigma}^{\beta}.
\label{7.3}
\ee
In eq.(\ref{7.3}), the definition of strength $A_{\mu \nu}$ is not
given by eq.(\ref{7.2}), it is given by
\be
\begin{array}{rcl}
A_{\mu \nu} &=& D_{\mu} A_{\nu}
- D_{\nu} A_{\mu}  \\
&=&  \partial_{\mu} A_{\nu} - \partial_{\nu} A_{\mu}
- g C_{\mu}^{\alpha} \partial_{\alpha} A_{\nu}
+ g C_{\nu}^{\alpha} \partial_{\alpha} A_{\mu},
\end{array}
\label{7.4}
\ee
where $D_{\mu}$ is the gravitational gauge
 covariant derivative, whose definition
is given by eq.(\ref{4.4}). The full Lagrangian ${\cal L}$ is given by,
\be
{\cal L} = J(C) {\cal L}_0.
\label{7.5}
\ee
The action $S$ is defined by
\be
S = \int {\rm d}^4 x ~ {\cal L}.
\label{7.6}
\ee
\\

The Lagrangian ${\cal L}$ can be separated into two parts:
the free Lagrangian ${\cal L}_F$ and interaction Lagrangian
${\cal L}_I$. The explicit forms of them are
\be
{\cal L}_F =
- \frac{1}{4} \eta^{\mu \rho} \eta^{\nu \sigma}
A_{0 \mu \nu} A_{0 \rho \sigma}
- \frac{m^2}{2}  \eta^{\mu \nu} A_{\mu} A_{\nu}
- \frac{1}{4} \eta^{\mu \rho} \eta^{\nu \sigma} \eta_{\alpha \beta }
F_{0 \mu \nu}^{\alpha} F_{0 \rho \sigma}^{\beta},
\label{7.7}
\ee
\be
\begin{array}{rcl}
{\cal L}_I &=&
{\cal L}_F \cdot ( J(C) - 1 ) 
- \frac{1}{4} \eta^{\mu \rho} \eta^{\nu \sigma} 
(J(C) g_{\alpha\beta} - \eta_{\alpha \beta} )
F_{0 \mu \nu}^{\alpha} F_{0 \rho \sigma}^{\beta} \\
&&\\
&& + g J(C) \eta^{\mu \rho} \eta^{\nu \sigma}
A_{0 \mu \nu} C_{\rho}^{\alpha} \partial_{\alpha} A_{\sigma} \\
&&\\
&& - \frac{g^2}{2} J(C) \eta^{\mu \rho} \eta^{\nu \sigma}
( C_{\mu}^{\alpha} C_{\rho}^{\beta} (\partial_{\alpha} A_{\nu} )
(\partial_{\beta} A_{\sigma} )
- C_{\nu}^{\alpha} C_{\rho}^{\beta} (\partial_{\alpha} A_{\mu} )
(\partial_{\beta} A_{\sigma} )  )  \\
&&\\
&&  + g J(C) \eta^{\mu \rho} \eta^{\nu \sigma}
g_{\alpha \beta} (\partial_{\mu} C_{\nu }^{\alpha}
- \partial_{\nu} C_{\mu}^{\alpha})
C_{\rho}^{\delta} \partial_{\delta} C_{\sigma}^{\beta} \\
&&\\
&&  - \frac{1}{2} g^2 J(C) \eta^{\mu \rho} \eta^{\nu \sigma}
g_{\alpha \beta}
(C_{\mu}^{\delta} \partial_{\delta} C_{\nu}^{\alpha}
- C_{\nu}^{\delta} \partial_{\delta} C_{\mu}^{\alpha} )
C_{\rho}^{\epsilon} \partial_{\epsilon} C_{\sigma}^{\beta},
\end{array}
\label{7.8}
\ee
where
$A_{0 \mu \nu} = \partial_{\mu} A_{\nu} - \partial_{\nu} A_{\mu}$
The first three lines of ${\cal L}_I$ contain interactions between
vector field and gravitational gauge fields. It can be seen that
the vector field can also directly couple to arbitrary number of
gravitational gauge fields, which is one of the most important
properties of gravitational gauge interactions. This interaction
property is required and determined by local gravitational
gauge symmetry.  \\

Under Lorentz transformations, group index and Lorentz index have
the same behavior. Therefor every term in the Lagrangian ${\cal L}$
are Lorentz scalar, and the whole Lagrangian ${\cal L}$ and
action $S$ have Lorentz symmetry. \\

Under gravitational gauge transformations, vector field $A_{\mu}$
transforms as
\be
A_{\mu} (x) \to A'_{\mu}(x) = (\ehat A_{\mu} (x)).
\label{7.9}
\ee
$D_{\mu} A_{\nu}$ and $A_{\mu \nu}$ transform covariantly,
\be
D_{\mu} A_{\nu} \to D'_{\mu} A'_{\nu} =
(\ehat D_{\mu} A_{\nu}) ,
\label{7.10}
\ee
\be
A_{\mu \nu} \to A'_{\mu \nu} =
(\ehat A_{\mu \nu}).
\label{7.11}
\ee
So, the gravitational gauge transformations of ${\cal L}_0$
and ${\cal L}$ respectively are
\be
{\cal L}_0 \to {\cal L}'_0 = (\ehat {\cal L}_0 ),
\label{7.12}
\ee
\be
{\cal L} \to {\cal L}'  =  J (\ehat {\cal L}_0 ).
\label{7.13}
\ee
The action of the system is gravitational gauge invariant.
\\

The global gravitational gauge transformation gives out conserved
current of gravitational gauge  symmetry.
Under infinitesimal global gravitational gauge
transformation, the vector field $A_{\mu}$ transforms as
\be
\delta A_{\mu} = - \epsilon^{\alpha} \partial_{\alpha} A_{\mu}.
\label{7.14}
\ee
The first order variation of action is
\be
\delta S = \int {\rm d}^4 x
\epsilon^{\alpha} \partial_{\mu} T_{i \alpha}^{\mu},
\label{7.15}
\ee
where $ T_{i \alpha}^{\mu}$ is the inertial energy-momentum
tensor whose definition is,
\be
T_{i \alpha}^{\mu} = J(C)
\left( - \frac{\partial {\cal L}_0}{\partial \partial_{\mu} A_{\nu}}
\partial_{\alpha} A_{\nu}
- \frac{\partial {\cal L}_0}{\partial \partial_{\mu} C_{\nu}^{\beta}}
\partial_{\alpha} C_{\nu}^{\beta} +
\delta^{\mu}_{\alpha} {\cal L}_0 \right).
\label{7.16}
\ee
$T_{i \alpha}^{\mu}$ is a conserved current. The space integration
of its time component gives out inertial energy-momentum of the
system,
\be
H = - P_{i~0} =  \int {\rm d}^3 \svec{x} J(C)
( \pi^{\mu} \tdot{A_{\mu}} +
\pi_{\alpha}^{\mu} \tdot{C _{\mu}^{\alpha}} - {\cal L}_0),
\label{7.17}
\ee
\be
P^i  = P_{i~i} =  \int {\rm d}^3 \svec{x} J(C)
( - \pi^{\mu} \partial_i A_{\mu} -
\pi_{\alpha}^{\mu} \partial_i {C} _{\mu}^{\alpha} ),
\label{7.18}
\ee
where
\be
\pi^{\mu} = \frac{\partial {\cal L}_0}{\partial \tdot{A_{\mu}}}.
\label{7.19}
\ee
\\

The equation of motion for vector field is
\be
\frac{\partial {\cal L}_0}{\partial A_{\nu}}
- \partial_{\mu}
\frac{\partial {\cal L}_0}{\partial \partial_{\mu} A_{\nu}}
- g G_{\alpha}^{-1 \lambda} 
( \partial_{\mu} C_{\lambda}^{\alpha})
\frac{\partial {\cal L}_0}{\partial \partial_{\mu} A_{\nu}} = 0.
\label{7.20}
\ee
From eq.(\ref{7.3}), we can obtain
\be
\frac{\partial {\cal L}_0}{\partial \partial_{\mu} A_{\nu}}
= - \eta^{\lambda \rho} \eta^{\nu \sigma}
G_{\lambda}^{\mu} A_{\rho \sigma},
\label{7.21}
\ee
\be
\frac{\partial {\cal L}_0}{\partial A_{\nu}}
= - m^2 \eta^{\lambda \nu} A_{\lambda}.
\label{7.22}
\ee
Then, eq.(\ref{7.20}) is changed into
\be
\eta^{\mu \rho} \eta^{\nu \sigma} D_{\mu} A_{\rho \sigma}
- m^2 \eta^{\mu \nu} A_{\mu}
= - \eta^{\lambda \rho} \eta^{\nu \sigma}
( \partial_{\mu} G_{\lambda}^{\mu} ) A_{\rho \sigma}
- g \eta^{\mu \rho} \eta^{\nu \sigma}
 A_{\rho \sigma} G_{\alpha}^{-1 \mu}
( D_{\mu} C_{\mu}^{\alpha}).
\label{7.23}
\ee
The equation of motion of gravitational gauge field is
\be
\partial_{\mu} (\eta^{\mu \lambda} \eta^{\nu \tau} g_{\alpha \beta}
F_{\lambda \tau}^{\beta} )  = - g T_{g \alpha}^{\nu},
\label{7.24}
\ee
where $ T_{g \alpha}^{\nu}$ is the gravitational energy-momentum
tensor,
\be
\begin{array}{rcl}
T_{g \alpha}^{\nu}&=&
 - \frac{\partial {\cal L}_0}{\partial D_{\nu} C_{\mu}^{\beta}}
\partial_{\alpha} C_{\mu}^{\beta}
- \frac{\partial {\cal L}_0}{\partial D_{\nu} A_{\mu}}
\partial_{\alpha} A_{\mu}
+ G_{\alpha}^{-1 \nu} {\cal L}_0  \\
&&\\
&& -  G_{\gamma}^{-1 \rho} (\partial_{\mu} C_{\rho}^{\gamma})
\frac{\partial {\cal L}_0}{\partial \partial_{\mu} C_{\nu}^{\alpha}}
 + \partial_{\mu} (\eta^{\nu \lambda}
\eta^{\sigma \tau} g_{\alpha \beta}
F_{\lambda \tau}^{\beta} C_{\sigma}^{\mu})\\
&&\\
&&- \frac{1}{2} \eta^{\mu\rho} \eta^{\lambda\sigma}
g_{\alpha\beta} G^{-1\nu}_{\gamma}
F^{\gamma}_{\mu\lambda} F^{\beta}_{\rho\sigma}.
\end{array}
\label{7.25}
\ee
$T_{g \alpha}^{\nu}$ is also a conserved current. The space integration
of its time component gives out the gravitational energy-momentum
which is the source of gravitational interactions.
It can be also seen that inertial energy-momentum tensor and
gravitational energy-momentum tensor are not equivalent.   \\

\section{Gravitational Interactions of Gauge Fields}
\setcounter{equation}{0}

It is know that QED, QCD and unified electroweak theory are
all gauge theories. In this chapter, we will
discuss how to unify these gauge theories with gravitational
gauge theory, and how to unify four different kinds of
fundamental interactions formally. \\

First, let's discuss QED theory. As an example, let's discuss
electromagnetic interactions of Dirac field. The traditional
electromagnetic interactions between Dirac field $\psi$ and
electromagnetic field $A_{\mu}$ is
\be
- \frac{1}{4} \eta^{\mu \rho} \eta^{\nu \sigma}
A_{\mu \nu} A_{\rho \sigma}
- \bar{\psi}
( \gamma^{\mu} ( \partial_{\mu} - i e A_{\mu}  ) + m ) \psi.
\label{8.1}
\ee
The Lagrangian that describes gravitational gauge interactions
between gravitational gauge field and Dirac field or electromagnetic
field and describes electromagnetic interactions between Dirac field
and electromagnetic field is
\be
{\cal L}_0 =
- \bar{\psi} (\gamma^{\mu} (D_{\mu} - i e A_{\mu}  ) + m ) \psi
- \frac{1}{4} \eta^{\mu \rho} \eta^{\nu \sigma}
{\mathbf A}_{\mu \nu} {\mathbf A}_{\rho \sigma}
- \frac{1}{4} \eta^{\mu \rho} \eta^{\nu \sigma}
g_{\alpha \beta } F_{\mu\nu}^{\alpha} F_{\rho \sigma}^{\beta},
\label{8.2}
\ee
where $D_{\mu}$ is the gravitational gauge covariant derivative
which is given by eq.(\ref{4.4}) and the strength of electromagnetic
field $A_{\mu}$ is
\be
{\mathbf A}_{\mu \nu} = A_{\mu \nu} + g G^{-1 \lambda}_{\alpha}
A_{\lambda} F_{\mu \nu}^{\alpha},
\label{8.3}
\ee
where $A_{\mu \nu}$ is given by eq.(\ref{7.4}) and $G^{-1}$ is given by
eq.(\ref{4.703}). The full Lagrangian density and the action of the
system are respectively given by,
\be
{\cal L} = J(C) {\cal L}_0,
\label{8.4}
\ee
\be
S = \int {\rm d}^4 x ~ {\cal L}.
\label{8.5}
\ee
\\

The system given by above Lagrangian has both $U(1)$ gauge
symmetry and gravitational gauge symmetry. Under $U(1)$ gauge
transformations,
\be
\psi(x) \to \psi'(x) = e^{-i \alpha(x)} \psi(x), 
\label{8.6}
\ee
\be
A_{\mu}(x) \to A'_{\mu}(x) = A_{\mu}(x) - \frac{1}{e} D_{\mu}
\alpha(x), \label{8.7}
\ee
\be
C_{\mu}^{\alpha}(x) \to C^{\prime \alpha}_{\mu}(x) =
C_{\mu}^{\alpha}(x). \label{8.8}
\ee
It can be proved that the Lagrangian ${\cal L}$ is invariant
under $U(1)$ gauge transformation.
Under gravitational gauge transformations,
\be
\psi(x) \to \psi'(x) = (\ehat \psi(x)), 
\label{8.9}
\ee
\be
A_{\mu}(x) \to A'_{\mu}(x) = (\ehat A_{\mu}(x)), 
\label{8.10}
\ee
\be
C_{\mu}(x) \to  C'_{\mu}(x) = \ehat (x) C_{\mu} (x) \ehat^{-1} (x)
+ \frac{i}{g} \ehat (x) (\partial_{\mu} \ehat^{-1} (x)).
\label{8.11}
\ee
The action $S$ given by eq.(\ref{8.4}) is invariant under gravitational
gauge transformation. \\

Lagrangian ${\cal L}$ can be separated into free Lagrangian ${\cal L}_F$
and interaction Lagrangian ${\cal L}_I$,
\be
{\cal L} = {\cal L}_F + {\cal L}_I,
\label{8.12}
\ee
where
\be {\cal L}_F = - \frac{1}{4} \eta^{\mu \rho} \eta^{\nu
\sigma} A_{0 \mu \nu} A_{0 \rho \sigma} - \bar{\psi} (
\gamma^{\mu}
\partial_{\mu}  + m ) \psi 
- \frac{1}{4} \eta^{\mu \rho} \eta^{\nu \sigma} 
\eta_{\alpha \beta } F_{0 \mu \nu}^{\alpha} 
F_{0 \rho \sigma}^{\beta},
\label{8.13}
\ee
\be
\begin{array}{rcl}
{\cal L}_I &=&
{\cal L}_F \cdot ( J(C) - 1 )
+ i e \cdot J(C) \bar{\psi} \gamma^{\mu} \psi A_{\mu} \\
&&\\
&&- \frac{1}{4} \eta^{\mu \rho} \eta^{\nu \sigma}
(J(C)  g_{\alpha \beta} - \eta_{\alpha \beta })
F_{0 \mu \nu}^{\alpha} F_{0 \rho \sigma}^{\beta}  \\
&&\\
&& + g J(C) \bar\psi \gamma^{\mu} \partial_{\alpha} \psi
C_{\mu}^{\alpha}
 + g J(C) \eta^{\mu \rho} \eta^{\nu \sigma}
A_{0 \mu \nu} C_{\rho}^{\alpha} \partial_{\alpha} A_{\sigma} \\
&&\\
&&- \frac{g}{2} J(C) \eta^{\mu \rho} \eta^{\nu \sigma} A_{\mu
\nu} G^{-1 \lambda }_{\alpha} A_{\lambda} F_{\rho \sigma}^{\alpha}\\
&&\\
&& -\frac{g^2}{4} J(C) \eta^{\mu \rho} \eta^{\nu \sigma} G^{-1
\kappa}_{\alpha} G^{-1 \lambda }_{\beta} A_{\kappa} A_{\lambda}
F_{\mu \nu}^{\alpha} F_{\rho \sigma}^{\beta} \\
&&\\
&& - \frac{g^2}{2} J(C) \eta^{\mu \rho} \eta^{\nu \sigma}
( C_{\mu}^{\alpha} C_{\rho}^{\beta} (\partial_{\alpha} A_{\nu} )
(\partial_{\beta} A_{\sigma} )
- C_{\nu}^{\alpha} C_{\rho}^{\beta} (\partial_{\alpha} A_{\mu} )
(\partial_{\beta} A_{\sigma} )  )  \\
&&\\
&&  + g J(C) \eta^{\mu \rho} \eta^{\nu \sigma}
g_{\alpha \beta} (\partial_{\mu} C_{\nu }^{\alpha}
- \partial_{\nu} C_{\mu}^{\alpha})
C_{\rho}^{\delta} \partial_{\delta} C_{\sigma}^{\beta} \\
&&\\
&&  - \frac{1}{2} g^2 J(C) \eta^{\mu \rho} \eta^{\nu \sigma}
g_{\alpha \beta}
(C_{\mu}^{\delta} \partial_{\delta} C_{\nu}^{\alpha}
- C_{\nu}^{\delta} \partial_{\delta} C_{\mu}^{\alpha} )
C_{\rho}^{\epsilon} \partial_{\epsilon} C_{\sigma}^{\beta}.
\end{array}
\label{8.14}
\ee
\\

The traditional Lagrangian for QCD is
\be
- \sum_n \bar{\psi}_n \lbrack \gamma^{\mu} (\partial_{\mu} -i g_c
A^i_{\mu } \frac{\lambda_i}{2} ) + m_n \rbrack \psi_n -\frac{1}{4}
\eta^{\mu \rho} \eta^{\nu \sigma} A^i_{\mu \nu } A^i_{\rho \sigma
},
\label{8.15}
\ee
where $\psi_n$ is the quark color triplet of the $n$th
flavor, $A_{\mu \alpha}$ is the color gauge vector
potential, $A_{\alpha \mu \nu }$ is the color gauge
covariant field strength tensor, $g_c$ is the strong
coupling constant, $\lambda_{\alpha}$ is the Gell-Mann
matrix and $m_n$ is the quark mass of the $n$th flavor.
In gravitational gauge theory, this Lagrangian should be
changed into
\be
\begin{array}{rcl}
{\cal L}_0 &=&
- \sum_n \bar{\psi}_n
\lbrack \gamma^{\mu} (D_{\mu} -i g_c A^i_{\mu}
\frac{\lambda^i}{2} ) + m_n
\rbrack \psi_n
-\frac{1}{4} \eta^{\mu \rho} \eta^{\nu \sigma}
{\mathbf A}^i_{\mu \nu } {\mathbf A}^i_{ \rho \sigma } \\
&&\\
&&- \frac{1}{4} \eta^{\mu \rho} \eta^{\nu \sigma} g_{\alpha \beta }
F_{\mu \nu}^{\alpha} F_{\rho \sigma}^{\beta},
\end{array}
\label{8.16}
\ee
where
\be
{\mathbf A}^i_{\mu \nu } = A^i_{\mu \nu} + g G^{-1
\lambda}_{\sigma} A^i_{\lambda} F^{\sigma}_{\mu\nu},
\label{8.17}
\ee
\be
A^i_{\mu \nu } =
D_{\mu} A^i_{\nu } - D_{\nu} A^i_{ \mu }
+ g_c f_{i j k}
A^j_{ \mu } A^k_{\nu }.
\label{8.18}
\ee
It can be proved that this system has both $SU(3)_c$
gauge symmetry and gravitational gauge symmetry.
The unified electroweak model can be discussed in
similar way. \\

Now, let's try to construct a theory which can describe
all kinds of fundamental interactions in Nature. First we
know that the fundamental particles that we know are
fundamental fermions(such as leptons and quarks),
gauge bosons(such as photon, gluons, gravitons and
intermediate gauge bosons $W^{\pm}$ and $Z^0$), and
possible Higgs bosons. According to the Standard Model,
leptons form left-hand doublets and right-hand singlets.
Let's denote
\be
\psi^{(1)}_L =\left (
\begin{array}{c}
\nu_e  \\
e
\end{array}
\right )_L
~~~,~~~
\psi^{(2)}_L =\left (
\begin{array}{c}
\nu_{\mu}  \\
\mu
\end{array}
\right )_L
~~~,~~~
\psi^{(3)}_L =\left (
\begin{array}{c}
\nu_{\tau}  \\
\tau
\end{array}
\right )_L ,
\label{8.19}
\ee
\be
\psi^{(1)}_R=e_R
~~~,~~~
\psi^{(2)}_R=\mu_R
~~~,~~~
\psi^{(3)}_R=\tau_R.
\label{8.20}
\ee
Neutrinos have no right-hand singlets. The weak hypercharge
for left-hand doublets $\psi^{(i)}_L$ is $-1$ and for right-hand
singlet $\psi^{(i)}_R$ is $-2$. All leptons carry no color
charge. In order to define the wave
function for quarks, we have to introduce
Kabayashi-Maskawa mixing matrix first, whose general form is,
\be
K =
\left (
\begin{array}{ccc}
c_1 & s_1 c_3 & s_1 s_3 \\
-s_1 c_2 & c_1 c_2 c_3 - s_2 s_3 e^{i \delta}
& c_1 c_2 s_3 + s_2 c_3 e^{i \delta}   \\
s_1 s_2 & -c_1 s_2 c_3 -c_2 s_3 e^{i \delta}
& -c_1 s_2 s_3 +c_2 c_3 e^{i \delta}
\end{array}
\right )
\label{8.21}
\ee
where
\be
c_i = {\rm cos} \theta_i ~~,~~~~ s_i = {\rm sin} \theta_i ~~~(i=1,2,3)
\label{8.22}
\ee
and $\theta_i$ are generalized Cabibbo angles. The mixing between
three different quarks $d,s$ and $b$ is given by
\be
\left (
\begin{array}{c}
d_{\theta} \\
s_{\theta}  \\
b_{\theta}
\end{array}
\right )
= K
\left (
\begin{array}{c}
d\\
s\\
b
\end{array}
\right ).
\label{8.23}
\ee
Quarks also form left-hand doublets and right-hand singlets,
\be
q_L^{(1)a} =
\left (
\begin{array}{c}
u_L^a \\
d_{\theta L}^a
\end{array}
\right ) ,~~
q_L^{(2)a} =
\left (
\begin{array}{c}
c_L^a \\
s_{\theta L}^a
\end{array}
\right ),~~
q_L^{(3)a} =
\left (
\begin{array}{c}
t_L^a \\
b_{\theta L}^a
\end{array}
\right ) ,
\label{8.24}
\ee
\be
\begin{array}{ccc}
q_u^{(1)a}= u_R^a
& q_u^{(2)a}= c_R^a
& q_u^{(3)a}= t_R^a \\
q_{ \theta d}^{(1)a}= d_{\theta R}^a
& q_{ \theta d}^{(2)a}= s_{\theta R}^a
& q_{ \theta d}^{(3)a}= b_{\theta R}^a,
\end{array}
\label{8.25}
\ee
where index $a$ is color index.
It is known that left-hand doublets have weak isospin
$\frac{1}{2}$ and weak hypercharge  $\frac{1}{3}$,
right-hand singlets  have no weak isospin, $q_u^{(j)a}$s
have weak hypercharge $\frac{4}{3}$ and $q_{\theta d}^{(j)a}$s
have weak hypercharge $ - \frac{2}{3}$. \\

For gauge bosons, gravitational gauge field is also denoted
by $C_{\mu}^{\alpha}$. The gluon field is denoted $A_{\mu}$,
\be
A_{\mu} =
A_{\mu}^i \frac{\lambda^i}{2}.
\label{8.26}
\ee
The color gauge covariant field strength tensor is also
given by eq.(\ref{8.18}). The $U(1)_Y$ gauge field is denoted
by $B_{\mu}$ and $SU(2)$ gauge field is denoted by
$F_{\mu}$
\be
F_{\mu} =
F^n_{\mu} \frac{\sigma_n}{2},
\label{8.27}
\ee
where $\sigma_n$ is the Pauli matrix. The $U(1)_Y$ gauge field strength
tensor is given by
\be
{\mathbf B}_{\mu\nu} = B_{\mu\nu} + g G^{-1 \lambda}_{\alpha}
B_{\lambda} F^{\alpha}_{\mu\nu},
\label{8.28a}
\ee
where
\be B_{ \mu \nu} = D_{\mu} B_{ \nu} - D_{\nu} B_{\mu},
\label{8.28b}
\ee
and the $SU(2)$ gauge field strength tensor is given by
\be
{\mathbf F}^n_{\mu\nu} = F^n_{\mu\nu} + g G^{-1 \lambda}_{\alpha}
F^n_{\lambda} F^{\alpha}_{\mu\nu}, 
\label{8.29a}
\ee
\be
F_{\mu \nu}^n = D_{\mu} F_{\nu}^n - D_{\nu} F_{\mu}^n + g_w
\epsilon _{lmn} F_{\mu}^l    F_{\nu}^m,
\label{8.29b}
\ee
where $g_w$ is the coupling constant for $SU(2)$ gauge interactions
and the coupling constant for $U(1)_Y$ gauge interactions is $g'_w$.
\\

If there exist Higgs particles in Nature, the Higgs fields is
represented by a complex scalar $SU(2)$ doublet,
\be
\phi =
\left (
\begin{array}{c}
\phi^{\dagger} \\
\phi^0
\end{array}
\right ).
\label{8.30}
\ee
The hypercharge of Higgs field $\phi$ is $1$. \\

The Lagrangian ${\cal L}_0$ that describes four kinds of fundamental
interactions is given by
\be
\begin{array}{rcl}
{\cal L}_0 &=&
-\sum_{j=1}^{3} \overline{\psi}_L^{(j)} \gamma ^{\mu}
(D_{\mu}+ \frac{i}{2} g'_w  B_{\mu} -ig_w F_{\mu} ) \psi_L^{(j)} \\
&&\\
&&- \sum_{j=1}^{3}\overline{e}_R^{(j)} \gamma ^{\mu}
(D_{\mu}+ ig'_w  B_{\mu} ) e_R^{(j)}  \\
&&\\
&&-\sum_{j=1}^{3} \overline{q}_L^{(j)a} \gamma^{\mu}
\left( (D_{\mu}-ig_w F_{\mu}- \frac{i}{6}g'_w B_{\mu} )\delta_{ab}
-i g_c A^k_{\mu} (\frac{\lambda^k}{2})_{ab} \right) q_L^{(j)b} \\
&&\\
&&-\sum_{j=1}^{3} \overline{q}_u^{(j)a} \gamma^{\mu}
\left( (D_{\mu}-i \frac{2}{3} g'_w B_{\mu} )\delta_{ab}
-i g_c A^k_{\mu} (\frac{\lambda^k}{2})_{ab} \right) q_u^{(j)b}  \\
&&\\
&& -\sum_{j=1}^{3} \overline{q}_{\theta d}^{(j)a} \gamma^{\mu}
\left( (D_{\mu} + i \frac{1}{3} g'_w B_{\mu} )\delta_{ab}
-i g_c A^k_{\mu} (\frac{\lambda^k}{2})_{ab} \right) q_{\theta d}^{(j)b} \\
&& \\
&&-\frac{1}{4}  \eta^{\mu \rho} \eta^{\nu \sigma}
{\mathbf F}^{n}_{ \mu \nu} {\mathbf F}^n_{\rho \sigma}
-\frac{1}{4} \eta^{\mu \rho} \eta^{\nu \sigma}
{\mathbf B}_{\mu \nu} {\mathbf B}_{\rho \sigma} \\
&& \\
&& -\frac{1}{4} \eta^{\mu \rho} \eta^{\nu \sigma}
{\mathbf A}^i_{\mu \nu } {\mathbf A}^i_{ \rho \sigma }
- \frac{1}{4} \eta^{\mu \rho} \eta^{\nu \sigma} g_{\alpha \beta }
F_{\mu \nu}^{\alpha} F_{\rho \sigma}^{\beta} \\
&&\\
&& -\left\lbrack (D_{\mu}- \frac{i}{2}
g'_w  B_{\mu} -ig_w F_{\mu}) \phi \right\rbrack ^{\dagger}
\cdot \left\lbrack (D_{\mu}- \frac{i}{2}
g'_w  B_{\mu} -ig_w F_{\mu}) \phi \right\rbrack \\
&&\\
&&   - \mu^2 \phi^{\dagger} \phi
+ \lambda (\phi^{\dagger} \phi)^2  \\
&&\\
&& - \sum_{j=1}^{3} f^{(j)}
\left(\overline{e}_R^{(j)} \phi^{\dag} \psi_L^{(j)}
+\overline{\psi}_L^{(j)} \phi  e_R^{(j)}\right)  \\
&&\\
&& -\sum_{j=1}^{3} \left( f_u^{(j)} \overline{q}_L^{(j)a}
\overline{\phi}
q_u^{(j)a} + f_u^{(j) \ast} \overline{q}_u^{(j)a}
\overline{\phi}^{\dag} q_L^{(j)a} \right)   \\
&&\\
&&-\sum_{j,k=1}^{3} \left( f_d^{(jk)} \overline{q}_L^{(j)a} \phi
q_{\theta d}^{(k)a}
+ f_d^{(jk) \ast} \overline{q}_{\theta d}^{(k)a}
\phi^{\dag} q_L^{(j)a} \right),
\end{array}
\label{8.31}
\ee
where
\be
\overline{\phi} = i \sigma_{2} \phi^{\ast} =
\left (
\begin{array}{c}
\phi^{0 \dag} \\
- \phi
\end{array}
\right ).
\label{8.32}
\ee
The full Lagrangian is given by
\be
{\cal L} = J(C) {\cal L}_0.
\label{8.33}
\ee
This Lagrangian describes four kinds of  fundamental interactions
in Nature. It has ($SU(3) \times SU(2) \times U(1)) \otimes_s
Gravitational~ Gauge ~Group$ symmetry\cite{7}. Four kinds of fundamental
interactions are formally unified in this Lagrangian. However,
this unification is not a genuine unification. Finally, an
important and fundamental problem is that, can we genuine unify
four kinds of fundamental interactions in a single group, in which
there is only one coupling constant for all kinds
of fundamental interactions? This theory may exist. \\

\section{Classical Limit of Quantum Gravity}
\setcounter{equation}{0}

In this chapter, we mainly discuss leading order approximation
of quantum gauge theory of gravity, which will give out 
classical Newton's theory of gravity. \\

First, we discuss an important problem qualitatively. It is 
know that, in usual gauge theory, such as QED, the coulomb 
force between two objects which carry like electric charges 
is always mutual repulsive. Gravitational gauge theory is 
also a kind of gauge theory, is the force between two static
massive objects attractive or repulsive? For the sake of 
simplicity, we use Dirac field as an example to discuss 
this problem. The discussions for
other kinds of fields can be proceeded similarly.  \\

Suppose that the gravitational field is very weak, so both the gravitational
field and the gravitational coupling constant are first order infinitesimal
quantities. Then in leading order approximation, both inertial
energy-momentum tensor and gravitational energy-momentum tensor
give the same results, which we denoted as
\be
T^{\mu}_{\alpha} =
\bar{\psi} \gamma^{\mu} \partial_{\alpha} \psi.
\label{9.1}
\ee
The time component of the current is
\be
T^{0}_{\alpha} =
-i \psi^{\dagger} \partial_{\alpha} \psi = \psi^{\dagger} \hat{P}_{\alpha} \psi.
\label{9.2}
\ee
Its space integration  gives out the energy-momentum of
the system. The interaction Lagrangian between Dirac field and
gravitational field is given by eq.(\ref{6.7}). After considering the
equation of motion of Dirac field, the coupling
between Dirac field and Gravitational gauge field in the leading order is:
\be
{\cal L}_I \approx g T_{\alpha}^{\mu} C_{\mu}^{\alpha}.
\label{9.3}
\ee
The the leading order interaction Hamiltonian density is given by
\be
{\cal H}_I  \approx - {\cal L}_I \approx
 - g T_{\alpha}^{\mu} C_{\mu}^{\alpha}.
\label{9.4}
\ee
The equation of motion of gravitational gauge field in the leading order is:
\be
\partial_{\lambda} \partial^{\lambda}
( \eta^{\nu \tau} \eta_{\alpha \beta}  C_{\tau}^{\beta})
- \partial^{\lambda} \partial^{\nu}
(  \eta_{\alpha \beta}  C_{\lambda}^{\beta})
= - g T^{\nu}_{\alpha}.
\label{9.5}
\ee
As a classical limit approximation,
let's consider static gravitational interactions
between two static objects. In this case,
the leading order component of energy-momentum tensor is $T^0_0$, other
components of energy-momentum tensor is a first order infinitesimal
quantity. So, we only need to consider the equation of motion of
$\nu = \alpha = 0$ of eq.(\ref{9.5}), which now becomes
\be
\partial_{\lambda} \partial^{\lambda} C_0^0
- \partial^{\lambda} \partial^0 C_{\lambda}^{0}
= - g T^0_0.
\label{9.6}
\ee
For static problems, all time derivatives vanish. Therefor,
the above equation is changed into
\be
\nabla ^2 C_0^0 = - g T^0_0.
\label{9.7}
\ee
This is just the Newton's equation of gravitational field. Suppose
that there is only one point object at the origin of the coordinate
system. Because $T^0_0$ is the negative value of energy density,
we can let
\be
T^0_0 = - M \delta(\svec{x}) .
\label{9.8}
\ee
Applying
\be
\nabla ^2  \frac{1}{r} = - 4 \pi \delta(\svec{x}),
\label{9.9}
\ee
with $r =  |\svec{x} |$, we get
\be
C^0_0  = - \frac{gM}{4 \pi r}.
\label{9.10}
\ee
This is just the gravitational potential which is expected in Newton's
theory of gravity. \\

Suppose that there is another point object at the position of
point $\svec{x}$ with mass $m$. The gravitational potential energy
between these two objects is that
\be
V(r) = \int {\rm d}^3 \svec{y} {\cal H}_I
=  -g \int {\rm d}^3 \svec{y} T^0_{2~0} (\svec{x}) C_0^0,
\label{9.11}
\ee
with $C^0_0$ is the gravitational potential generated by the first
point object, and $T^0_{2~0}$ is the $(0,0)$ component of the
energy-momentum tensor of the second object,
\be
T^0_{2~0} (\svec{y})= - m  \delta(\svec{y} - \svec{x}) .
\label{9.12}
\ee
The final result for gravitational potential energy between two
point objects is
\be
V(r) = - \frac{g^2 M m}{4 \pi r}.
\label{9.13}
\ee
The gravitational potential energy between two point objects is always
negative, which is what expected by Newton's theory of gravity and is
the inevitable result of the attractive nature of gravitational
interactions.
\\

The gravitational force that the first point object acts on the second
point object is
\be
\svec{f} = - \nabla V(r) = - \frac{g^2 M m}{4 \pi r^2} \hat{r},
\label{9.14}
\ee
where $\hat r = \svec{r} / r $. This is the famous formula
of Newton's gravitational force.  Therefore, in the classical limit,
the gravitational gauge theory can return to Newton's theory of
gravity. Besides, from eq.(\ref{9.14}), we can clearly see that the
gravitational interaction force between two point objects is
attractive.
\\

Now, we want to ask a problem: why in QED, the force between 
two like electric charges is always repulsive, while in 
gravitational gauge theory, the force between two like 
gravitational charges can be attractive?
A simple answer to this fundamental problem is that the attractive 
nature of the gravitational force is an inevitable result of
the global Lorentz symmetry of the system. Because of the requirement
of global Lorentz symmetry, the Lagrangian
function given by eq.(\ref{4.20}) must use  $g_{\alpha \beta}$ ,
 can not use the ordinary $\delta$-function $\delta_{\alpha \beta}$.
It can be easily prove that, if we use $\delta_{\alpha \beta}$ 
instead of $g_{\alpha \beta}$ in eq.(\ref{4.20}), the Lagrangian 
of pure gravitational gauge field is not invariant under global 
Lorentz transformation.  On the other hand,
if we use $\delta_{\alpha \beta}$ instead of  $\eta_{\alpha \beta}$
in eq.(\ref{4.20}), the gravitational force will be repulsive which obviously
contradicts with experiment results. In QED, $\delta_{a b}$ is
used to construct the Lagrangian for electromagnetic fields, therefore,
the interaction force between two like electric charges is
always repulsive.
\\

One fundamental influence of using the metric $g_{\alpha \beta}$
in the Lagrangian of pure gravitational field is that the kinematic
energy term of gravitation field $C_{\mu}^0$ is always negative.
According to  eq.(\ref{4.20}), the free lagrangian of 
pure gravitational  gauge field is
\be
{\cal L}_{0F} = - \frac{1}{4} \eta^{\mu \rho} \eta^{\nu \sigma}
\eta_{ \alpha \beta}
F_{\mu \nu}^{\alpha} F_{\rho \sigma}^{\beta}.
\label{9.1401}
\ee
In above relation, $\eta^{00}$ is negative which causes
that the kinematic energy  of gravitation field $C_{\mu}^0$ 
is  negative. This result is novels, but it is not surprising, 
for gravitational interaction energy is always negative.
In a meaning, it is the reflection of the
negative nature of graviton's kinematic energy. Though the kinematic
energy term of gravitation field $C_{\mu}^0$ is always negative,
the kinematic energy term of gravitation field $C_{\mu}^i$ is always
positive. The negative energy problem of graviton does not cause any
trouble in quantum gauge theory of gravity. Contrarily, it will help us
to understand some puzzle phenomena of Nature. From theoretical
point of view, the negative nature of graviton's kinematic energy
is essentially an inevitable result of global Lorentz symmetry.
Global Lorentz symmetry of the system,
attractive nature of gravitational interaction
force and negative nature of graviton's kinematical energy are essentially
related to each other, and they have the same origin in nature.
\\

In general relativity, gravitational field obeys Einstein field equation, which
is usually written in the following form,
\be
R_{\mu \nu} -\frac{1}{2} g_{\mu \nu} R + \lambda g_{\mu \nu}
= - 8 \pi G T_{\mu \nu},
\label{9.15}
\ee
where $R_{\mu \nu}$ is Ricci tensor, $R$ is curvature, $G$ is Newton
gravitational constant and $\lambda$ is cosmology constant. The classical
limit of Einstein field equation is
\be
\nabla^2 g_{00} =
  - 8 \pi G T_{0 0}.
\label{9.16}
\ee
Compare this equation with eq.(\ref{9.7}) and use 
eq.(\ref{4.707}), we get
\be
g^2 = 4 \pi G.
\label{9.17}
\ee
In order to get eq.(\ref{9.17}), the following relations are used
\be
T_{0 0} = - T^0_0,~~~~
g_{00} \simeq -(1 + 2 g C_0^0).
\label{9.18}
\ee
\\

In general relativity, Einstein field equation transforms covariantly
under general coordinates transformation, in other words, it is a general
covariant equation.  In gravitational gauge theory, the system has
local gravitational gauge symmetry. From mathematical point of view,
general coordinates transformation is equivalent to local gravitational
gauge transformation. Therefore, it seems that two theories have
the same symmetry. On the other hand, both theories have
global Lorentz symmetry.\\

\section{Path Integral Quantization of Gravitational\\ Gauge Fields}
\setcounter{equation}{0}

For the sake of simplicity, in this chapter and the next chapter, we only
discuss pure gravitational gauge field. For pure gravitational gauge
field, its Lagrangian function is
\be
{\cal L} = - \frac{1}{4} \eta^{\mu \rho} \eta^{\nu \sigma}
g_{\alpha \beta} J(C)
F_{\mu \nu}^{\alpha} F_{\rho \sigma}^{\beta}.
\label{10.1}
\ee
Its space-time integration gives out the action of the system
\be
S = \int {\rm d}^4 x  {\cal L}.
\label{10.2}
\ee
This action has local gravitational gauge symmetry. Gravitational
gauge field $C_{\mu}^{\alpha}$ has $4 \times 4 = 16$ degrees of
freedom. But, if gravitons are massless, the system has only
$2 \times 4 = 8$ degrees of freedom. There are gauge degrees
of freedom in the theory. Because only physical degrees of
freedom can be quantized, in order to quantize the system, we
have to introduce gauge conditions to eliminate un-physical degrees
of freedom. For the sake of convenience, we take temporal gauge
conditions
\be
C_0^{\alpha} = 0, ~~~(\alpha = 0,1,2,3).
\label{10.3}
\ee
\\

In temporal gauge, the generating functional $W\lbrack J \rbrack$ is
given by
\be
W\lbrack J \rbrack = N \int \lbrack {\cal D} C\rbrack
\left(\prod_{\alpha, x} \delta( C_0^{\alpha}(x))\right)
exp \left\lbrace i \int {\rm d}^4 x ( {\cal L}
+ J^{\mu}_{\alpha} C^{\alpha}_{\mu}),
\right\rbrace
\label{10.4}
\ee
where $N$ is the normalization constant, $J^{\mu}_{\alpha}$
is a fixed external source and $ \lbrack {\cal D} C\rbrack $ is the
integration measure,
\be
\lbrack {\cal D} C\rbrack
= \prod_{\mu=0}^{3} \prod_{\alpha= 0}^{3} \prod_j
\left(\varepsilon {\rm d}C_{\mu}^{\alpha} (\tau_j)
/ \sqrt{2 \pi i \hbar} \right).
\label{10.5}
\ee
We use this generation functional as our starting
point of the path integral quantization of gravitational gauge field. \\

Generally speaking, the action of the system has local gravitational gauge
symmetry, but the gauge condition has no local gravitational gauge
symmetry. If we make a local gravitational gauge transformations,
the action of the system is kept unchanged while gauge condition will
be changed. Therefore, through local gravitational gauge transformation,
we can change one gauge condition to another gauge condition. The most
general gauge condition is
\be
f^{\alpha} (C(x)) - \varphi^{\alpha} (x) = 0,
\label{10.6}
\ee
where $\varphi^{\alpha}(x)$ is an arbitrary space-time function.
The Fadeev-Popov determinant $\Delta_f(C)$ \cite{f01} is defined by
\be
\Delta_f^{-1} (C) \equiv
\int \lbrack {\cal D} g \rbrack
\prod_{x, \alpha} \delta \left(f^{\alpha} ( ^gC(x))
- \varphi^{\alpha}(x) \right),
\label{10.7}
\ee
where $g$ is an element of gravitational gauge group, $^gC$ is the gravitational
gauge field after gauge transformation $g$ and
$\lbrack {\cal D} g \rbrack $ is the integration measure on
gravitational gauge group
\be
\lbrack {\cal D} g \rbrack
= \prod_x {\rm d}^4 \epsilon(x),
\label{10.8}
\ee
where $\epsilon (x)$ is the transformation parameter of $\ehat$.
Both $\lbrack {\cal D} g \rbrack $ and $\lbrack {\cal D} C \rbrack $
are not invariant under gravitational gauge transformation. Suppose
that,
\be
\lbrack {\cal D} (gg') \rbrack
= J_1(g') \lbrack {\cal D} g \rbrack,
\label{10.9}
\ee
\be
\lbrack {\cal D} ~^gC \rbrack
= J_2(g) \lbrack {\cal D} C \rbrack.
\label{10.10}
\ee
$J_1(g)$ and $J_2(g)$ satisfy the following relations
\be
J_1(g) \cdot J_1(g^{-1}) = 1,
\label{10.11}
\ee
\be
J_2(g) \cdot J_2(g^{-1}) = 1.
\label{10.12}
\ee
It can be proved that, under gravitational gauge transformations,
the Fadeev-Popov determinant transforms as
\be
\Delta_f^{-1} ( ^{g'}C ) = J_1^{-1}(g') \Delta_f^{-1} (C).
\label{10.13}
\ee
\\

Insert eq.(\ref{10.7}) into eq.(\ref{10.4}), we get
\be
\begin{array}{rcl}
W\lbrack J \rbrack &=& N \int \lbrack {\cal D} g \rbrack
\int \lbrack {\cal D} C\rbrack~~
\left\lbrack \prod_{\alpha, y}
\delta( C_0^{\alpha}(y)) \right\rbrack \cdot
\Delta_f  (C) \\
&&\\
&&\cdot \left\lbrack \prod_{\beta, z}
\delta ( f^{\beta} ( ^gC(z))- \varphi^{\beta}(z) )\right\rbrack
 \cdot exp \left\lbrace i \int {\rm d}^4 x ( {\cal L}
+ J^{\mu}_{\alpha} C^{\alpha}_{\mu})
\right\rbrace .
\end{array}
\label{10.14}
\ee
Make a gravitational gauge transformation,
\be
C(x)~~ \to ~~^{g^{-1}} C(x),
\label{10.15}
\ee
then,
\be
^g C(x)~~ \to
~~^{gg^{-1}} C(x).
\label{10.16}
\ee
After this transformation, the generating functional is changed
into
\be
\begin{array}{rcl}
W\lbrack J \rbrack &=& N \int \lbrack {\cal D} g \rbrack
\int \lbrack {\cal D} C\rbrack~~
J_1(g) J_2(g^{-1}) \cdot \left\lbrack \prod_{\alpha, y}
\delta( ^{g^{-1}}C_0^{\alpha}(y)) \right\rbrack \cdot
\Delta_f  (C) \\
&&\\
&&\cdot \left\lbrack \prod_{\beta, z}
\delta ( f^{\beta} ( C(z))- \varphi^{\beta}(z) )\right\rbrack
 \cdot exp \left\lbrace i \int {\rm d}^4 x ( {\cal L}
+ J^{\mu}_{\alpha} \cdot  ^{g^{-1}} \!\!\! C^{\alpha}_{\mu})
\right\rbrace.
\end{array}
\label{10.17}
\ee
\\

Suppose that the gauge transformation $g_0(C)$ transforms general
gauge condition $f^{\beta}(C) - \varphi^{\beta} = 0$ to temporal
gauge condition $C_0^{\alpha} = 0$, and suppose that this transformation
$g_0(C)$ is unique. Then two $\delta$-functions in eq.(\ref{10.17}) require
that the integration on gravitational gauge group must be in the
neighborhood of $g^{-1}_0(C)$. Therefore eq.(\ref{10.17}) is changed into
\be
\begin{array}{rcl}
W\lbrack J \rbrack &=& N \int \lbrack {\cal D} C\rbrack~~
\Delta_f  (C)  \cdot \left\lbrack \prod_{\beta, z}
\delta ( f^{\beta} ( C(z))- \varphi^{\beta}(z) )\right\rbrack \\
&&\\
&& \cdot exp \left\lbrace i \int {\rm d}^4 x ( {\cal L}
+ J^{\mu}_{\alpha} \cdot  ^{g_0}\!C^{\alpha}_{\mu})
\right\rbrace \\
&&\\
&& \cdot J_1(g_0^{-1}) J_2(g_0) \cdot \int \lbrack {\cal D} g \rbrack
\left\lbrack \prod_{\alpha, y}
\delta( ^{g^{-1}}C_0^{\alpha}(y))\right\rbrack.
\end{array}
\label{10.18}
\ee
The last line in eq.(\ref{10.18}) will cause no trouble in renormalization,
and if we consider the contribution from ghost fields which will
be introduced below, it will becomes a quantity which is independent
of gravitational gauge field. So, we put it into normalization constant
$N$ and still denote the new normalization constant as $N$. We also
change $J^{\mu}_{\alpha} ~  ^{g_0}\!C^{\alpha}_{\mu}$
into $J^{\mu}_{\alpha} C^{\alpha}_{\mu}$, this will cause no
trouble in renormalization. Then we get
\be
\begin{array}{rcl}
W\lbrack J \rbrack &=& N \int \lbrack {\cal D} C\rbrack~~
\Delta_f  (C)  \cdot \lbrack \prod_{\beta, z}
\delta ( f^{\beta} ( C(z))- \varphi^{\beta}(z) )\rbrack \\
&& \cdot exp \lbrace i \int {\rm d}^4 x ( {\cal L}
+ J^{\mu}_{\alpha} C^{\alpha}_{\mu}) \rbrace.
\end{array}
\label{10.19}
\ee
In fact, we can use this formula as our start-point of path integral
quantization of gravitational gauge field, so we need not worried
about the influences of the third lines of eq.(\ref{10.18}).
\\

Use another functional
\be
exp \left\lbrace - \frac{i}{2 \alpha}
\int {\rm d}^4 x \eta_{\alpha \beta}
\varphi^{\alpha}(x) \varphi^{\beta}(x) \right\rbrace,
\label{10.20}
\ee
times both sides of eq.(\ref{10.19}) and then make functional
integration
$\int \lbrack {\cal D} \varphi  \rbrack$,
we get
\be
W\lbrack J \rbrack  = N \int \lbrack {\cal D} C\rbrack~~
\Delta_f (C) \cdot exp \left\lbrace i \int {\rm d}^4 x ( {\cal L}
- \frac{1}{2 \alpha} \eta_{\alpha \beta} f^{\alpha} f^{\beta}
+ J^{\mu}_{\alpha} C^{\alpha}_{\mu}) \right\rbrace.
\label{10.21}
\ee
Now, let's discuss the contribution from $\Delta_f (C)$ which
is related to the ghost fields. Suppose that $g = \ehat$
is an infinitesimal gravitational gauge transformation. Then
eq.(\ref{4.12}) gives out
\be
^gC_{\mu}^{\alpha} (x)
= C_{\mu}^{\alpha} (x)
- \frac{1}{g} {\mathbf D}_{\mu~\sigma}^{\alpha} \epsilon^{\sigma},
\label{10.22}
\ee
where
\be
{\mathbf D}_{\mu~\sigma}^{\alpha}
=\delta^{\alpha}_{\sigma} \partial_{\mu}
- g \delta^{\alpha}_{\sigma} C_{\mu}^{\beta} \partial_{\beta}
+ g \partial_{\sigma} C_{\mu}^{\alpha}.
\label{10.23}
\ee
In order to deduce eq.(\ref{10.22}), the following relation is used
\be
\Lambda^{\alpha}_{~\beta}
= \delta^{\alpha}_{\beta}
+ \partial_{\beta} \epsilon^{\alpha}
+ o( \epsilon^2).
\label{10.24}
\ee
${\mathbf D}_{\mu}$ can be regarded as the covariant derivate
in adjoint representation, for
\be
{\mathbf D}_{\mu} \epsilon
= \lbrack D_{\mu} ~~~,~~~ \epsilon \rbrack,
\label{10.25}
\ee
\be
({\mathbf D}_{\mu} \epsilon)^{\alpha}
= {\mathbf D}_{\mu~\sigma}^{\alpha} \epsilon^{\sigma}.
\label{10.26}
\ee
Using all these relations,  we have,
\be
f^{\alpha} (^gC(x)) = f^{\alpha} (C)
- \frac{1}{g} \int {\rm d}^4 y
\frac{\delta f^{\alpha}(C(x))}{\delta C_{\mu}^{\beta}(y)}
{\mathbf D}_{\mu~\sigma}^{\beta}(y) \epsilon^{\sigma}(y)
+ o(\epsilon^2).
\label{10.27}
\ee
Therefore, according to eq.(\ref{10.7}) and eq.(\ref{10.6}), we get
\be
\Delta_f^{-1} (C) =
\int \lbrack {\cal D} \epsilon \rbrack
\prod_{x, \alpha}
\delta \left( - \frac{1}{g}     \int {\rm d}^4 y
\frac{\delta f^{\alpha}(C(x))}{\delta C_{\mu}^{\beta}(y)}
{\mathbf D}_{\mu~\sigma}^{\beta}(y) \epsilon^{\sigma}(y) \right).
\label{10.28}
\ee
Define
\be
\begin{array}{rcl}
{\mathbf M}^{\alpha}_{~\sigma}(x,y) &=& -g
\frac{\delta}{\delta \epsilon^{\sigma}(y)}
f^{\alpha}(^gC(x)) \\
&&\\
&=&\int {\rm d}^4 z
\frac{\delta f^{\alpha}(C(x))}{\delta C_{\mu}^{\beta}(z)}
{\mathbf D}_{\mu~\sigma}^{\beta}(z) \delta(z-y) .
\end{array}
\label{10.29}
\ee
Then eq.(\ref{10.28}) is changed into
\be
\begin{array}{rcl}
\Delta_f^{-1} (C) &=&
\int \lbrack {\cal D} \epsilon \rbrack
\prod_{x, \alpha}
\delta \left( - \frac{1}{g} \int {\rm d}^4 y
{\mathbf M}^{\alpha}_{~\sigma}(x,y) \epsilon^{\sigma}(y)
\right)  \\
&&\\
&=& const. \times (det {\mathbf M} )^{-1}.
\end{array}
\label{10.30}
\ee
Therefore,
\be
\Delta_f (C) = const. \times det {\mathbf M}.
\label{10.31}
\ee
Put this constant into normalization constant, then generating
functional eq.(\ref{10.21}) is changed into
\be
W\lbrack J \rbrack  = N \int \lbrack {\cal D} C\rbrack~~
det {\mathbf M} \cdot exp \left\lbrace i \int {\rm d}^4 x ( {\cal L}
- \frac{1}{2 \alpha} \eta_{\alpha \beta} f^{\alpha} f^{\beta}
+ J^{\mu}_{\alpha} C^{\alpha}_{\mu}) \right\rbrace.
\label{10.32}
\ee
\\

In order to evaluate the contribution from $det {\mathbf M}$,
we introduce ghost fields $\eta^{\alpha}(x)$
and $\bar{\eta}_{\alpha}(x)$. Using the following relation
\be
\int \lbrack {\cal D} \eta \rbrack
\lbrack {\cal D}\bar{\eta} \rbrack
exp \left\lbrace i \int {\rm d}^4 x {\rm d}^4 y ~
\bar{\eta}_{\alpha} (x) {\mathbf M}^{\alpha}_{~\beta}(x,y)
\eta^{\beta} (y) \right\rbrace
 = const. \times det {\mathbf M}
\label{10.33}
\ee
and put the constant into the normalization constant, we can get
\be
W\lbrack J \rbrack  = N \int \lbrack {\cal D} C\rbrack
\lbrack {\cal D} \eta \rbrack
\lbrack {\cal D}\bar{\eta} \rbrack
exp \left\lbrace i \int {\rm d}^4 x ( {\cal L}
- \frac{1}{2 \alpha} \eta_{\alpha \beta} f^{\alpha} f^{\beta}
+ \bar{\eta} {\mathbf M } \eta
+ J^{\mu}_{\alpha} C^{\alpha}_{\mu}) \right\rbrace,
\label{10.34}
\ee
where $\int {\rm d}^4x \bar{\eta} {\mathbf M } \eta$ is a
simplified notation, whose explicit expression is
\be
\int {\rm d}^4x \bar{\eta} {\mathbf M } \eta
= \int {\rm d}^4 x {\rm d}^4 y ~
\bar{\eta}_{\alpha} (x) {\mathbf M}^{\alpha}_{~\beta}(x,y)
\eta^{\beta} (y).
\label{10.35}
\ee
The appearance of the non-trivial ghost fields is a inevitable
result of the non-Able nature of the gravitational gauge group.
\\

Now, let's take Lorentz covariant gauge condition,
\be
f^{\alpha} (C) = \partial^{\mu} C_{\mu}^{\alpha} .
\label{10.36}
\ee
Then
\be
\int {\rm d}^4x \bar{\eta} {\mathbf M } \eta =
- \int {\rm d}^4x \left( \partial^{\mu}
\bar{\eta}_{\alpha} (x) \right)
{\mathbf D}_{\mu~\beta}^{\alpha}(x) \eta^{\beta} (x).
\label{10.37}
\ee
And eq.(\ref{10.34}) is changed into
\be
\begin{array}{rcl}
W\lbrack J, \beta, \bar{\beta} \rbrack
& = & N \int \lbrack {\cal D} C\rbrack
\lbrack {\cal D} \eta \rbrack
\lbrack {\cal D}\bar{\eta} \rbrack
exp \left\lbrace i \int {\rm d}^4 x ( {\cal L}
- \frac{1}{2 \alpha}
\eta_{\alpha \beta} f^{\alpha} f^{\beta} \right. \\
&&\\
&& \left. - (\partial^{\mu}\bar{\eta}_{\alpha} )
{\mathbf D}_{\mu~\sigma}^{\alpha} \eta^{\sigma}
+ J^{\mu}_{\alpha} C^{\alpha}_{\mu}
+ \bar{\eta}_{\alpha} \beta^{\alpha}
+ \bar{\beta}_{\alpha} \eta^{\alpha}
) \right\rbrace,
\end{array}
\label{10.38}
\ee
where we have introduced external sources $\eta^{\alpha}(x)$
and $\bar{\eta}_{\alpha}(x)$ of ghost fields. \\

The effective Lagrangian ${\cal L}_{eff}$
is defined by
\be
{\cal L}_{eff} \equiv
{\cal L} - \frac{1}{2 \alpha}
\eta_{\alpha \beta} f^{\alpha} f^{\beta}
- (\partial^{\mu}\bar{\eta}_{\alpha} )
{\mathbf D}_{\mu~\sigma}^{\alpha} \eta^{\sigma}.
\label{10.39}
\ee
${\cal L}_{eff}$ can separate into free Lagrangian
${\cal L}_F$  and interaction Lagrangian ${\cal L}_I$,
\be
{\cal L}_{eff} = {\cal L}_F + {\cal L}_I,
\label{10.40}
\ee
where
\be
\begin{array}{rcl}
{\cal L}_F &=& - \frac{1}{2}
\eta^{\mu \rho} \eta^{\nu \sigma}  g_{\alpha \beta}
\left\lbrack (\partial_{\mu} C_{\nu}^{\alpha})
(\partial_{\rho} C_{\sigma}^{\beta})
-(\partial_{\mu} C_{\nu}^{\alpha})
(\partial_{\sigma} C_{\rho}^{\beta})
\right\rbrack   \\
&&\\
&&  -\frac{1}{2 \alpha} \eta_{\alpha \beta}
(\partial^{\mu} C_{\mu}^{\alpha})
(\partial^{\nu} C_{\nu}^{\beta})
- (\partial^{\mu} \bar{\eta}_{\alpha})
(\partial_{\mu} \eta^{\alpha}),
\end{array}
\label{10.41}
\ee
\be
\begin{array}{rcl}
{\cal L}_I &=& - \frac{1}{2} 
\left(J(C) g_{\alpha \beta}  - \eta_{\alpha \beta} \right)
\eta^{\mu \rho} \eta^{\nu \sigma} 
\left\lbrack (\partial_{\mu} C_{\nu}^{\alpha})
(\partial_{\rho} C_{\sigma}^{\beta})
-(\partial_{\mu} C_{\nu}^{\alpha})
(\partial_{\sigma} C_{\rho}^{\beta})
\right\rbrack   \\
&&\\
&&   + g J(C) \eta^{\mu \rho} \eta^{\nu \sigma}
g_{\alpha \beta} (\partial_{\mu} C_{\nu }^{\alpha}
- \partial_{\nu} C_{\mu}^{\alpha})
C_{\rho}^{\delta} \partial_{\delta} C_{\sigma}^{\beta} \\
&&\\
&&  - \frac{1}{2} g^2 J(C) \eta^{\mu \rho} \eta^{\nu \sigma}
g_{\alpha \beta}
(C_{\mu}^{\delta} \partial_{\delta} C_{\nu}^{\alpha}
- C_{\nu}^{\delta} \partial_{\delta} C_{\mu}^{\alpha} )
C_{\rho}^{\epsilon} \partial_{\epsilon} C_{\sigma}^{\beta}\\
&&\\
&& + g(\partial^{\mu} \bar{\eta}_{\alpha})
C_{\mu}^{\beta} (\partial_{\beta} \eta^{\alpha})
- g(\partial^{\mu} \bar{\eta}_{\alpha})
(\partial_{\sigma} C_{\mu}^{\alpha}) \eta^{\sigma}.
\end{array}
\label{10.42}
\ee
From the interaction Lagrangian, we can see that ghost fields
do not couple to $J(C)$. This is the reflection of the fact
that ghost fields are not physical fields, they are virtual fields.
Besides, the gauge fixing term does not couple to $J(C)$ either.
Using effective Lagrangian ${\cal L}_{eff}$, the generating
functional $W\lbrack J, \beta, \bar{\beta} \rbrack$ can be
simplified to
\be
W\lbrack J, \beta, \bar{\beta} \rbrack
=  N \int \lbrack {\cal D} C\rbrack
\lbrack {\cal D} \eta \rbrack
\lbrack {\cal D}\bar{\eta} \rbrack
exp \left\lbrace i \int {\rm d}^4 x ( {\cal L}_{eff}
+ J^{\mu}_{\alpha} C^{\alpha}_{\mu}
+ \bar{\eta}_{\alpha} \beta^{\alpha}
+ \bar{\beta}_{\alpha} \eta^{\alpha}
) \right\rbrace,
\label{10.43}
\ee
\\

Use eq.(\ref{10.41}), we can deduce propagator of gravitational gauge
fields and ghost fields. First, we change its form to
\be
\int {\rm d}^4x {\cal L}_F =
\int {\rm d}^4x \left\lbrace \frac{1}{2}
C_{\nu}^{\alpha} \left\lbrack
\eta_{\alpha \beta} \left(
\eta^{\mu \nu} \partial^2 - (1 - \frac{1}{\alpha})
\partial^{\mu} \partial^{\nu} \right)
\right\rbrack C_{\nu}^{\beta}
+ \bar{\eta}_{\alpha}
\partial^2 \eta^{\alpha} \right\rbrace.
\label{10.44}
\ee
Denote the propagator of gravitational gauge field as
\be
-i \Delta_{F \mu \nu}^{\alpha \beta} (x),
\label{10.45}
\ee
and denote the propagator of ghost field as
\be
-i \Delta_{F \beta}^{\alpha } (x).
\label{10.46}
\ee
They satisfy the following equation,
\be
- \left\lbrack
\eta_{ \alpha \beta} \left(
\eta^{\mu \nu} \partial^2 - (1 - \frac{1}{\alpha})
\partial^{\mu} \partial^{\nu} \right)
\right\rbrack \Delta_{F \nu \rho}^{\beta \gamma} (x)
= \delta(x) \delta^{\gamma}_{\alpha} \delta^{\mu}_{\rho},
\label{10.47}
\ee
\be
- \partial^2 \Delta_{F \beta}^{\alpha } (x)
= \delta_{\beta}^{\alpha } \delta(x).
\label{10.48}
\ee
Make Fourier transformations to momentum space
\be
\Delta_{F \mu \nu}^{\alpha \beta} (x)
= \int \frac{{\rm d}^4k}{(2 \pi)^4}
\swav{\Delta}_{F \mu \nu}^{\alpha \beta}(k) \cdot  e^{ikx},
\label{10.49}
\ee
\be
\Delta_{F \beta}^{\alpha } (x)
= \int \frac{{\rm d}^4k}{(2 \pi)^4}
\swav{\Delta}_{F \beta}^{\alpha }(k) \cdot e^{ikx},
\label{10.50}
\ee
where $\swav{\Delta}_{F \mu \nu}^{\alpha \beta}(k)$
and $\swav{\Delta}_{F \beta}^{\alpha }(k)$ are corresponding
propagators in momentum space. They satisfy the following
equations,
\be
\eta_{\alpha \beta}
\left\lbrack
k^2 \eta^{\mu \nu} - (1 - \frac{1}{\alpha}) k^{\mu}k^{\nu}
\right\rbrack
\swav{\Delta}_{F \nu \rho}^{\beta \gamma}(k)
= \delta^{\gamma}_{\alpha} \delta^{\mu}_{\rho},
\label{10.51}
\ee
\be
k^2 \swav{\Delta}_{F \beta}^{\alpha }(k)
= \delta^{\alpha}_{\beta}.
\label{10.52}
\ee
The solutions to these two equations give out the propagators
in momentum space,
\be
-i \swav{\Delta}_{F \mu \nu}^{\alpha \beta}(k)
= \frac{-i}{k^2 - i \epsilon}
\eta^{\alpha  \beta}
\left\lbrack
\eta_{\mu \nu} - (1 - \alpha)
\frac{k_{\mu} k_{\nu}}{k^2 - i \epsilon},
\right\rbrack
\label{10.53}
\ee
\be
-i \swav{\Delta}_{F \beta}^{\alpha }(k) =
\frac{-i}{k^2 - i \epsilon} \delta^{\alpha}_{\beta}.
\label{10.54}
\ee
It can be seen that the forms of these propagators are quite
similar to those in traditional non-Able gauge theory. The only
difference is that the metric is different. \\

The interaction Lagrangian ${\cal L}_I$ is a function of
gravitational gauge field $C_{\mu}^{\alpha}$ and ghost
fields $\eta^{\alpha}$ and $\bar \eta_{\alpha}$,
\be
{\cal L}_I =
{\cal L}_I ( C, \eta, \bar\eta ).
\label{10.55}
\ee
Then eq.(\ref{10.43}) is changed into,
\be
\begin{array}{rcl}
W\lbrack J, \beta, \bar{\beta} \rbrack
& = & N \int \lbrack {\cal D} C\rbrack
\lbrack {\cal D} \eta \rbrack
\lbrack {\cal D}\bar{\eta} \rbrack
~exp \left\lbrace i \int {\rm d}^4 x
 {\cal L}_I ( C, \eta, \bar\eta ) \right\rbrace \\
&&\\
&&\cdot exp \left\lbrace i \int {\rm d}^4 x ( {\cal L}_F
+ J^{\mu}_{\alpha} C^{\alpha}_{\mu}
+ \bar{\eta}_{\alpha} \beta^{\alpha}
+ \bar{\beta}_{\alpha} \eta^{\alpha}
) \right\rbrace  \\
&&\\
&=& exp \left\lbrace i \int {\rm d}^4 x
 {\cal L}_I ( \frac{1}{i}\frac{\delta}{\delta J},
\frac{1}{i}\frac{\delta}{\delta \bar \beta},
\frac{1}{-i}\frac{\delta}{\delta \beta} ) \right\rbrace
\cdot W_0\lbrack J, \beta, \bar{\beta} \rbrack ,
\end{array}
\label{10.56}
\ee
where
\be
\begin{array}{rcl}
W_0\lbrack J, \beta, \bar{\beta} \rbrack
&=&  N \int \lbrack {\cal D} C\rbrack
\lbrack {\cal D} \eta \rbrack
\lbrack {\cal D}\bar{\eta} \rbrack
exp \left\lbrace i \int {\rm d}^4 x ( {\cal L}_F
+ J^{\mu}_{\alpha} C^{\alpha}_{\mu}
+ \bar{\eta}_{\alpha} \beta^{\alpha}
+ \bar{\beta}_{\alpha} \eta^{\alpha}
) \right\rbrace  \\
&&\\
&=&  exp \left\lbrace
\frac{i}{2} \int\int {\rm d}^4 x {\rm d}^4 y
\left\lbrack J^{\mu}_{\alpha} (x)
\Delta_{F \mu \nu}^{\alpha \beta} (x-y)
J^{\nu}_{\beta} (y) \right. \right.  \\
&&\\
&& \left.\left.~~+ \bar\eta_{\alpha}(x)
\Delta_{F \beta}^{\alpha } (x-y) \eta^{\beta}(y).
\right\rbrack  \right\rbrace
\end{array}
\label{10.57}
\ee
\\

Finally, let's discuss Feynman rules. Here, we only give
out the lowest order interactions in gravitational gauge
theory. It is known that, a vertex can involve arbitrary number
of gravitational gauge fields. Therefore, it is impossible
to list all Feynman rules for all kinds of vertex. \\

The interaction
Lagrangian between gravitational gauge field and ghost field
is
\be
+ g(\partial^{\mu} \bar{\eta}_{\alpha})
C_{\mu}^{\beta} (\partial^{\beta} \eta^{\alpha})
- g(\partial^{\mu} \bar{\eta}_{\alpha})
(\partial_{\sigma} C_{\mu}^{\alpha}) \eta^{\sigma}.
\label{10.58}
\ee
This vertex belongs to
$C_{\mu}^{\alpha}(k) \bar{\eta}_{\beta}(q) \eta^{\delta}(p) $
three body interactions, its Feynman rule is
\be
-i g \delta^{\beta}_{\delta} q^{\mu} p_{\alpha}
+ i g \delta^{\beta}_{\alpha} q^{\mu} k_{\delta}.
\label{10.59}
\ee
The lowest order interaction Lagrangian between gravitational gauge
field and Dirac field is
\be
g \bar \psi \gamma^{\mu} \partial_{\alpha} \psi C_{\mu}^{\alpha}
- g \eta_{1 \alpha}^{\mu} \bar \psi
\gamma^{\nu} \partial_{\nu} \psi C_{\mu}^{\alpha}
- g m \delta_{\alpha}^{\mu} \bar \psi \psi C_{\mu}^{\alpha}.
\label{10.60}
\ee
This vertex belongs to
$C_{\mu}^{\alpha}(k) \bar \psi(q) \psi(p)$
three body interactions, its Feynman rule is
\be
- g \gamma^{\mu} p_{\alpha}
+ g \eta_{1 \alpha}^{\mu} \gamma^{\nu} p_{\nu}
- i m g \delta_{\alpha}^{\mu}.
\label{10.61}
\ee
The lowest order interaction Lagrangian between gravitational
gauge field and real scalar field is
\be
g \eta^{\mu \nu} C_{\mu}^{\alpha}
(\partial_{\nu} \phi) (\partial_{\alpha} \phi)
- \frac{1}{2} g \delta_{\alpha}^{\mu}
C_{\mu}^{\alpha} ( (\partial^{\nu} \phi)
(\partial_{\nu} \phi) + m^2 \phi^2 ).
\label{10.62}
\ee
This  vertex  belongs to
$C_{\mu}^{\alpha}(k) \phi(q) \phi(p)$
three body interactions, its Feynman rule is
\be
- i g \eta^{\mu \nu}
( p_{\nu} q_{\alpha} +q_{\nu} p_{\alpha}  )
-i g \delta_{\alpha}^{\mu} ( -p^{\nu} q_{\nu} + m^2 ).
\label{10.63}
\ee
The lowest order interaction Lagrangian between gravitational
gauge field and complex scalar field is
\be
g \eta^{\mu \nu} C_{\mu}^{\alpha}
((\partial_{\alpha} \phi) (\partial_{\nu} \phi^*)
+ (\partial_{\alpha} \phi^*) (\partial_{\nu} \phi))
- g \delta_{\alpha}^{\mu} C_{\mu}^{\alpha}
( (\partial^{\nu} \phi) (\partial_{\nu} \phi^*)
+ m^2 \phi \phi^*  ).
\label{10.64}
\ee
This vertex belongs to
$C_{\mu}^{\alpha}(k) \phi^*(-q) \phi(p)$
three body interactions, its Feynman rule is
\be
i g \eta^{\mu \nu}
( p_{\nu} q_{\alpha} +q_{\nu} p_{\alpha}  )
-i g \delta_{\alpha}^{\mu} ( p^{\nu} q_{\nu} + m^2 ).
\label{10.65}
\ee
The lowest order coupling between vector field and gravitational
gauge field is
\be
\begin{array}{l}
 g \eta^{\mu \rho} \eta^{\nu \sigma} A_{0 \mu \nu}
C_{\rho}^{\alpha} \partial_{\alpha} A_{\sigma} \\
\\
+ (g \delta_{\tau}^{\lambda} C_{\lambda}^{\tau} )
( - \frac{1}{4} \eta^{\mu \rho} \eta^{\nu \sigma}
A_{0 \mu \nu} A_{0 \rho \sigma}
- \frac{m^2}{2} \eta^{\mu \nu} A_{\mu} A_{\nu} ),
\end{array}
\label{10.66}
\ee
where
\be
A_{0 \mu \nu} =
\partial_{\mu} A_{\nu} - \partial_{\nu} A_{\mu} .
\label{10.67}
\ee
This vertex belongs to
$C_{\mu}^{\alpha}(k) A_{\rho}(p) A_{\sigma}(q) $
three body interactions. Its Feynman rule is
\be
\begin{array}{l}
- ig \eta^{\mu \beta} \eta^{\rho \sigma}
(p_{\beta} q_{\alpha} +  p_{\alpha} q_{\beta})
+ ig \eta^{\mu \rho} \eta^{\sigma \beta} p_{\beta} q_{\alpha}
+ ig \eta^{\mu \sigma} \eta^{\rho \beta} q_{\beta} p_{\alpha}\\
\\
+ \frac{i}{2} g \delta_{\alpha}^{\mu} \eta^{\lambda \beta}
   \eta^{\rho \sigma}
   ( p_{\lambda} q_{\beta} +  q_{\lambda} p_{\beta} ) \\
\\
- \frac{i}{2} g \delta_{\alpha}^{\mu} \eta^{\rho \beta}
   \eta^{\nu \sigma}  p_{\nu} q_{\beta}
- \frac{i}{2} g \delta_{\alpha}^{\mu} \eta^{\rho \nu}
   \eta^{\beta \sigma}  q_{\nu} p_{\beta}
-i g m^2 \delta_{\alpha}^{\mu} \eta^{\rho \sigma}.
\end{array}
\label{10.68}
\ee
The lowest order self coupling of gravitational gauge fields
is
\be
\begin{array}{l}
 g \eta^{\mu \rho} \eta^{\nu \sigma} \eta_{\alpha \beta}
\lbrack (\partial_{\mu} C_{\nu}^{\alpha} ) C_{\rho}^{\beta_1}
  (\partial_{\beta_1} C_{\sigma}^{\beta} )
- (\partial_{\mu} C_{\nu}^{\alpha} ) C_{\sigma}^{\beta_1}
  (\partial_{\beta_1} C_{\rho}^{\beta} )
\rbrack  \\
\\
- \frac{1}{4} (g \delta_{\tau}^{\lambda} C_{\lambda}^{\tau} )
  \eta^{\mu \rho} \eta^{\nu \sigma} \eta_{\alpha \beta}
  F_{0 \mu \nu}^{\alpha} F_{0 \rho \sigma}^{\beta}
- \frac{g}{2} 
  \eta^{\mu \rho} \eta^{\nu \sigma} \eta_{\tau \beta}
  C_{\alpha}^{\tau}
  F_{0 \mu \nu}^{\alpha} F_{0 \rho \sigma}^{\beta}.
\end{array}
\label{10.69}
\ee
This vertex belongs to
$C_{\nu}^{\alpha}(p) C_{\sigma}^{\beta}(q) C_{\rho}^{\gamma}(r)$
three body interactions. Its Feynman rule is
\be
\begin{array}{l}
-i g \lbrack \eta^{\mu \rho} \eta^{\nu \sigma} \eta_{\alpha \beta}
     ( p_{\mu} q_{\gamma} +  q_{\mu} p_{\gamma} )
+     \eta^{\mu \sigma} \eta^{\nu \rho} \eta_{\alpha \gamma}
     ( p_{\mu} r_{\beta} +  r_{\mu} p_{\beta}) \\
\\
+     \eta^{\mu \nu} \eta^{\rho \sigma} \eta_{\gamma \beta}
     (q_{\mu} r_{\alpha} + r_{\mu} q_{\alpha}) \rbrack \\
\\
+ ig \lbrack \eta^{\mu \sigma} \eta^{\nu \rho} \eta_{ \alpha \beta}
      p_{\mu} q_{\gamma}
+    \eta^{\mu \nu} \eta^{\rho \sigma} \eta_{ \alpha \beta}
      q_{\mu} p_{\gamma}
+     \eta^{\mu \rho} \eta^{\nu \sigma} \eta_{\alpha \gamma}
      p_{\mu} r_{\beta} \\
\\
+      \eta^{\mu \nu} \eta^{\rho \sigma} \eta_{\alpha \gamma}
      r_{\mu} p_{\beta}
+     \eta^{\mu \rho} \eta^{\nu \sigma} \eta_{\beta \gamma}
      q_{\mu} r_{\alpha}
+     \eta^{\mu \sigma} \eta^{\nu \rho} \eta_{\beta \gamma}
      r_{\mu} q_{\alpha}  \rbrack  \\
\\
+     ig \delta_{\gamma}^{\rho}  \eta_{\alpha \beta}
      ( p_{\mu} q^{\mu} \eta^{\nu \sigma} - p^{\sigma} q^{\nu} )
+     ig \delta_{\alpha}^{\nu}  \eta_{\beta \gamma}
      ( q_{\mu} r^{\mu} \eta^{\rho \sigma} - r^{\sigma} q^{\rho} )\\
\\
+   ig \delta_{\beta}^{\sigma}  \eta_{\alpha \gamma}
      ( r_{\mu} p^{\mu} \eta^{\rho \nu} - p^{\rho} r^{\nu} )  \\
\\
  \cdots
\end{array}
\label{10.70}
\ee
\\

It could be found that all Feynman rules for vertex is
proportional to energy-momenta of one or more particles,
which is one of the most important properties of gravitational
interactions. In fact, this interaction property is expected
for gravitational interactions, for energy-momentum
is the source of gravity. \\

\section{Renormalization }
\setcounter{equation}{0}

In gravitational gauge theory, the gravitational coupling
constant has the dimensionality of negative powers of mass.
According to traditional theory of power counting law, it seems
that the gravitational gauge theory is a kind of non-renormalizable
theory. But this result is not correct. The power counting
law does not work here. General speaking, power counting law
does not work when a theory has gauge symmetry. If a theory has
gauge symmetry, the constraints from gauge symmetry will make
some divergence cancel each other. In gravitational gauge theory,
this mechanism works very well. In this chapter, we will give a
strict formal proof on the renormalization of the gravitational
gauge theory. We will find that the effect of renormalization is
just a scale transformation of the original theory. Though
there are infinite number of divergent vertexes in the gravitational
gauge theory, we need not introduce infinite number of interaction
terms that do not exist in the original Lagrangian and infinite
number of parameters. All the divergent vertex can find its
correspondence in the original Lagrangian. Therefor, in
renormalization, what we need to do is not to introduce extra
interaction terms to cancel divergent terms, but to redefine
the fields, coupling constants and some other parameters of
the original theory. Because most of counterterms come
from the factor $J(C)$, this factor is key important for
renormalization. Without this factor, the theory is
non-renormalizable. In a word, the gravitational gauge theory
is a renormalizable gauge theory. Now, let's start our discussion
on renormalization from the generalized BRST transformations.
Our proof is quite similar to the proof of the renormalizablility
of non-Able gauge field theory.\cite{c01,c02,c03,c04,c05,c06}\\

The generalized BRST transformations are 
\be  
\delta C_{\mu}^{\alpha} 
= -  {\mathbf D}_{\mu~\beta}^{\alpha} \eta^{\beta} 
\delta \lambda, 
\label{11.1} 
\ee 
\be  
\delta \eta^{\alpha} = g \eta^{\sigma} 
(\partial_{\sigma}\eta^{\alpha}) \delta \lambda, 
\label{11.2} 
\ee 
\be  
\delta \bar\eta_{\alpha} = \frac{1}{\alpha} 
\eta_{\alpha \beta} f^{\beta} \delta \lambda, 
\label{11.3} 
\ee 
\be  
\delta \eta^{\mu \nu} = 0, 
\label{11.4} 
\ee 
\be  
\delta g_{\alpha \beta} = g \left ( 
g_{\alpha \sigma} (\partial_{\beta} \eta^{\sigma}) 
+ g_{\sigma \beta} (\partial_{\alpha} \eta^{\sigma}) 
+\eta^{\sigma} (\partial_{\sigma} g_{\alpha\beta})
\right ) \delta \lambda, 
\label{11.5} 
\ee  
where $\delta \lambda$ is an infinitesimal Grassman constant.  
It can be strict proved that the generalized BRST transformations 
for fields $C_{\mu}^{\alpha}$ and $\eta^{\alpha}$ are nilpotent: 
\be  
\delta( {\mathbf D}_{\mu~\beta}^{\alpha} \eta^{\beta} ) 
=0, 
\label{11.6} 
\ee 
\be  
\delta(\eta^{\sigma}(\partial_{\sigma}\eta^{\alpha})) 
=0. 
\label{11.7} 
\ee 
It means that all second order variations of fields vanish.\\ 
 
Using the above transformation rules, it can be strictly proved 
the generalized BRST transformation for gauge field strength tensor 
$F^{\alpha}_{\mu \nu}$ is 
\be  
\delta F^{\alpha}_{\mu \nu} = 
g \left(  
- (\partial_{\sigma} \eta^{\alpha}) F^{\sigma}_{\mu \nu} 
+ \eta^{\sigma} (\partial_{\sigma} F^{\alpha}_{\mu \nu}) 
\right) \delta \lambda, 
\label{11.8} 
\ee  
and the transformation for the factor $J(C)$ is 
\be  
\delta J(C) = 
g \left( 
(\partial_{\alpha} \eta^{\alpha}) J(C) 
+ \eta^{\alpha} (\partial_{\alpha} J(C) ) 
\right) \delta \lambda. 
\label{11.9} 
\ee  
Therefore, under generalized BRST transformations,  
the Lagrangian ${\cal L}$ given by eq.(\ref{10.1}) transforms as 
\be  
\delta {\cal L} = g (\partial_{\alpha} 
(\eta^{\alpha}  {\cal L})) \delta \lambda. 
\label{11.10} 
\ee 
It is a total derivative term, its space-time integration 
vanish, i.e., the action of eq.(\ref{10.2}) is invariant under 
generalized BRST transformations, 
\be  
\delta (\int {\rm d}^4x {\cal L}) = \delta S =0. 
\label{11.11} 
\ee  
On the other hand, it can be strict proved that 
\be  
\delta \left( 
- \frac{1}{2 \alpha} 
\eta_{\alpha \beta} f^{\alpha} f^{\beta} 
+ \bar{\eta}_{\alpha} \partial^{\mu} 
{\mathbf D}_{\mu~\sigma}^{\alpha} \eta^{\sigma} 
\right) =0. 
\label{11.12} 
\ee 
\\ 
 
The non-renormalized effective Lagrangian  is denoted 
as $ {\cal L}_{eff}^{\lbrack 0 \rbrack} $. It is given by 
\be  
{\cal L}_{eff}^{\lbrack 0 \rbrack} = 
{\cal L} - \frac{1}{2 \alpha} 
\eta_{\alpha \beta} f^{\alpha} f^{\beta} 
+ \bar{\eta}_{\alpha} \partial^{\mu} 
{\mathbf D}_{\mu~\sigma}^{\alpha} \eta^{\sigma}. 
\label{11.13} 
\ee  
The effective action is defined by 
\be  
S_{eff}^{\lbrack 0 \rbrack}  
= \int {\rm d}^4x {\cal L}_{eff}^{\lbrack 0 \rbrack}. 
\label{11.14} 
\ee 
Using eqs.(\ref{10.11} - \ref{11.12}), we can prove that this effective 
action is invariant under generalized BRST transformations, 
\be  
\delta S_{eff}^{\lbrack 0 \rbrack} =0. 
\label{11.15} 
\ee  
This is a strict relation without any approximation. 
It is known that BRST symmetry plays key  role 
in the renormalization of gauge theory, for it ensures 
the validity of the Ward-Takahashi identity. \\ 
 
Before we go any further, we have to do another important 
work, i.e., to prove that the functional integration measure 
$\lbrack {\cal D}C \rbrack \lbrack {\cal D}\eta \rbrack 
\lbrack {\cal D}\bar\eta \rbrack $ is also generalized BRST 
invariant.  We have said before that the functional integration  
measure $\lbrack {\cal D}C \rbrack$ is not a gauge invariant 
measure, therefore, it is highly important to prove that 
$\lbrack {\cal D}C \rbrack \lbrack {\cal D}\eta \rbrack 
\lbrack {\cal D}\bar\eta \rbrack $ is a generalized BRST 
invariant measure. BRST transformation is a kind of  
transformation which involves both bosonic fields and 
fermionic fields. For the sake of simplicity, let's formally  
denote all bosonic fields as $B = \lbrace B_i \rbrace$ 
and denote all fermionic fields as $F = \lbrace F_i \rbrace$. 
All fields that are involved in generalized BRST transformation 
are simply denoted by $(B,F)$. Then, generalized BRST 
transformation is formally expressed as 
\be  
(B,F) ~~~ \to ~~~(B',F'). 
\label{11.16} 
\ee  
The transformation matrix of this transformation is 
\be 
J =   
\left ( 
\begin{array}{cc} 
\frac{\partial B_i}{\partial B'_j} & 
\frac{\partial B_i}{\partial F'_l} \\ 
\frac{\partial F_k}{\partial B'_j} & 
\frac{\partial F_k}{\partial F'_l}  
\end{array} 
\right ) 
= 
\left ( 
\begin{array}{cc} 
a & 
\alpha \\ 
\beta &  
b    
\end{array} 
\right )  , 
\label{11.17} 
\ee  
where 
\be  
a = \left( \frac{\partial B_i}{\partial B'_j} \right), 
\label{11.18} 
\ee 
\be  
b = \left( \frac{\partial F_k}{\partial F'_l} \right), 
\label{11.19} 
\ee  
\be  
\alpha = \left( \frac{\partial B_i}{\partial F'_l} \right), 
\label{11.20} 
\ee  
\be  
\beta = \left( \frac{\partial F_k}{\partial B'_j} \right). 
\label{11.21} 
\ee  
Matrixes $a$ and $b$ are bosonic square matrix while  
$\alpha$ and $\beta$ generally are not square matrix. In order 
to calculate the Jacobian $det(J)$.  
we realize the transformation (\ref{11.16})  
in two steps. The first step is a bosonic transformation 
\be  
(B,F) ~~~ \to ~~~(B',F). 
\label{11.22} 
\ee  
The transformation matrix of this transformation is denoted as $J_1$, 
\be 
J_1 =   
\left ( 
\begin{array}{cc} 
a - \alpha b^{-1} \beta & 
\alpha b^{-1} \\ 
0 & 
1  
\end{array} 
\right )  .   
\label{11.23} 
\ee  
Its Jacobian is 
\be  
det~ J_1 = det( a - \alpha b^{-1} \beta ). 
\label{11.24} 
\ee  
Therefore, 
\be 
\int \prod_i {\rm d}B_i \prod_k {\rm d}F_k 
= \int \prod_i {\rm d}B'_i \prod_k {\rm d}F_k 
\cdot det( a - \alpha b^{-1} \beta ). 
\label{11.25} 
\ee 
The second step is a fermionic transformation, 
\be  
(B',F) ~~~ \to ~~~(B',F'). 
\label{11.26} 
\ee  
Its transformation matrix is denoted as $J_2$, 
\be 
J_2 =   
\left ( 
\begin{array}{cc} 
1  & 
0  \\ 
\beta & 
b   
\end{array} 
\right )  .    
\label{11.27} 
\ee  
Its Jacobian is the inverse of the determinant of the transformation  
matrix, 
\be 
(det~ J_2)^{-1} = (det~b)^{-1}. 
\label{11.28} 
\ee  
Using this relation, eq.(\ref{11.25}) is changed into 
\be 
\int \prod_i {\rm d}B_i \prod_k {\rm d}F_k 
= \int \prod_i {\rm d}B'_i \prod_k {\rm d}F'_k 
\cdot det( a - \alpha b^{-1} \beta ) (det~b)^{-1} . 
\label{11.29} 
\ee 
For generalized BRST transformation, all non-diagonal matrix elements 
are proportional to Grassman constant $\delta \lambda$. Non-diagonal  
matrix $\alpha$ and $\beta$ contains only non-diagonal matrix elements, 
so, 
\be  
\alpha b^{-1} \beta  \propto (\delta \lambda)^2 = 0. 
\label{11.30} 
\ee  
It means that 
\be    
\int \prod_i {\rm d}B_i \prod_k {\rm d}F_k 
= \int \prod_i {\rm d}B'_i \prod_k {\rm d}F'_k 
\cdot det( a  ) \cdot (det~b)^{-1} . 
\label{11.31} 
\ee 
Generally speaking, $C_{\mu}^{\alpha}$ and  
$\partial_{\nu} C_{\mu}^{\alpha}$ are independent degrees of freedom, 
so are $\eta^{\alpha}$ and $\partial_{\nu} \eta^{\alpha}$. Using 
eqs.(\ref{11.1} - \ref{11.3}), we obtain 
\be  
\begin{array}{rcl} 
(det ~ a^{-1}) & = & 
det \left\lbrack 
( \delta^{\alpha}_{\beta} + 
g (\partial_{\beta} \eta^{\alpha} )  
\delta \lambda ) \delta^{\mu}_{\nu} 
\right\rbrack  \\ 
&&  \\ 
& = & 
\prod_{\mu,\alpha,x} 
\left\lbrack 
( \delta^{\alpha}_{\alpha} + 
g (\partial_{\alpha} \eta^{\alpha} ) 
\delta \lambda ) \delta^{\mu}_{\nu} 
\right\rbrack  \\ 
&&\\ 
&=&   
\prod_{x} 
( 1 + g (\partial_{\alpha} \eta^{\alpha} ) 
\delta \lambda). 
\end{array} 
\label{11.32} 
\ee  
\be  
\begin{array}{rcl} 
(det ~ b^{-1}) & = & 
det \left( 
 \delta^{\alpha}_{\beta} + 
g (\partial_{\beta} \eta^{\alpha} ) 
\delta \lambda \right)  \\ 
&&  \\ 
& = & 
\prod_{x}            
( 1 + g (\partial_{\alpha} \eta^{\alpha} )  
\delta \lambda) . 
\end{array} 
\label{11.33} 
\ee  
In the second line of eq.(\ref{11.32}), there is no summation  
over the repeated $\alpha$ index. Using these two relations, we have 
\be  
det( a  ) \cdot (det~b)^{-1} ~=~ \prod_x~{\rm\bf 1} ~=~1. 
\label{11.34} 
\ee    
Therefore, under generalized BRST transformation, functional 
integrational measure  
$\lbrack {\cal D}C \rbrack \lbrack {\cal D}\eta \rbrack 
\lbrack {\cal D}\bar\eta \rbrack $  
is invariant, 
\be  
\lbrack {\cal D}C \rbrack \lbrack {\cal D}\eta \rbrack 
\lbrack {\cal D}\bar\eta \rbrack  ~=~ 
\lbrack {\cal D}C' \rbrack \lbrack {\cal D}\eta' \rbrack 
\lbrack {\cal D}\bar\eta' \rbrack . 
\label{11.35} 
\ee 
Though both $\lbrack {\cal D}C \rbrack$ and  
$\lbrack {\cal D}\eta \rbrack$ are not invariant under generalized 
BRST transformation, their product is invariant under generalized 
BRST transformation. This result is interesting and important. \\ 
 
The generating functional  
$W^{\lbrack 0 \rbrack} \lbrack J \rbrack $ is 
\be   
W^{\lbrack 0 \rbrack} \lbrack J \rbrack 
=  N \int \lbrack {\cal D} C\rbrack 
\lbrack {\cal D} \eta \rbrack             
\lbrack {\cal D}\bar{\eta} \rbrack    
exp \left\lbrace i \int {\rm d}^4 x  
( {\cal L}^{\lbrack 0 \rbrack} _{eff} 
+ J^{\mu}_{\alpha} C^{\alpha}_{\mu} 
) \right\rbrace. 
\label{10.36} 
\ee 
Because  
\be  
\int {\rm d}\eta^{\beta} {\rm d}\bar\eta^{\sigma} 
\cdot \bar\eta^{\alpha} \cdot f(\eta,\bar\eta ) = 0, 
\label{11.37} 
\ee 
where $f(\eta,\bar\eta )$ is a bilinear function of  
$\eta$ and $\bar\eta$, we have 
\be  
\int \lbrack {\cal D} C\rbrack 
\lbrack {\cal D} \eta \rbrack       
\lbrack {\cal D}\bar{\eta} \rbrack  
\cdot \bar\eta^{\alpha}(x) \cdot 
exp \left\lbrace i \int {\rm d}^4 y 
( {\cal L}^{\lbrack 0 \rbrack}_{eff}(y) 
+ J^{\mu}_{\alpha}(y) C^{\alpha}_{\mu}(y)   
) \right\rbrace = 0. 
\label{11.38} 
\ee  
If all fields are the fields after generalized BRST transformation, 
eq.(\ref{11.38}) still holds, i.e. 
\be  
\int \lbrack {\cal D} C' \rbrack 
\lbrack {\cal D} \eta' \rbrack  
\lbrack {\cal D}\bar{\eta}' \rbrack 
\cdot \bar\eta^{\prime \alpha}(x) \cdot   
exp \left\lbrace i \int {\rm d}^4 y 
( {\cal L}^{\prime \lbrack 0 \rbrack}_{eff}(y) 
+ J^{\mu}_{\alpha}(y) C^{\prime \alpha}_{\mu}(y) 
) \right\rbrace = 0, 
\label{11.39} 
\ee  
where ${\cal L}^{\prime \lbrack 0 \rbrack}_{eff} $ is the 
effective Lagrangian after generalized BRST transformation. 
Both functional integration measure and effective action  
$\int {\rm d}^4 y {\cal L}^{\prime \lbrack 0 \rbrack}_{eff}(y)$ 
are generalized BRST invariant, so, using 
eqs.(\ref{11.1} - \ref{11.3}), we get 
\be  
\begin{array}{ll} 
\int \lbrack {\cal D} C \rbrack 
\lbrack {\cal D} \eta \rbrack 
\lbrack {\cal D}\bar{\eta} \rbrack 
& \left\lbrack 
\frac{1}{\alpha} f^{\alpha} (C(x)) \delta\lambda 
- i \bar\eta^{\alpha} (C(x)) 
\int {\rm d}^4z ( J^{\mu}_{\beta}(z) 
{\mathbf D}_{\mu \sigma}^{\beta}(z)  
\eta^{\sigma}(z) \delta\lambda ) 
\right\rbrack  \\ 
&\\ 
&\cdot exp \left\lbrace i \int {\rm d}^4 y 
( {\cal L}^{ \lbrack 0 \rbrack}_{eff}(y) 
+ J^{\mu}_{\alpha}(y) C^{\alpha}_{\mu}(y) 
) \right\rbrace = 0. 
\end{array} 
\label{11.40} 
\ee 
This equation will lead to  
\be  
\frac{1}{\alpha} f^{\alpha}  
\left(\frac{1}{i} \frac{\delta}{\delta J(x)} \right)  
W^{\lbrack 0 \rbrack} \lbrack J \rbrack 
- \int {\rm d}^4y ~ J^{\mu}_{\beta}(y) 
{\mathbf D}_{\mu \sigma}^{\beta} 
\left( \frac{1}{i} \frac{\delta}{\delta J(x)} \right) 
W^{\lbrack 0 \rbrack \sigma \alpha} \lbrack y,x, J \rbrack =0, 
\label{11.41} 
\ee 
where 
\be   
W^{\lbrack 0 \rbrack\sigma \alpha} \lbrack y,x,J \rbrack 
=  N i  \int \lbrack {\cal D} C\rbrack 
\lbrack {\cal D} \eta \rbrack       
\lbrack {\cal D}\bar{\eta} \rbrack 
\bar\eta^{\alpha}(x) \eta^{\sigma}(y) 
exp \left\lbrace i \int {\rm d}^4 z 
( {\cal L}^{\lbrack 0 \rbrack} _{eff} 
+ J^{\mu}_{\alpha} C^{\alpha}_{\mu} 
) \right\rbrace. 
\label{10.42} 
\ee 
This is the generalized Ward-Takahashi identity for 
generating functional $W^{\lbrack 0 \rbrack} \lbrack J \rbrack$.\\ 
 
Now, let's introduce the external sources of ghost fields, then 
the generation functional becomes 
\be   
W^{\lbrack 0 \rbrack}\lbrack J, \beta, \bar{\beta} \rbrack  
=  N \int \lbrack {\cal D} C\rbrack 
\lbrack {\cal D} \eta \rbrack 
\lbrack {\cal D}\bar{\eta} \rbrack 
exp \left\lbrace i \int {\rm d}^4 x  
( {\cal L}_{eff}^{\lbrack 0 \rbrack} 
+ J^{\mu}_{\alpha} C^{\alpha}_{\mu} 
+ \bar{\eta}_{\alpha} \beta^{\alpha} 
+ \bar{\beta}_{\alpha} \eta^{\alpha} 
) \right\rbrace, 
\label{11.43} 
\ee   
In renormalization of the theory, we have to introduce external 
sources $K^{\mu}_{\alpha}$ and $L_{\alpha}$ 
of the following composite operators, 
\be  
{\mathbf D}_{\mu \beta}^{\alpha} \eta^{\beta} 
~~,~~g \eta^{\sigma} (\partial_{\sigma} \eta^{\alpha} ). 
\label{11.44} 
\ee  
Then the effective Lagrangian becomes 
\be               
\begin{array}{rcl}                                        
{\swav{\cal L}}^{\lbrack 0 \rbrack} (C,\eta,\bar\eta,K,L) 
&=& {\cal L} - \frac{1}{2 \alpha} 
\eta_{\alpha \beta} f^{\alpha} f^{\beta} 
+ \bar{\eta}_{\alpha} \partial^{\mu} 
{\mathbf D}_{\mu~\sigma}^{\alpha} \eta^{\sigma}  
+K^{\mu}_{\alpha} {\mathbf D}_{\mu \beta}^{\alpha} \eta^{\beta} 
+ g L_{\alpha} \eta^{\sigma} (\partial_{\sigma} \eta^{\alpha})  \\ 
&&\\ 
&=& {\cal L}_{eff}^{\lbrack 0 \rbrack} 
+K^{\mu}_{\alpha} {\mathbf D}_{\mu \beta}^{\alpha} \eta^{\beta} 
+ g L_{\alpha} \eta^{\sigma} (\partial_{\sigma} \eta^{\alpha}) . 
\end{array} 
\label{11.45} 
\ee  
Then, 
\be  
\swav{S}^{\lbrack 0 \rbrack} \lbrack C,\eta,\bar\eta,K,L \rbrack  
= \int {\rm d}^4x {\cal \swav{L}}^{\lbrack 0 \rbrack} 
(C,\eta,\bar\eta,K,L). 
\label{11.46} 
\ee 
It is easy to deduce that 
\be           
\frac{\delta \swav{S}^{\lbrack 0 \rbrack}} 
{\delta K^{\mu}_{\alpha} } =  
{\mathbf D}_{\mu \beta}^{\alpha} \eta^{\beta}, 
\label{11.47} 
\ee  
\be           
\frac{\delta \swav{S}^{\lbrack 0 \rbrack}} 
{\delta L_{\alpha} } = 
g \eta^{\sigma} (\partial_{\sigma} \eta^{\alpha}). 
\label{11.48} 
\ee 
The generating functional now becomes, 
\be   
W^{\lbrack 0 \rbrack}\lbrack J, \beta, \bar{\beta},K,L \rbrack 
=  N \int \lbrack {\cal D} C\rbrack 
\lbrack {\cal D} \eta \rbrack 
\lbrack {\cal D}\bar{\eta} \rbrack 
exp \left\lbrace i \int {\rm d}^4 x                   
( \swav{{\cal L}}^{\lbrack 0 \rbrack} 
+ J^{\mu}_{\alpha} C^{\alpha}_{\mu} 
+ \bar{\eta}_{\alpha} \beta^{\alpha} 
+ \bar{\beta}_{\alpha} \eta^{\alpha} 
) \right\rbrace. 
\label{11.49} 
\ee   
In previous discussion, we have already proved that  
$S_{eff}^{\lbrack 0 \rbrack}$ is generalized BRST 
invariant. External sources $K^{\mu}_{\alpha}$ 
and $L_{\alpha}$ keep unchanged under generalized BRST 
transformation. Using nilpotent property of generalized 
BRST transformation, it is easy to prove that the two new 
terms   
$K^{\mu}_{\alpha} {\mathbf D}_{\mu \beta}^{\alpha} \eta^{\beta}$ 
and 
$g L_{\alpha} \eta^{\sigma} (\partial_{\sigma} \eta^{\alpha})$ 
in ${\swav{\cal L}}^{\lbrack 0 \rbrack}$ are also 
generalized BRST invariant, 
\be  
\delta (K^{\mu}_{\alpha}  
{\mathbf D}_{\mu \beta}^{\alpha} \eta^{\beta}) 
=0, 
\label{11.50} 
\ee  
\be  
\delta( g L_{\alpha} \eta^{\sigma}  
(\partial_{\sigma} \eta^{\alpha}) ) =0. 
\label{11.51} 
\ee 
Therefore, the action given by (\ref{11.46}) are generalized  
BRST invariant, 
\be  
\delta \swav{S}^{\lbrack 0 \rbrack} = 0. 
\label{11.52} 
\ee 
It gives out 
\be  
\begin{array}{l} 
\int {\rm d}^4x \left\lbrace 
 - ({\mathbf D}_{\mu \beta}^{\alpha} \eta^{\beta}(x)) 
\delta\lambda \frac{\delta}{\delta C_{\mu}^{\alpha}(x)} 
+ g \eta^{\sigma}(x) (\partial_{\sigma} \eta^{\alpha}(x))  
\delta\lambda  \frac{\delta}{\delta \eta^{\alpha}(x)} \right.  \\ 
\\ 
 \left. + \frac{1}{\alpha} f^{\alpha}(C(x)) \delta\lambda  
\frac{\delta}{\delta \bar\eta_{\alpha}(x)} 
\right\rbrace \swav{S}^{\lbrack 0 \rbrack} = 0. 
\end{array} 
\label{11.53} 
\ee  
Using relations (\ref{11.47} - \ref{11.48}), we can get 
\be  
\int {\rm d}^4x \left\lbrace 
\frac{\delta \swav{S}^{\lbrack 0 \rbrack}} 
{\delta K^{\mu}_{\alpha}(x) } 
\frac{\delta \swav{S}^{\lbrack 0 \rbrack}} 
{\delta C_{\mu}^{\alpha}(x) } 
+ \frac{\delta \swav{S}^{\lbrack 0 \rbrack}} 
{\delta L_{\alpha}(x) } 
\frac{\delta \swav{S}^{\lbrack 0 \rbrack}} 
{\delta \eta^{\alpha}(x) } 
+ \frac{1}{\alpha} f^{\alpha}(C(x)) 
\frac{\delta \swav{S}^{\lbrack 0 \rbrack}} 
{\delta \bar\eta_{\alpha}(x) } 
\right\rbrace  = 0. 
\label{11.54} 
\ee  
On the other hand, from (\ref{11.45} - \ref{11.46}), we can obtain that 
\be  
\frac{\delta \swav{S}^{\lbrack 0 \rbrack}} 
{\delta \bar\eta_{\alpha}(x) } 
=\partial^{\mu}  
\left( {\mathbf D}_{\mu \beta}^{\alpha} \eta^{\beta}(x)\right). 
\label{11.55} 
\ee 
Combine (\ref{11.47}) with (\ref{11.55}), we get 
\be 
\frac{\delta \swav{S}^{\lbrack 0 \rbrack}} 
{\delta \bar\eta_{\alpha}(x) } 
= \partial^{\mu} \left( \frac{\delta \swav{S}^{\lbrack 0 \rbrack}} 
{\delta K^{\mu}_{\alpha}(x) } \right). 
\label{11.56} 
\ee  
\\ 
 
In generation functional  
$W^{\lbrack 0 \rbrack}\lbrack J, \beta, \bar{\beta},K,L \rbrack$, 
all fields are integrated, so, if we set all fields to the fields 
after generalized BRST transformations, the final result should 
not be changed, i.e. 
\be   
\begin{array}{l} 
\swav{W}^{\lbrack 0 \rbrack}\lbrack J, \beta, \bar{\beta},K,L \rbrack 
=  N \int \lbrack {\cal D} C' \rbrack 
\lbrack {\cal D} \eta' \rbrack 
\lbrack {\cal D} \bar{\eta}' \rbrack  \\ 
\\ 
~~~~~\cdot exp \left\lbrace i \int {\rm d}^4 x               
( {\cal L}^{\lbrack 0 \rbrack} (C',\eta',\bar\eta',K,L) 
+ J^{\mu}_{\alpha} C^{\prime \alpha}_{\mu} 
+ \bar{\eta}'_{\alpha} \beta^{\alpha} 
+ \bar{\beta}_{\alpha} \eta^{\prime \alpha} 
) \right\rbrace. 
\end{array} 
\label{11.57} 
\ee  
Both action (\ref{11.46}) and functional integration  
measure $\lbrack {\cal D} C \rbrack 
\lbrack {\cal D} \eta \rbrack 
\lbrack {\cal D} \bar{\eta} \rbrack$ 
are generalized BRST invariant, so, the above relation 
gives out 
\be 
\begin{array}{l} 
\int \lbrack {\cal D} C \rbrack 
\lbrack {\cal D} \eta \rbrack 
\lbrack {\cal D} \bar{\eta} \rbrack 
\left\lbrace 
 i \int {\rm d}^4x \left( 
J^{\mu}_{\alpha} \frac{\delta \swav{S}^{\lbrack 0 \rbrack}} 
{\delta K^{\mu}_{\alpha}(x) } 
- \bar\beta_{\alpha} \frac{\delta \swav{S}^{\lbrack 0 \rbrack}} 
{\delta L_{\alpha}(x) } 
+ \frac{1}{\alpha} \eta_{\alpha \sigma} f^{\alpha} \beta^{\sigma}  
\right)\right\rbrace \\ 
\\ 
\cdot exp \left\lbrace i \int {\rm d}^4 y 
( {\cal L}^{\lbrack 0 \rbrack} (C,\eta,\bar\eta,K,L) 
+ J^{\mu}_{\alpha} C^{\alpha}_{\mu} 
+ \bar{\eta}_{\alpha} \beta^{\alpha} 
+ \bar{\beta}_{\alpha} \eta^{\alpha} 
) \right\rbrace = 0. 
\end{array} 
\label{11.58} 
\ee  
On the other hand, because the ghost field $\bar\eta_{\alpha}$ 
was integrated in  
$W^{\lbrack 0 \rbrack}\lbrack J, \beta, \bar{\beta},K,L \rbrack$, 
if we use $\bar\eta'_{\alpha}$ in the in functional integration, 
it will not change the generating functional. That is 
\be   
\begin{array}{l} 
\swav{W}^{\lbrack 0 \rbrack}\lbrack J, \beta, \bar{\beta},K,L \rbrack 
=  N \int \lbrack {\cal D} C \rbrack 
\lbrack {\cal D} \eta \rbrack 
\lbrack {\cal D} \bar{\eta}' \rbrack  \\ 
\\ 
~~~~~\cdot exp \left\lbrace i \int {\rm d}^4 x 
( {\cal L}^{\lbrack 0 \rbrack} (C,\eta,\bar\eta',K,L) 
+ J^{\mu}_{\alpha} C^{\alpha}_{\mu} 
+ \bar{\eta}'_{\alpha} \beta^{\alpha} 
+ \bar{\beta}_{\alpha} \eta^{\alpha} 
) \right\rbrace. 
\end{array} 
\label{11.59} 
\ee  
Suppose that  
\be  
\bar{\eta}'_{\alpha} = \bar{\eta}_{\alpha} 
+ \delta \bar{\eta}_{\alpha}. 
\label{11.60} 
\ee 
Then (\ref{11.59}) and (\ref{11.49}) will gives out 
\be  
\begin{array}{l} 
\int \lbrack {\cal D} C \rbrack 
\lbrack {\cal D} \eta \rbrack 
\lbrack {\cal D} \bar{\eta} \rbrack 
\left\lbrace  
\int {\rm d}^4x \delta \bar{\eta}_{\alpha} 
( \frac{\delta \swav{S}^{\lbrack 0 \rbrack}} 
{\delta \bar\eta_{\alpha}(x) } + \beta^{\alpha} (x)  
)\right\rbrace \\ 
\\ 
\cdot exp \left\lbrace i \int {\rm d}^4 y 
( {\cal L}^{\lbrack 0 \rbrack} (C,\eta,\bar\eta,K,L) 
+ J^{\mu}_{\alpha} C^{\alpha}_{\mu} 
+ \bar{\eta}_{\alpha} \beta^{\alpha} 
+ \bar{\beta}_{\alpha} \eta^{\alpha} 
) \right\rbrace = 0. 
\end{array} 
\label{11.61} 
\ee   
Because $\delta \bar{\eta}_{\alpha}$ is an arbitrary 
variation, from (\ref{11.61}), we will get 
\be  
\begin{array}{l} 
\int \lbrack {\cal D} C \rbrack 
\lbrack {\cal D} \eta \rbrack 
\lbrack {\cal D} \bar{\eta} \rbrack  
\left( \frac{\delta \swav{S}^{\lbrack 0 \rbrack}} 
{\delta \bar\eta_{\alpha}(x) } + \beta^{\alpha} (x)  
\right)  \\ 
\\ 
\cdot exp \left\lbrace i \int {\rm d}^4 y 
( {\cal L}^{\lbrack 0 \rbrack} (C,\eta,\bar\eta,K,L) 
+ J^{\mu}_{\alpha} C^{\alpha}_{\mu} 
+ \bar{\eta}_{\alpha} \beta^{\alpha} 
+ \bar{\beta}_{\alpha} \eta^{\alpha} 
) \right\rbrace = 0. 
\end{array} 
\label{11.62} 
\ee   
\\ 
 
The generating functional of connected Green function is given by 
\be 
\swav{Z}^{\lbrack 0 \rbrack}\lbrack J,  
\beta, \bar{\beta},K,L \rbrack 
= - i~ {\rm ln}~  
\swav{W}^{\lbrack 0 \rbrack}\lbrack J,  
\beta, \bar{\beta},K,L \rbrack. 
\label{11.63} 
\ee  
After Legendre transformation, we will get the generating 
functional of irreducible vertex  
$\swav{\Gamma}^{\lbrack 0 \rbrack}\lbrack C , 
\bar\eta,\eta,K,L \rbrack$, 
\be 
\begin{array}{rcl} 
\swav{\Gamma}^{\lbrack 0 \rbrack}\lbrack C, 
\bar\eta,\eta,K,L \rbrack 
&=& \swav{Z}^{\lbrack 0 \rbrack}\lbrack J, 
\beta, \bar{\beta},K,L \rbrack  \\ 
&&\\ 
&& - \int {\rm d}^4 x 
\left( J^{\mu}_{\alpha} C^{\alpha}_{\mu} 
+ \bar{\eta}_{\alpha} \beta^{\alpha} 
+ \bar{\beta}_{\alpha} \eta^{\alpha} 
\right) . 
\end{array} 
\label{11.64} 
\ee  
Functional derivative of the generating functional  
$\swav{Z}^{\lbrack 0 \rbrack}$ gives out the classical fields 
$C_{\mu}^{\alpha}$, $\eta^{\alpha}$ and $\bar\eta_{\alpha}$, 
\be  
C_{\mu}^{\alpha} =  
\frac{\delta \swav{Z}^{\lbrack 0 \rbrack}} 
{\delta J^{\mu}_{\alpha}}, 
\label{11.65} 
\ee  
\be  
\eta^{\alpha} =  
\frac{\delta \swav{Z}^{\lbrack 0 \rbrack}} 
{\delta \bar\beta_{\alpha}}, 
\label{11.66} 
\ee 
\be  
\bar\eta_{\alpha} =     
- \frac{\delta \swav{Z}^{\lbrack 0 \rbrack}} 
{\delta \beta^{\alpha}}. 
\label{11.67} 
\ee  
Then, functional derivative of the generating functional 
$\swav{\Gamma}^{\lbrack 0 \rbrack}$ gives out external 
sources $J^{\mu}_{\alpha}$, $\bar\beta_{\alpha}$ and 
$\beta^{\alpha}$, 
\be  
\frac{\delta \swav{\Gamma}^{\lbrack 0 \rbrack}} 
{\delta C_{\mu}^{\alpha}} 
= - J^{\mu}_{\alpha}, 
\label{11.68} 
\ee 
\be  
\frac{\delta \swav{\Gamma}^{\lbrack 0 \rbrack}} 
{\delta \eta^{\alpha}} 
=  \bar\beta_{\alpha}, 
\label{11.69} 
\ee  
\be  
\frac{\delta \swav{\Gamma}^{\lbrack 0 \rbrack}} 
{\delta \bar\eta_{\alpha}}    
=  - \beta^{\alpha}. 
\label{11.70} 
\ee  
Besides, there are two other relations which can be strictly 
deduced from (\ref{11.64}), 
\be  
\frac{\delta\swav{\Gamma}^{\lbrack 0 \rbrack}} 
{\delta K^{\mu}_{\alpha}} 
= \frac{\delta\swav{Z}^{\lbrack 0 \rbrack}} 
{\delta K^{\mu}_{\alpha}}, 
\label{11.71} 
\ee  
\be  
\frac{\delta\swav{\Gamma}^{\lbrack 0 \rbrack}} 
{\delta L_{\alpha}} 
= \frac{\delta \swav{Z}^{\lbrack 0 \rbrack}} 
{\delta L_{\alpha}}.      
\label{11.72} 
\ee  
\\ 
 
It is easy to prove that 
\be 
\begin{array}{rcl} 
&& i \frac{\delta \swav{S}^{\lbrack 0 \rbrack}} 
{\delta K^{\mu}_{\alpha}(x) } 
exp \left\lbrace i \int {\rm d}^4 y 
( {\cal L}^{\lbrack 0 \rbrack} (C,\eta,\bar\eta,K,L) 
+ J^{\mu}_{\alpha} C^{\alpha}_{\mu}  
+ \bar{\eta}_{\alpha} \beta^{\alpha} 
+ \bar{\beta}_{\alpha} \eta^{\alpha} 
) \right\rbrace  \\ 
&&\\ 
&=& \frac{\delta} {\delta K^{\mu}_{\alpha}(x) }  
exp \left\lbrace i \int {\rm d}^4 y 
( {\cal L}^{\lbrack 0 \rbrack} (C,\eta,\bar\eta,K,L) 
+ J^{\mu}_{\alpha} C^{\alpha}_{\mu}  
+ \bar{\eta}_{\alpha} \beta^{\alpha} 
+ \bar{\beta}_{\alpha} \eta^{\alpha} 
) \right\rbrace, 
\end{array} 
\label{11.73} 
\ee  
\be  
\begin{array}{rcl} 
&& i \frac{\delta \swav{S}^{\lbrack 0 \rbrack}} 
{\delta L_{\alpha}(x) } 
exp \left\lbrace i \int {\rm d}^4 y 
( {\cal L}^{\lbrack 0 \rbrack} (C,\eta,\bar\eta,K,L) 
+ J^{\mu}_{\alpha} C^{\alpha}_{\mu}  
+ \bar{\eta}_{\alpha} \beta^{\alpha} 
+ \bar{\beta}_{\alpha} \eta^{\alpha} 
) \right\rbrace  \\ 
&&\\ 
&=& \frac{\delta} {\delta L_{\alpha}(x) } 
exp \left\lbrace i \int {\rm d}^4 y 
( {\cal L}^{\lbrack 0 \rbrack} (C,\eta,\bar\eta,K,L) 
+ J^{\mu}_{\alpha} C^{\alpha}_{\mu}  
+ \bar{\eta}_{\alpha} \beta^{\alpha} 
+ \bar{\beta}_{\alpha} \eta^{\alpha} 
) \right\rbrace. 
\end{array} 
\label{11.74} 
\ee  
Use these two relations, we can change (\ref{11.58}) into 
\be 
\begin{array}{l} 
\int \lbrack {\cal D} C \rbrack 
\lbrack {\cal D} \eta \rbrack   
\lbrack {\cal D} \bar{\eta} \rbrack 
\left\lbrace 
 \int {\rm d}^4x \left( 
J^{\mu}_{\alpha}(x) \frac{\delta } {\delta K^{\mu}_{\alpha}(x) } 
- \bar\beta_{\alpha}(x) \frac{\delta }{\delta L_{\alpha}(x) } 
+ \frac{i}{\alpha} \eta_{\alpha \sigma}  
f^{\alpha}(\frac{1}{i}\frac{\delta }{\delta J^{\mu}_{\gamma}(x) } ) 
\beta^{\sigma}(x)   \right)\right\rbrace \\ 
\\ 
\cdot exp \left\lbrace i \int {\rm d}^4 y 
( {\cal L}^{\lbrack 0 \rbrack} (C,\eta,\bar\eta,K,L) 
+ J^{\mu}_{\alpha} C^{\alpha}_{\mu} 
+ \bar{\eta}_{\alpha} \beta^{\alpha} 
+ \bar{\beta}_{\alpha} \eta^{\alpha} 
) \right\rbrace = 0. 
\end{array} 
\label{11.75} 
\ee  
Using relations (\ref{11.68} -  \ref{11.70}) 
and definition of generating functional 
(\ref{11.57}), we can rewrite this equation into 
\be  
\int {\rm d}^4x \left\lbrace 
\frac{\delta \swav{W}^{\lbrack 0 \rbrack} }  
{\delta K^{\mu}_{\alpha}(x) } 
\frac{\delta \swav{\Gamma}^{\lbrack 0 \rbrack} }       
{\delta C_{\mu}^{\alpha}(x) } 
+ \frac{\delta \swav{W}^{\lbrack 0 \rbrack} }       
{\delta L_{\alpha}(x) } 
\frac{\delta \swav{\Gamma}^{\lbrack 0 \rbrack} }  
{\delta \eta^{\alpha}(x) }  
+ \frac{i}{\alpha} \eta_{\alpha \sigma} 
f^{\alpha}(\frac{1}{i}\frac{\delta }{\delta J^{\mu}_{\rho}(x) } ) 
\swav{W}^{\lbrack 0 \rbrack} 
\frac{\delta \swav{\Gamma}^{\lbrack 0 \rbrack} } 
{\delta \bar\eta_{\sigma}(x) } 
\right\rbrace = 0. 
\label{11.76} 
\ee 
Using (\ref{11.63}), we can obtain that 
\be  
\frac{\delta \swav{W}^{\lbrack 0 \rbrack} } 
{\delta K^{\mu}_{\alpha}(x) } 
= i \frac{\delta \swav{\Gamma}^{\lbrack 0 \rbrack} } 
{\delta K^{\mu}_{\alpha}(x) } \cdot 
\swav{W}^{\lbrack 0 \rbrack}, 
\label{11.77} 
\ee  
\be  
\frac{\delta \swav{W}^{\lbrack 0 \rbrack} } 
{\delta L_{\alpha}(x) } 
= i \frac{\delta \swav{\Gamma}^{\lbrack 0 \rbrack} } 
{\delta L_{\alpha}(x) } \cdot 
\swav{W}^{\lbrack 0 \rbrack}. 
\label{11.78} 
\ee  
Then (\ref{11.76}) is changed into 
\be  
\int {\rm d}^4x \left\lbrace 
\frac{\delta \swav{\Gamma}^{\lbrack 0 \rbrack} }  
{\delta K^{\mu}_{\alpha}(x) } 
\frac{\delta \swav{\Gamma}^{\lbrack 0 \rbrack} }       
{\delta C_{\mu}^{\alpha}(x) } 
+ \frac{\delta \swav{\Gamma}^{\lbrack 0 \rbrack} } 
{\delta L_{\alpha}(x) } 
\frac{\delta \swav{\Gamma}^{\lbrack 0 \rbrack} }     
{\delta \eta^{\alpha}(x) }  
+ \frac{i}{\alpha} \eta_{\alpha \sigma} 
f^{\alpha} 
\frac{\delta \swav{\Gamma}^{\lbrack 0 \rbrack} } 
{\delta \bar\eta_{\sigma}(x) } 
\right\rbrace = 0. 
\label{11.79} 
\ee 
Using (\ref{11.56}) and (\ref{11.73}), (\ref{11.62}) becomes 
\be  
\begin{array}{l} 
\int \lbrack {\cal D} C \rbrack 
\lbrack {\cal D} \eta \rbrack 
\lbrack {\cal D} \bar{\eta} \rbrack 
\left\lbrack  - i \partial^{\mu} 
\frac{\delta } {\delta K^{\mu}_{\alpha}(x) }  
+ \beta^{\alpha} (x)   \right\rbrack  \\ 
\\ 
\cdot exp \left\lbrace i \int {\rm d}^4 y        
( {\cal L}^{\lbrack 0 \rbrack} (C,\eta,\bar\eta,K,L) 
+ J^{\mu}_{\alpha} C^{\alpha}_{\mu} 
+ \bar{\eta}_{\alpha} \beta^{\alpha} 
+ \bar{\beta}_{\alpha} \eta^{\alpha} 
) \right\rbrace = 0. 
\end{array}   
\label{11.80} 
\ee   
In above equation, the factor $- i \partial^{\mu} 
\frac{\delta } {\delta K^{\mu}_{\alpha}(x) }  
+ \beta^{\alpha} (x)$ 
can move out of functional integration, then (\ref{11.80}) 
gives out 
\be  
\partial^{\mu} 
\frac{\delta \swav{\Gamma}^{\lbrack 0 \rbrack}  }  
{\delta K^{\mu}_{\alpha}(x) } 
= \frac{\delta \swav{\Gamma}^{\lbrack 0 \rbrack}  } 
{\delta \bar\eta_{\alpha}(x) }. 
\label{11.81}  
\ee   
In order to obtain this relation, (\ref{11.57}), (\ref{11.77}) and 
(\ref{11.70}) are used. \\ 
 
Define  
\be  
\bar{\Gamma}^{\lbrack 0 \rbrack} 
\lbrack C, \bar\eta, \eta, K, L \rbrack   
= \swav{\Gamma}^{\lbrack 0 \rbrack} 
\lbrack C, \bar\eta, \eta, K, L \rbrack 
+ \frac{1}{2 \alpha} 
\int {\rm d}^4x \eta_{\alpha \beta} 
f^{\alpha} f^{\beta}  
\label{11.82} 
\ee 
It is easy to prove that 
\be  
\frac{\delta \bar{\Gamma}^{\lbrack 0 \rbrack}  } 
{\delta K^{\mu}_{\alpha}(x) } 
= \frac{\delta \swav{\Gamma}^{\lbrack 0 \rbrack}  } 
{\delta K^{\mu}_{\alpha}(x) } 
\label{11.83}, 
\ee 
\be  
\frac{\delta \bar{\Gamma}^{\lbrack 0 \rbrack}  }  
{\delta L_{\alpha}(x) } 
= \frac{\delta \swav{\Gamma}^{\lbrack 0 \rbrack}  } 
{\delta L_{\alpha}(x) }, 
\label{11.84} 
\ee 
\be  
\frac{\delta \bar{\Gamma}^{\lbrack 0 \rbrack}  }  
{\delta \eta^{\alpha}(x) }       
= \frac{\delta \swav{\Gamma}^{\lbrack 0 \rbrack}  } 
{\delta \eta^{\alpha}(x) }    ,   
\label{11.85} 
\ee  
\be  
\frac{\delta \bar{\Gamma}^{\lbrack 0 \rbrack}  } 
{\delta \bar\eta_{\alpha}(x) } 
= \frac{\delta \swav{\Gamma}^{\lbrack 0 \rbrack}  } 
{\delta \bar\eta_{\alpha}(x) }    , 
\label{11.86} 
\ee  
\be  
\frac{\delta \bar{\Gamma}^{\lbrack 0 \rbrack}  } 
{\delta C_{\mu}^{\alpha}(x) } 
= \frac{\delta \swav{\Gamma}^{\lbrack 0 \rbrack}  } 
{\delta C_{\mu}^{\alpha}(x) } 
- \frac{1}{\alpha} \eta_{\alpha \beta} 
\partial^{\mu}  f^{\beta}. 
\label{11.87} 
\ee 
Using these relations, (\ref{11.81}) and (\ref{11.79}) are changed into 
\be  
\partial^{\mu} 
\frac{\delta \bar{\Gamma}^{\lbrack 0 \rbrack}  } 
{\delta K^{\mu}_{\alpha}(x) } 
= \frac{\delta \bar{\Gamma}^{\lbrack 0 \rbrack}  } 
{\delta \bar\eta_{\alpha}(x) }, 
\label{11.88}  
\ee   
\be  
\int {\rm d}^4x \left\lbrace 
\frac{\delta \bar{\Gamma}^{\lbrack 0 \rbrack} } 
{\delta K^{\mu}_{\alpha}(x) } 
\frac{\delta \bar{\Gamma}^{\lbrack 0 \rbrack} } 
{\delta C_{\mu}^{\alpha}(x) }    
+ \frac{\delta \bar{\Gamma}^{\lbrack 0 \rbrack} } 
{\delta L_{\alpha}(x) } 
\frac{\delta \bar{\Gamma}^{\lbrack 0 \rbrack} } 
{\delta \eta^{\alpha}(x) }  
\right\rbrace = 0.   
\label{11.89} 
\ee 
Eqs.(\ref{11.88} - \ref{11.89}) are generalized 
Ward-Takahashi identities 
of generating functional of regular vertex. It is the foundations  
of the renormalization of the gravitational gauge theory.\\ 
 
Generating functional $\swav{\Gamma}^{\lbrack 0 \rbrack}$  
is the generating functional of regular vertex with external 
sources, which is constructed from the Lagrangian 
$\swav{\cal L}_{eff}^{\lbrack 0 \rbrack}$. It is a functional 
of physical field, therefore, we can make a functional  
expansion 
\be  
\begin{array}{rcl} 
\swav{\Gamma}^{\lbrack 0 \rbrack} &=& 
\sum_n \int \frac{\delta^n \swav{\Gamma}^{\lbrack 0 \rbrack}} 
{\delta C^{\alpha_1}_{\mu_1}(x_1) \cdots  
\delta C^{\alpha_n}_{\mu_n}(x_n) } |_{C=\eta=\bar\eta=0} 
C^{\alpha_1}_{\mu_1}(x_1) \cdots C^{\alpha_n}_{\mu_n}(x_n) 
{\rm d}^4x_1 \cdots {\rm d}^4x_n  \\ 
&&\\ 
&& + \sum_n \int \frac{\delta^2} 
{\delta \bar\eta_{\beta_1}(y_1) \delta\eta^{\beta_2}(y_2) } 
\frac{\delta^n \swav{\Gamma}^{\lbrack 0 \rbrack}} 
{\delta C^{\alpha_1}_{\mu_1}(x_1) \cdots    
\delta C^{\alpha_n}_{\mu_n}(x_n) } |_{C=\eta=\bar\eta=0}  \\ 
&&\\ 
&&~~~ \cdot \bar\eta_{\beta_1}(y_1) \eta^{\beta_2}(y_2) 
C^{\alpha_1}_{\mu_1}(x_1) \cdots C^{\alpha_n}_{\mu_n}(x_n) 
{\rm d}^4y_1 {\rm d}^4y_2 {\rm d}^4x_1 \cdots {\rm d}^4x_n\\ 
&&\\  
&& + \cdots. 
\end{array} 
\label{11.90} 
\ee  
In this functional expansion, the expansion coefficients are 
regular vertexes with external sources. Before renormalization, 
these coefficients contain divergences. If we calculate these 
divergences in the methods of dimensional regularization,  
the form of these divergence will not violate gauge symmetry 
of the theory\cite{t01,t02}.  
In other words, in the method of dimensional  
regularization, gravitational gauge symmetry is not 
violated and the generating functional of regular vertex 
satisfies Ward-Takahashi identities 
(\ref{11.88} - \ref{11.89}). In order to eliminate 
the ultraviolet divergences of the theory, we need to introduce  
counterterms into Lagrangian. All these counterterms are 
formally denoted by $\delta {\cal L}$. Then the renormalized  
Lagrangian is 
\be  
\swav{\cal L}_{eff} =  
\swav{\cal L}_{eff}^{\lbrack 0 \rbrack} 
+ \delta {\cal L}. 
\label{11.91} 
\ee  
Because $\delta {\cal L}$ contains all counterterms,  
$\swav{\cal L}_{eff}$ is the Lagrangian density after complete 
renormalization. The generating functional of regular vertex 
which is calculate from $\swav{\cal L}_{eff}$ is denoted 
by $\swav{\Gamma}$. The regular vertexes calculated from 
this generating functional $\swav{\Gamma}$ contain no 
ultraviolet divergence anymore. Then let external sources 
$K^{\mu}_{\alpha}$ and $L_{\alpha}$ vanish, we will get 
generating functional $\Gamma$ of regular vertex without external 
sources, 
\be  
\Gamma = \swav{\Gamma} |_{K=L=0} . 
\label{11.92} 
\ee  
The regular vertexes which are generated from $\Gamma$ will 
contain no ultraviolet divergence either. Therefore, the 
S-matrix for all physical process are finite. For a renormalizable 
theory, the counterterm $\delta {\cal L}$ only contain finite 
unknown parameters which are needed to be determined by  
experiments. If conterterm $\delta {\cal L}$ contains infinite 
unknown parameters, the theory will lost its predictive power 
and it is conventionally regarded as a non-renormalizable  
theory. Now, the main task for us is to prove that 
the conterterm $\delta {\cal L}$ for the gravitational gauge 
theory only contains a few unknown parameters. If we do this, 
we will have proved that the gravitational gauge theory 
is renormalizable. \\ 
 
Now, we use inductive method to prove the renormalizability 
of the gravitational gauge theory.  
In the previous discussion, we have proved that the generating 
functional of regular vertex before renormalization satisfies  
Ward-Takahashi identities (\ref{11.88} - \ref{11.89}).  
The effective Lagrangian density that contains all counterterms  
which cancel all divergences of $l$-loops ($0 \leq l \leq L $)  
is denoted by $\swav{\cal  L}^{\lbrack L \rbrack}$.  
$\swav{\Gamma}^{\lbrack L \rbrack}$ is the generating functional 
of regular vertex which is calculated from 
$\swav{\cal  L}^{\lbrack L \rbrack}$. The regular vertex which 
is generated by $\swav{\Gamma}^{\lbrack L \rbrack}$ will contain 
no divergence if the number of the loops of the diagram is not 
greater than $L$. We have proved that the generating functional  
$\swav{\Gamma}^{\lbrack L \rbrack}$ satisfies Ward-Takahashi 
identities if $L=0$. Hypothesis that Ward-Takahashi identities 
are also satisfied when $L=N$, that is 
\be  
\partial^{\mu} 
\frac{\delta \bar{\Gamma}^{\lbrack N \rbrack}  } 
{\delta K^{\mu}_{\alpha}(x) } 
= \frac{\delta \bar{\Gamma}^{\lbrack N \rbrack}  } 
{\delta \bar\eta_{\alpha}(x) }, 
\label{11.93} 
\ee   
\be  
\int {\rm d}^4x \left\lbrace 
\frac{\delta \bar{\Gamma}^{\lbrack N \rbrack} } 
{\delta K^{\mu}_{\alpha}(x) } 
\frac{\delta \bar{\Gamma}^{\lbrack N \rbrack} } 
{\delta C_{\mu}^{\alpha}(x) } 
+ \frac{\delta \bar{\Gamma}^{\lbrack N \rbrack} } 
{\delta L_{\alpha}(x) } 
\frac{\delta \bar{\Gamma}^{\lbrack N \rbrack} } 
{\delta \eta^{\alpha}(x) }  
\right\rbrace = 0.   
\label{11.94} 
\ee 
Our goal is to prove that  Ward-Takahashi identities 
are also satisfied when $L=N+1$.\\ 
 
Now, let's introduce a special product which is defined by 
\be  
A * B \equiv \int {\rm d}^4x \left\lbrace 
\frac{\delta A }{\delta K^{\mu}_{\alpha}(x) } 
\frac{\delta B }{\delta C_{\mu}^{\alpha}(x) } 
+ \frac{\delta A }{\delta L_{\alpha}(x) } 
\frac{\delta B }{\delta \eta^{\alpha}(x) }  
\right\rbrace . 
\label{11.95} 
\ee 
Then (\ref{11.94}) will be simplified to 
\be   
\bar{\Gamma}^{\lbrack N \rbrack} * 
\bar{\Gamma}^{\lbrack N \rbrack} = 0. 
\label{11.96} 
\ee  
$\bar{\Gamma}^{\lbrack N \rbrack}$ contains all contributions from 
all possible diagram with arbitrary loops. The contribution from 
$l$-loop diagram is proportional to $\hbar^l$. We can expand 
$\bar{\Gamma}^{\lbrack N \rbrack}$ as a power serials of $\hbar^l$, 
\be  
\bar{\Gamma}^{\lbrack N \rbrack} 
= \sum_M \hbar^M \bar{\Gamma}^{\lbrack N \rbrack}_M, 
\label{11.97} 
\ee 
where $\bar{\Gamma}^{\lbrack N \rbrack}_M$ is the contribution 
from all $M$-loop diagrams. According to our inductive hypothesis, 
all $\bar{\Gamma}^{\lbrack N \rbrack}_M$ are finite is $M \leq N$. 
Therefore, divergence first appear in  
$\bar{\Gamma}^{\lbrack N \rbrack}_{N+1}$. 
Substitute (\ref{11.97}) into (\ref{11.96}), we will get 
\be   
\sum_{M,L}  \hbar^{M+L} 
\bar{\Gamma}_M^{\lbrack N \rbrack} * 
\bar{\Gamma}_L^{\lbrack N \rbrack} = 0. 
\label{11.98} 
\ee  
The $(L+1)$-loop contribution of (\ref{11.98}) is 
\be   
\sum_{M=0}^{N+1}  
\bar{\Gamma}_M^{\lbrack N \rbrack} *  
\bar{\Gamma}_{N-M+1}^{\lbrack N \rbrack} = 0. 
\label{11.99} 
\ee  
$\bar{\Gamma}_{N+1}^{\lbrack N \rbrack}$ can separate into 
two parts: finite part  
$\bar{\Gamma}_{N+1,F}^{\lbrack N \rbrack}$ 
and divergent part  
$\bar{\Gamma}_{N+1,div}^{\lbrack N \rbrack}$, that is 
\be  
\bar{\Gamma}_{N+1}^{\lbrack N \rbrack} 
= \bar{\Gamma}_{N+1,F}^{\lbrack N \rbrack} 
+ \bar{\Gamma}_{N+1,div}^{\lbrack N \rbrack}. 
\label{11.100} 
\ee  
$\bar{\Gamma}_{N+1,div}^{\lbrack N \rbrack}$ is a divergent  
function of $(4-D)$ if we calculate loop diagrams in dimensional 
regularization. In other words, all terms in  
$\bar{\Gamma}_{N+1,div}^{\lbrack N \rbrack}$ 
are divergent terms when $(4-D)$ approaches zero.   
Substitute (\ref{11.100}) into (\ref{11.99}), if we only 
concern divergent  terms, we will get 
\be   
\bar{\Gamma}_{N+1,div}^{\lbrack N \rbrack} * 
\bar{\Gamma}_0^{\lbrack N \rbrack}  + 
\bar{\Gamma}_0^{\lbrack N \rbrack} * 
\bar{\Gamma}_{N+1,div}^{\lbrack N \rbrack} = 0. 
\label{11.101} 
\ee  
$\bar{\Gamma}_{N+1,F}^{\lbrack N \rbrack}$ has no contribution 
to the divergent part. Because $\bar{\Gamma}_0^{\lbrack N \rbrack}$ 
represents contribution from tree diagram and counterterms has 
no contribution to tree diagram, we have 
\be  
\bar{\Gamma}_0^{\lbrack N \rbrack} 
= \bar{\Gamma}_0^{\lbrack 0 \rbrack}. 
\label{11.102} 
\ee 
Denote 
\be  
\bar{\Gamma}_0 
= \bar{\Gamma}_0^{\lbrack N \rbrack} 
= \swav{S}^{\lbrack 0 \rbrack} 
+ \frac{1}{2 \alpha} 
\int {\rm d}^4x \eta_{\alpha \beta} 
f^{\alpha} f^{\beta}. 
\label{11.103} 
\ee 
Then (\ref{11.101}) is changed into 
\be   
\bar{\Gamma}_{N+1,div}^{\lbrack N \rbrack} * 
\bar{\Gamma}_0  + 
\bar{\Gamma}_0 * 
\bar{\Gamma}_{N+1,div}^{\lbrack N \rbrack} = 0. 
\label{11.104} 
\ee  
Substitute (\ref{11.97}) into (\ref{11.93}), we get 
\be  
\partial^{\mu} 
\frac{\delta \bar{\Gamma}_{N+1}^{\lbrack N \rbrack}  } 
{\delta K^{\mu}_{\alpha}(x) } 
= \frac{\delta \bar{\Gamma}_{N+1}^{\lbrack N \rbrack}  } 
{\delta \bar\eta_{\alpha}(x) }. 
\label{11.105} 
\ee  
The finite part $\bar{\Gamma}_{N+1,F}^{\lbrack N \rbrack}$ 
has no contribution to the divergent part, so we have 
\be  
\partial^{\mu} 
\frac{\delta \bar{\Gamma}_{N+1,div}^{\lbrack N \rbrack}  } 
{\delta K^{\mu}_{\alpha}(x) }  
= \frac{\delta \bar{\Gamma}_{N+1,div}^{\lbrack N \rbrack}  } 
{\delta \bar\eta_{\alpha}(x) }. 
\label{11.106} 
\ee  
\\ 
 
The operator $\hat{g}$ is defined by 
\be  
\begin{array}{rcl} 
\hat{g} &\equiv& \int {\rm d}^4x \left\lbrace 
\frac{\delta \bar{\Gamma}_0 }{\delta C_{\mu}^{\alpha}(x) } 
\frac{\delta  }{\delta K^{\mu}_{\alpha}(x) }   
+ \frac{\delta \bar{\Gamma}_0 }{\delta L_{\alpha}(x) } 
\frac{\delta  }{\delta \eta^{\alpha}(x) }  \right. \\ 
&&\\ 
&& \left. 
+ \frac{\delta \bar{\Gamma}_0 }{\delta K^{\mu}_{\alpha}(x) } 
\frac{\delta  }{\delta C_{\mu}^{\alpha}(x) }  
+ \frac{\delta \bar{\Gamma}_0 }{\delta \eta^{\alpha}(x) } 
\frac{\delta  }{\delta  L_{\alpha}(x) } 
\right\rbrace . 
\end{array} 
\label{11.107} 
\ee 
Using this definition, (\ref{11.104}) simplifies to 
\be  
\hat{g} \bar{\Gamma}_{N+1,div}^{\lbrack N \rbrack} = 0. 
\label{11.108} 
\ee 
\\ 
 
It can be strictly proved that operator $\hat{g}$ is a  
nilpotent operator, i.e. 
\be  
\hat{g}^2=0. 
\label{11.109}   
\ee 
Suppose that $f\lbrack C \rbrack$ is an arbitrary functional 
of gravitational gauge field $C_{\mu}^{\alpha}$ which is invariant 
under local gravitational gauge transformation.  
$f\lbrack C \rbrack$ is invariant under generalized BRST 
transformation. The generalized BRST transformation  
of $f\lbrack C \rbrack$ is 
\be  
\delta f\lbrack C \rbrack = 
- \int {\rm d}^4x  
\frac{\delta f\lbrack C \rbrack} 
{\delta C_{\mu}^{\alpha}(x)} 
{\mathbf D}_{\mu \beta}^{\alpha} \eta^{\beta} \delta \lambda. 
\label{11.110} 
\ee 
Because $\delta \lambda$ is an arbitrary Grassman variable,  
(\ref{11.110}) gives out 
\be  
\delta f\lbrack C \rbrack = 
- \int {\rm d}^4x 
\frac{\delta f\lbrack C \rbrack}   
{\delta C_{\mu}^{\alpha}(x)} 
{\mathbf D}_{\mu \beta}^{\alpha} \eta^{\beta}. 
\label{11.111} 
\ee 
Because $f \lbrack C \rbrack$ is a functional of only gravitational 
gauge fields, its functional derivatives to other fields vanish 
\be  
\frac{\delta f \lbrack C \rbrack } 
{\delta K^{\mu}_{\alpha}(x) } =0, 
\label{11.112} 
\ee 
\be  
\frac{\delta f \lbrack C \rbrack } 
{\delta L_{\alpha}(x) } =0, 
\label{11.113} 
\ee 
\be  
\frac{\delta f \lbrack C \rbrack } 
{\delta \eta^{\alpha}(x) } =0.       
\label{11.114} 
\ee 
Using these relations, (\ref{11.111}) is changed into 
\be  
\delta f\lbrack C \rbrack = 
\hat{g} f\lbrack C \rbrack. 
\label{11.115} 
\ee 
The generalized BRST symmetry of $f\lbrack C \rbrack$ gives 
out the following important property of operator $\hat{g}$, 
\be  
\hat{g} f\lbrack C \rbrack = 0. 
\label{11.116} 
\ee 
\\ 
 
Using two important properties of operator $\hat{g}$ which are 
shown in eq.(\ref{11.109}) 
and eq.(\ref{11.116}), we could see that the solution 
of eq.(\ref{11.108}) can be written in the following form 
\be 
\bar{\Gamma}_{N+1,div}^{\lbrack N \rbrack} 
= f\lbrack C \rbrack  
+ \hat{g} f'\lbrack C,\eta,\bar\eta,K,L \rbrack, 
\label{11.117} 
\ee  
$f\lbrack C \rbrack$ is a gauge invariant functional and 
$f'\lbrack C,\eta,\bar\eta,K,L \rbrack$ is an arbitrary functional 
of fields $C_{\mu}^{\alpha}(x)$, $\eta^{\alpha}(x)$, 
$\bar\eta_{\alpha}(x)$ and external sources  
$K^{\mu}_{\alpha}(x)$ and $L_{\alpha}(x)$. 
\\ 
 
Now, let's consider constrain from eq.(\ref{11.105}). 
Using eq.(\ref{11.112}) 
and eq.(\ref{11.114}), we can see that $f \lbrack C \rbrack$ satisfies 
eq.(\ref{11.105}), so eq.(\ref{11.105}) has no 
constrain on $f \lbrack C \rbrack$. 
Define a new variable 
\be  
B^{\mu}_{\alpha} 
= K^{\mu}_{\alpha} 
- \partial^{\mu} \bar\eta_{\alpha}. 
\label{11.118} 
\ee  
$f_1 \lbrack B \rbrack$ is an arbitrary functional of $B$. 
It can be proved that 
\be  
\frac{\delta f_1 \lbrack B \rbrack } 
{\delta B^{\mu}_{\alpha}(x) }  
= \frac{\delta f_1 \lbrack B \rbrack } 
{\delta K^{\mu}_{\alpha}(x) } , 
\label{11.119} 
\ee 
\be  
\frac{\delta f_1 \lbrack B \rbrack } 
{\delta \bar\eta_{\alpha}(x) }      
= \partial^{\mu} \frac{\delta f_1 \lbrack B \rbrack } 
{\delta B^{\mu}_{\alpha}(x) }. 
\label{11.120} 
\ee 
Combine these two relations, we will get 
\be  
\frac{\delta f_1 \lbrack B \rbrack } 
{\delta \bar\eta_{\alpha}(x) } 
= \partial^{\mu} \frac{\delta f_1 \lbrack B \rbrack } 
{\delta K^{\mu}_{\alpha}(x) }. 
\label{11.121} 
\ee  
There $f_1 \lbrack B \rbrack$ is a solution to eq.(\ref{11.106}). 
Suppose that there is another functional $f_2$ that is 
given by, 
\be  
f_2 \lbrack K,C,\eta,L \rbrack 
= \int {\rm d}^4x ~ K^{\mu}_{\alpha} 
T_{\mu}^{\alpha}(C,\eta,L), 
\label{11.122} 
\ee  
where $T_{\mu}^{\alpha}$ is a conserved current 
\be  
\partial^{\mu} T_{\mu}^{\alpha} =0.  
\label{11.123} 
\ee  
It can be easily proved that $f_2 \lbrack K,C,\eta,L \rbrack$ 
is also a solution of eq.(\ref{11.105}). Because $\bar{\Gamma}_0$ satisfies 
eq.(\ref{11.106}) (please see eq.(\ref{11.88})), operator $\hat{g}$ commutes 
with $\frac{\delta}{\delta \bar\eta_{\alpha}(x) } 
- \partial^{\mu} \frac{\delta }{\delta K^{\mu}_{\alpha}(x) } $. 
It means that functional $f'\lbrack C,\eta,\bar\eta,K,L \rbrack$ 
in eq.(\ref{11.117}) must satisfy eq.(\ref{11.106}). According to these 
discussion, the solution of $f'\lbrack C,\eta,\bar\eta,K,L \rbrack$ 
has the following form, 
\be            
f'\lbrack C,\eta,\bar\eta,K,L \rbrack 
= f_1 \lbrack C,\eta, 
K^{\mu}_{\alpha} - \partial^{\mu} \bar\eta_{\alpha}, L \rbrack 
+ \int {\rm d}^4x ~ K^{\mu}_{\alpha}                      
T_{\mu}^{\alpha}(C,\eta,L).                                        	  
\label{11.124} 
\ee  
\\ 
 
In order to determine $f'\lbrack C,\eta,\bar\eta,K,L \rbrack$, 
we need to study dimensions of various fields and external sources. 
Set the dimensionality of mass to $1$, i.e. 
\be  
D \lbrack \hat{P}_{\mu} \rbrack =1. 
\label{11.125} 
\ee 
 Then we have 
\be  
D \lbrack C_{\mu}^{\alpha} \rbrack =1,  
\label{11.126} 
\ee 
\be            
D \lbrack {\rm d}^4x \rbrack =-4, 
\label{11.127} 
\ee 
\be            
D \lbrack D_{\mu} \rbrack =1, 
\label{11.128} 
\ee 
\be            
D \lbrack \eta \rbrack  = D \lbrack \bar\eta \rbrack =1, 
\label{11.129} 
\ee 
\be            
D \lbrack K \rbrack  = D \lbrack L \rbrack =2, 
\label{11.130} 
\ee 
\be            
D \lbrack g \rbrack   = - 1, 
\label{11.131} 
\ee 
\be 
D \lbrack \bar{\Gamma}_{N+1,div}^{\lbrack N \rbrack} \rbrack   
= D \lbrack S \rbrack = 0. 
\label{11.132} 
\ee 
Using these relations, we can prove that 
\be  
D \lbrack \hat{g} \rbrack   =1, 
\label{11.133} 
\ee 
\be            
D \lbrack f' \rbrack   =-1. 
\label{11.134} 
\ee 
Define virtual particle number $N_g$ of ghost field 
$\eta$ is 1, and that of ghost field $\bar\eta$ is -1, i.e. 
\be 
N_g \lbrack \eta \rbrack = 1, 
\label{11.135} 
\ee 
\be 
N_g \lbrack \bar\eta \rbrack = - 1. 
\label{11.136} 
\ee  
The virtual particle number is a additive conserved quantity, so  
Lagrangian and action carry no virtual particle number, 
\be 
N_g \lbrack S \rbrack =  N_g \lbrack {\cal L} \rbrack    =  0. 
\label{11.137} 
\ee  
The virtual particle number $N_g$ of other fields and external 
sources are 
\be 
N_g \lbrack C \rbrack = N_g \lbrack D_{\mu} \rbrack =  0, 
\label{11.138} 
\ee  
\be 
N_g \lbrack g \rbrack =  0,   
\label{11.139} 
\ee  
\be 
N_g \lbrack \bar{\Gamma} \rbrack =  0,                                 
\label{11.140} 
\ee                                             
\be 
N_g \lbrack K \rbrack =  -1,                                 
\label{11.141} 
\ee                           
\be 
N_g \lbrack L \rbrack =  -2.                                 
\label{11.142} 
\ee                                                               
Using all these relations, we can determine the virtual particle  
number $N_g$ of $\hat g$ and $f'$, 
\be 
N_g \lbrack \hat g \rbrack =  1, 
\label{11.143} 
\ee              
\be 
N_g \lbrack f' \rbrack =  -1.  
\label{11.144} 
\ee             
According to eq.(\ref{11.134}) and eq.(\ref{11.144}), we know that the  
dimensionality of $f'$ is $-1$ and its virtual particle 
number is also $-1$. Besides, $f'$ must be a Lorentz scalar. 
Combine all these results, the only two possible solutions of 
$f_1 \lbrack C,\eta, K^{\mu}_{\alpha} - \partial^{\mu}  
\bar\eta_{\alpha}, L \rbrack$ in eq.(\ref{11.124}) are 
\be  
(K^{\mu}_{\alpha} - \partial^{\mu} \bar\eta_{\alpha}  ) 
C_{\mu}^{\alpha}, 
\label{11.145}   
\ee         
\be  
\bar\eta^{\alpha} L_{\alpha}. 
\label{11.146} 
\ee 
The only possible solution of $T_{\mu}^{\alpha}$ is  
$C_{\mu}^{\alpha}$. But in general gauge conditions, 
$C_{\mu}^{\alpha}$ does not satisfy the conserved  
equation  eq.(\ref{11.123}). Therefore, the solution to  
$f'\lbrack C,\eta,\bar\eta,K,L \rbrack$ 
is linear combination of (\ref{11.145}) and (\ref{11.146}), i.e. 
\be 
f'\lbrack C,\eta,\bar\eta,K,L \rbrack =  
\int {\rm d}^4x \left\lbrack 
\beta_{N+1}(\varepsilon) (K^{\mu}_{\alpha} - \partial^{\mu} 
\bar\eta_{\alpha}  ) C_{\mu}^{\alpha} 
+ \gamma_{N+1}(\varepsilon)  
\bar\eta^{\alpha} L_{\alpha} 
\right\rbrack, 
\label{11.147} 
\ee  
where $\varepsilon = (4-D)$, $\beta_{N+1}(\varepsilon)$ 
and $\gamma_{N+1}(\varepsilon)$ are divergent parameters when 
$\varepsilon$ approaches zero. 
Then using the definition of $\hat g$, 
we can obtain the following result, 
\be  
\begin{array}{rcl} 
\hat g f'\lbrack C,\eta,\bar\eta,K,L \rbrack 
&=& - \beta_{N+1}(\varepsilon) 
\int {\rm d}^4x \left\lbrack 
\frac{\delta \bar{\Gamma}_0 }{\delta C_{\mu}^{\alpha}(x) } 
C_{\mu}^{\alpha}(x) 
+ \frac{\delta \bar{\Gamma}_0 }{\delta K^{\mu}_{\alpha}(x) } 
K^{\mu}_{\alpha}(x) 
- \bar\eta_{\alpha} \partial^{\mu} 
{\mathbf D}_{\mu \beta}^{\alpha} \eta^{\beta} 
\right\rbrack  \\ 
&&\\ 
&&- \gamma_{N+1}(\varepsilon) 
\int {\rm d}^4x \left\lbrack 
\frac{\delta \bar{\Gamma}_0 }{\delta L_{\alpha}(x) } 
L_{\alpha}(x) 
+ \frac{\delta \bar{\Gamma}_0 }{\delta \eta^{\alpha}(x) }       
\eta^{\alpha}(x)       
\right\rbrack. 
\end{array} 
\label{11.148} 
\ee  
The only possible solution to $f \lbrack C \rbrack$ of eq.(\ref{11.117}) 
which is constructed only from gravitational gauge fields 
is the action $S\lbrack C \rbrack$ of gauge fields.  
Therefore, the most general solution of  
$\bar{\Gamma}_{N+1,div}^{\lbrack N \rbrack}$ is 
\be 
\begin{array}{rcl} 
\bar{\Gamma}_{N+1,div}^{\lbrack N \rbrack} 
&=& \alpha_{N+1}(\varepsilon)  S\lbrack C \rbrack 
- \int {\rm d}^4x \left\lbrack 
\beta_{N+1}(\varepsilon) \frac{\delta \bar{\Gamma}_0 } 
{\delta C_{\mu}^{\alpha}(x) } C_{\mu}^{\alpha}(x) 
+ \beta_{N+1}(\varepsilon) \frac{\delta \bar{\Gamma}_0 } 
{\delta K^{\mu}_{\alpha}(x) } K^{\mu}_{\alpha}(x)  
\right. \\ 
&&\\ 
&& \left. 
+ \gamma_{N+1}(\varepsilon) \frac{\delta \bar{\Gamma}_0 } 
{\delta  L_{\alpha}(x) } L_{\alpha}(x) 
+ \gamma_{N+1}(\varepsilon) \frac{\delta \bar{\Gamma}_0 } 
{\delta  \eta^{\alpha}(x) } \eta^{\alpha}(x) 
- \beta_{N+1}(\varepsilon) \bar\eta_{\alpha} \partial^{\mu} 
{\mathbf D}_{\mu \beta}^{\alpha} \eta^{\beta} 
\right\rbrack. 
\end{array} 
\label{11.149} 
\ee  
\\ 
 
In fact, the action $S\lbrack C \rbrack$ is a functional 
of pure gravitational gauge field. It also contains gravitational 
coupling constant $g$. So, we can denote it as 
$S\lbrack C,g \rbrack$. From eq.(\ref{4.20}), 
eq.(\ref{4.24}) and eq.(\ref{4.25}), 
we can prove that the action $S\lbrack C,g \rbrack$ has 
the following important properties, 
\be  
S\lbrack gC,1 \rbrack = g^2 S\lbrack C,g \rbrack. 
\label{11.150} 
\ee  
Differentiate both sides of eq.(\ref{11.150}) with respect to coupling 
constant $g$, we can get 
\be 
S\lbrack C,g \rbrack  
= \frac{1}{2} \int {\rm d}^4x  
\frac{\delta S\lbrack C,g \rbrack }{\delta C_{\mu}^{\alpha}(x) }  
C_{\mu}^{\alpha}(x) 
- \frac{1}{2} g \frac{\partial S\lbrack C,g \rbrack}{\partial g}. 
\label{11.151} 
\ee  
\\ 
 
It can be easily proved that 
\be  
\begin{array}{l} 
\int {\rm d}^4x C_{\mu}^{\alpha}(x) 
\frac{\delta  }{\delta C_{\mu}^{\alpha}(x) } 
\left\lbrack 
\int {\rm d}^4y \bar\eta_{\alpha}(y) \partial^{\mu} 
{\mathbf D}_{\mu \beta}^{\alpha} \eta^{\beta}(y) 
\right\rbrack  \\ 
\\ 
= \int {\rm d}^4x \left\lbrack 
(\partial^{\mu} \bar\eta_{\beta}(x)) 
(\partial_{\mu} \eta^{\beta}(x)) 
+ \bar\eta_{\alpha}(x) \partial^{\mu} 
{\mathbf D}_{\mu \beta}^{\alpha} \eta^{\beta}(x) 
\right\rbrack, 
\end{array} 
\label{11.152} 
\ee  
\be  
\begin{array}{l} 
\int {\rm d}^4x C_{\mu}^{\alpha}(x) 
\frac{\delta  }{\delta C_{\mu}^{\alpha}(x) } 
\left\lbrack 
\int {\rm d}^4y K^{\mu}_{\alpha}(y) 
{\mathbf D}_{\mu \beta}^{\alpha} \eta^{\beta}(y) 
\right\rbrack  \\ 
\\ 
= - \int {\rm d}^4x \left\lbrack 
K^{\mu}_{\alpha}(x) \partial_{\mu} \eta^{\alpha}(x) 
- K^{\mu}_{\alpha}(x) {\mathbf D}_{\mu \beta}^{\alpha} 
\eta^{\beta}(x) 
\right\rbrack, 
\end{array} 
\label{11.153} 
\ee 
\be  
\begin{array}{l} 
g \frac{\partial  }{\partial g } 
\left\lbrack 
\int {\rm d}^4x \bar\eta_{\alpha}(x) \partial^{\mu} 
{\mathbf D}_{\mu \beta}^{\alpha} \eta^{\beta}(x) 
\right\rbrack  \\ 
\\ 
= \int {\rm d}^4x \left\lbrack   
(\partial^{\mu} \bar\eta_{\beta}(x)) 
(\partial_{\mu} \eta^{\beta}(x)) 
+ \bar\eta_{\alpha}(x) \partial^{\mu} 
{\mathbf D}_{\mu \beta}^{\alpha} \eta^{\beta}(x) 
\right\rbrack, 
\end{array} 
\label{11.154} 
\ee 
\be 
\begin{array}{l} 
g \frac{\partial  }{\partial g } 
\left\lbrack 
\int {\rm d}^4x K^{\mu}_{\alpha}(x) 
{\mathbf D}_{\mu \beta}^{\alpha} \eta^{\beta}(x) 
\right\rbrack  \\ 
\\ 
= - \int {\rm d}^4x \left\lbrack 
K^{\mu}_{\alpha}(x) \partial_{\mu} \eta^{\alpha} 
- K^{\mu}_{\alpha}(x) {\mathbf D}_{\mu \beta}^{\alpha} 
\eta^{\beta} 
\right\rbrack, 
\end{array} 
\label{11.155} 
\ee 
\be 
\begin{array}{l}  
g \frac{\partial  }{\partial g } 
\left\lbrack 
\int {\rm d}^4x ~g L_{\alpha}(x) \eta^{\beta}(x) 
(\partial_{\beta} \eta^{\alpha}(x) ) 
\right\rbrack  \\ 
\\ 
=   
\int {\rm d}^4x~ g L_{\alpha}(x) \eta^{\beta}(x) 
(\partial_{\beta} \eta^{\alpha}(x) ), 
\end{array} 
\label{11.156} 
\ee

Using eqs.(\ref{11.152} - \ref{11.153}), 
eq.(\ref{11.103}) and eq.(\ref{11.45}), 
we can prove that 
\be  
\begin{array}{rcl} 
\int {\rm d}^4x  
\frac{\delta S\lbrack C,g \rbrack  }{\delta C_{\mu}^{\alpha}(x) } 
C_{\mu}^{\alpha}(x)  &=& 
\int {\rm d}^4x                     
\frac{\delta \bar{\Gamma}_0  }{\delta C_{\mu}^{\alpha}(x) } 
C_{\mu}^{\alpha}(x)   
+ \int {\rm d}^4x \left\lbrack   
-(\partial^{\mu} \bar\eta_{\alpha}(x)) 
(\partial_{\mu} \eta^{\alpha}(x)) \right. \\ 
&&\\ 
&&\left. - \bar\eta_{\alpha}(x) \partial^{\mu} 
{\mathbf D}_{\mu \beta}^{\alpha} \eta^{\beta}(x) 
+K^{\mu}_{\alpha}(x) \partial_{\mu} \eta^{\alpha}(x) 
- K^{\mu}_{\alpha}(x) {\mathbf D}_{\mu \beta}^{\alpha} 
\eta^{\beta}(x) 
\right\rbrack.  
\end{array} 
\label{11.157} 
\ee 
Similarly, we can get,  
\be  
\begin{array}{rcl} 
g \frac{\partial S\lbrack C,g \rbrack }{\partial g } 
&=&  g \frac{\partial \bar{\Gamma}_0 }{\partial g }   
+  \int {\rm d}^4x \left\lbrack   
- (\partial^{\mu} \bar\eta_{\alpha}(x)) 
(\partial_{\mu} \eta^{\alpha}(x))  
 - \bar\eta_{\alpha}(x) \partial^{\mu}    
{\mathbf D}_{\mu \beta}^{\alpha} \eta^{\beta}(x)  
+K^{\mu}_{\alpha}(x) \partial_{\mu} \eta^{\alpha}(x) \right.  \\ 
&&\\ 
&& \left. 
- K^{\mu}_{\alpha}(x) {\mathbf D}_{\mu \beta}^{\alpha} 
\eta^{\beta}(x)  
- g L_{\alpha}(x) \eta^{\beta}(x) 
(\partial_{\beta} \eta^{\alpha}(x) ) 
\right\rbrack. 
\end{array} 
\label{11.158} 
\ee 
Substitute eqs.(\ref{11.157} -\ref{11.158}) 
into eq.(\ref{11.151}), we will get 
\be 
S\lbrack C,g \rbrack 
= \frac{1}{2} \int {\rm d}^4x    
\frac{\delta \bar{\Gamma}_0 }{\delta C_{\mu}^{\alpha}(x) } 
C_{\mu}^{\alpha}(x) 
- \frac{1}{2} g \frac{\partial \bar{\Gamma}_0}{\partial g} 
+ \int {\rm d}^4x \left\lbrace 
\frac{1}{2} g L_{\alpha}(x) \eta^{\beta}(x) 
(\partial_{\beta} \eta^{\alpha}(x) \right\rbrace. 
\label{11.159} 
\ee  
Substitute eq.(\ref{11.159}) into eq.(\ref{11.149}). we will get 
\be 
\begin{array}{rcl} 
\bar{\Gamma}_{N+1,div}^{\lbrack N \rbrack} 
&=& \int {\rm d}^4x \left\lbrack 
(\frac{\alpha_{N+1}(\varepsilon)}{2}  
- \beta_{N+1}(\varepsilon) ) C_{\mu}^{\alpha}(x) 
\frac{\delta \bar{\Gamma}_0 }{\delta C_{\mu}^{\alpha}(x) }  
 \right. \\ 
&&\\ 
&&+ (\frac{\alpha_{N+1}(\varepsilon)}{2}  
- \gamma_{N+1}(\varepsilon) ) L_{\alpha}(x) 
\frac{\delta \bar{\Gamma}_0 }  
{\delta  L_{\alpha}(x) }   \\ 
&&\\ 
&& + \gamma_{N+1}(\varepsilon) \eta^{\alpha}(x)  
\frac{\delta \bar{\Gamma}_0 } {\delta  \eta^{\alpha}(x) }  
+ \beta_{N+1}(\varepsilon) \bar\eta_{\alpha}(x)  
\frac{\delta \bar{\Gamma}_0 } {\delta  \bar\eta_{\alpha}(x) } \\ 
&&\\ 
&& \left. + \beta_{N+1}(\varepsilon) K^{\mu}_{\alpha}(x) 
\frac{\delta \bar{\Gamma}_0 }{\delta K^{\mu}_{\alpha}(x) }     
\right\rbrack - \frac{\alpha_{N+1}(\varepsilon)}{2} 
g \frac{\partial \bar{\Gamma}_0 }{\partial g }  
\end{array} 
\label{11.160} 
\ee  
\\ 
 
On the other hand, we can prove the following relations 
\be 
\begin{array}{l} 
\int {\rm d}^4x  \eta^{\alpha}(x) 
\frac{\delta  }{\delta \eta^{\alpha}(x) } 
\left\lbrack 
\int {\rm d}^4y ~ \bar\eta_{\beta}(y) \partial^{\mu} 
{\mathbf D}_{\mu \sigma}^{\beta} \eta^{\sigma}(y) 
\right\rbrack   
= \int {\rm d}^4x~  
 \bar\eta_{\beta}(x) \partial^{\mu}  
{\mathbf D}_{\mu \sigma}^{\beta} \eta^{\sigma}(x), 
\end{array}     
\label{11.161} 
\ee 
\be 
\begin{array}{l} 
\int {\rm d}^4x  \eta^{\alpha}(x)   
\frac{\delta  }{\delta \eta^{\alpha}(x) }    
\left\lbrack 
\int {\rm d}^4y ~ K^{\mu}_{\beta}(y) 
{\mathbf D}_{\mu \sigma}^{\beta} \eta^{\sigma}(y) 
\right\rbrack   
=  \int {\rm d}^4x ~             
 K^{\mu}_{\beta}(x)   
{\mathbf D}_{\mu \sigma}^{\beta} \eta^{\sigma}(x), 
\end{array}     
\label{11.162} 
\ee 
\be 
\begin{array}{l} 
\int {\rm d}^4x  \eta^{\alpha}(x) 
\frac{\delta  }{\delta \eta^{\alpha}(x) } 
\left\lbrack 
\int {\rm d}^4y ~  g L_{\beta}(y)  
\eta^{\sigma}(y) (\partial_{\sigma} \eta^{\beta}(y) ) 
\right\rbrack   
= 2  \int {\rm d}^4x~ 
 g L_{\beta}(x)   
\eta^{\sigma}(x) (\partial_{\sigma} \eta^{\beta}(x) ), 
\end{array}     
\label{11.163} 
\ee 
 
\be  
\begin{array}{l} 
\int {\rm d}^4x  \eta^{\alpha}(x) 
\frac{\delta \bar\Gamma_0  }{\delta \eta^{\alpha}(x) } 
= \int {\rm d}^4x \left\lbrace 
 \bar\eta_{\beta}(x) \partial^{\mu}  
{\mathbf D}_{\mu \sigma}^{\beta} \eta^{\sigma}(x) \right. \\ 
\\ 
  \left. 
+  K^{\mu}_{\beta}(x)   
{\mathbf D}_{\mu \sigma}^{\beta} \eta^{\sigma}(x) 
+ 2  g L_{\beta}(x) 
\eta^{\sigma}(x) (\partial_{\sigma} \eta^{\beta}(x)) 
\right\rbrace. 
\end{array} 
\label{11.164} 
\ee 
Substitute eqs.(\ref{11.161} - \ref{11.163}) 
into eq.(\ref{11.164}), we will get 
\be  
\int {\rm d}^4x  \left\lbrace 
- \eta^{\alpha} 
\frac{\delta \bar\Gamma_0  }{\delta \eta^{\alpha} } 
+  \bar\eta_{\alpha}    
\frac{\delta \bar\Gamma_0  }{\delta \bar\eta_{\alpha} }   
+ K^{\mu}_{\alpha} 
\frac{\delta \bar\Gamma_0  }{\delta K^{\mu}_{\alpha} } 
+ 2 L_{\alpha} 
\frac{\delta \bar\Gamma_0  }{\delta L_{\alpha} } 
\right\rbrace = 0. 
\label{11.165} 
\ee  
Eq.(\ref{11.165}) times $\frac{\gamma_{N+1}-\beta_{N+1}}{2}$, then 
add up this results and eq.(\ref{11.160}), we will get 
\be   
\begin{array}{rcl} 
\bar{\Gamma}_{N+1,div}^{\lbrack N \rbrack} 
&=& \int {\rm d}^4x \left\lbrack 
(\frac{\alpha_{N+1}(\varepsilon)}{2}  
- \beta_{N+1}(\varepsilon) )\left( C_{\mu}^{\alpha}(x) 
\frac{\delta \bar{\Gamma}_0 }{\delta C_{\mu}^{\alpha}(x) } 
+  L_{\alpha}(x) 
\frac{\delta \bar{\Gamma}_0 } 
{\delta  L_{\alpha}(x) } \right) \right.  \\ 
&&\\ 
&& \left. 
\frac{\beta_{N+1}(\varepsilon)+\gamma_{N+1}(\varepsilon)}{2} 
 \left( \eta^{\alpha}(x) 
\frac{\delta \bar\Gamma_0  }{\delta \eta^{\alpha}(x) } 
+ \bar\eta_{\alpha}(x) 
\frac{\delta \bar{\Gamma}_0 } {\delta  \bar\eta_{\alpha}(x) } 
 +  K^{\mu}_{\alpha}(x) 
\frac{\delta \bar{\Gamma}_0 }{\delta K^{\mu}_{\alpha}(x) } 
\right) 
\right\rbrack \\ 
&&\\ 
&&  - \frac{\alpha_{N+1}(\varepsilon)}{2} 
g \frac{\partial \bar{\Gamma}_0 }{\partial g }. 
\end{array} 
\label{11.166} 
\ee  
This is the most general form of  
$\bar{\Gamma}_{N+1,div}^{\lbrack N \rbrack}$ 
which satisfies Ward-Takahashi identities. \\ 
 
According to minimal subtraction, the counterterm that cancel 
the divergent part of $\bar{\Gamma}_{N+1}^{\lbrack N \rbrack}$ 
is just $-\bar{\Gamma}_{N+1,div}^{\lbrack N \rbrack}$, that is 
\be  
\swav{S}^{\lbrack N+1 \rbrack} 
= \swav{S}^{\lbrack N \rbrack} 
- \bar{\Gamma}_{N+1,div}^{\lbrack N \rbrack} 
+ o(\hbar^{N+2}), 
\label{11.167} 
\ee  
where the term of $o(\hbar^{N+2})$ has no contribution to 
the divergences of $(N+1)$-loop diagrams. Suppose that  
$\bar{\Gamma}_{N+1}^{\lbrack N+1 \rbrack}$ is the generating 
functional of regular vertex which is calculated from  
$\swav{S}^{\lbrack N+1 \rbrack}$. It can be easily proved that 
\be   
\Gamma_{N+1}^{\lbrack N+1 \rbrack}  
= \Gamma_{N+1}^{\lbrack N \rbrack}   
- \bar{\Gamma}_{N+1,div}^{\lbrack N \rbrack}. 
\label{11.168} 
\ee 
Using eq.(\ref{11.100}), we can get 
\be   
\Gamma_{N+1}^{\lbrack N+1 \rbrack} 
= \bar{\Gamma}_{N+1,F}^{\lbrack N \rbrack}. 
\label{11.169} 
\ee 
$\Gamma_{N+1}^{\lbrack N+1 \rbrack}$ contains no divergence 
which is just what we expected. \\ 
 
Now, let's try to determine the form of  
$\swav{S}^{\lbrack N+1 \rbrack}$. Denote the non-renormalized  
action of the system as 
\be 
\swav{S}^{\lbrack 0 \rbrack} 
= \swav{S}^{\lbrack 0 \rbrack} 
\lbrack 
C_{\mu}^{\alpha}, \bar\eta_{\alpha}, \eta^{\alpha}, 
K^{\mu}_{\alpha}, L_{\alpha}, g, \alpha 
\rbrack.  
\label{11.170} 
\ee  
As one of the inductive hypothesis, we suppose that the 
action of the system after $\hbar^N$ order renormalization is  
\be  
\begin{array}{l} 
\swav{S}^{\lbrack N \rbrack}\lbrack 
C_{\mu}^{\alpha}, \bar\eta_{\alpha}, \eta^{\alpha}, 
K^{\mu}_{\alpha}, L_{\alpha}, g, \alpha \rbrack  \\ 
\\ 
= \swav{S}^{\lbrack 0 \rbrack} 
\lbrack 
\sqrt{Z_1^{\lbrack N \rbrack}}C_{\mu}^{\alpha},  
\sqrt{Z_2^{\lbrack N \rbrack}}\bar\eta_{\alpha},  
\sqrt{Z_3^{\lbrack N \rbrack}}\eta^{\alpha}, 
\sqrt{Z_4^{\lbrack N \rbrack}}K^{\mu}_{\alpha},  
\sqrt{Z_5^{\lbrack N \rbrack}}L_{\alpha},  
      Z_g^{\lbrack N \rbrack} g,  
      Z_{\alpha}^{\lbrack N \rbrack} \alpha 
\rbrack. 
\end{array} 
\label{11.171} 
\ee  
Substitute eq(\ref{11.166}) and 
eq.(\ref{11.171}) into eq.(\ref{11.167}), 
we obtain 
\be   
\begin{array}{rcl} 
\swav{S}^{\lbrack N+1 \rbrack} &=&  
\swav{S}^{\lbrack 0 \rbrack} 
\lbrack 
\sqrt{Z_1^{\lbrack N \rbrack}}C_{\mu}^{\alpha}, 
\sqrt{Z_2^{\lbrack N \rbrack}}\bar\eta_{\alpha},  
\sqrt{Z_3^{\lbrack N \rbrack}}\eta^{\alpha}, 
\sqrt{Z_4^{\lbrack N \rbrack}}K^{\mu}_{\alpha}, 
\sqrt{Z_5^{\lbrack N \rbrack}}L_{\alpha}, 
      Z_g^{\lbrack N \rbrack} g,  
      Z_{\alpha}^{\lbrack N \rbrack} \alpha 
\rbrack  \\ 
&&\\ 
&& - \int {\rm d}^4x \left\lbrack 
(\frac{\alpha_{N+1}(\varepsilon)}{2}  
- \beta_{N+1}(\varepsilon) )\left( C_{\mu}^{\alpha}(x) 
\frac{\delta \bar{\Gamma}_0 }{\delta C_{\mu}^{\alpha}(x) } 
+  L_{\alpha}(x) 
\frac{\delta \bar{\Gamma}_0 } 
{\delta  L_{\alpha}(x) } \right) \right.  \\ 
&&\\ 
&& \left. 
+ \frac{\beta_{N+1}(\varepsilon)+\gamma_{N+1}(\varepsilon)}{2} 
 \left( \eta^{\alpha}(x) 
\frac{\delta \bar\Gamma_0  }{\delta \eta^{\alpha}(x) } 
+ \bar\eta_{\alpha}(x) 
\frac{\delta \bar{\Gamma}_0 } {\delta  \bar\eta_{\alpha}(x) } 
 +  K^{\mu}_{\alpha}(x) 
\frac{\delta \bar{\Gamma}_0 }{\delta K^{\mu}_{\alpha}(x) } 
\right) 
\right\rbrack \\ 
&&\\ 
&&  + \frac{\alpha_{N+1}(\varepsilon)}{2} 
g \frac{\partial \bar{\Gamma}_0 }{\partial g } 
 + o(\hbar^{N+2}). 
\end{array} 
\label{11.172} 
\ee  
Using eq.(\ref{11.103}), we can prove that 
\be   
\int {\rm d}^4x C_{\mu}^{\alpha}(x) 
\frac{\delta \bar{\Gamma}_0}{\delta C_{\mu}^{\alpha}(x)} 
= \int {\rm d}^4x C_{\mu}^{\alpha}(x) 
\frac{\delta \swav{S}^{\lbrack 0 \rbrack}} 
{\delta C_{\mu}^{\alpha}(x)} 
+ 2 \alpha  
\frac{\partial  \swav{S}^{\lbrack 0 \rbrack}} 
{\partial \alpha}, 
\label{11.173} 
\ee  
\be   
\int {\rm d}^4x L_{\alpha}(x) 
\frac{\delta \bar{\Gamma}_0}{\delta L_{\alpha}(x)} 
= \int {\rm d}^4x L_{\alpha}(x) 
\frac{\delta \swav{S}^{\lbrack 0 \rbrack}} 
{\delta L_{\alpha}(x)}, 
\label{11.174} 
\ee 
\be   
\int {\rm d}^4x \bar\eta_{\alpha}(x)          
\frac{\delta \bar{\Gamma}_0}{\delta \bar\eta_{\alpha}(x)}       
= \int {\rm d}^4x   \bar\eta_{\alpha}(x)       
\frac{\delta \swav{S}^{\lbrack 0 \rbrack}} 
{\delta \bar\eta_{\alpha}(x)},      
\label{11.175}   
\ee 
\be   
\int {\rm d}^4x \eta^{\alpha}(x)   
\frac{\delta \bar{\Gamma}_0}{\delta \eta^{\alpha}(x)} 
= \int {\rm d}^4x   \eta^{\alpha}(x) 
\frac{\delta \swav{S}^{\lbrack 0 \rbrack}} 
{\delta \eta^{\alpha}(x)}, 
\label{11.176}   
\ee 
\be   
\int {\rm d}^4x K^{\mu}_{\alpha}(x) 
\frac{\delta \bar{\Gamma}_0}{\delta K^{\mu}_{\alpha}(x)} 
= \int {\rm d}^4x K^{\mu}_{\alpha}(x) 
\frac{\delta \swav{S}^{\lbrack 0 \rbrack}} 
{\delta K^{\mu}_{\alpha}(x)}, 
\label{11.177}   
\ee 
\be  
g \frac{\partial \bar{\Gamma}_0}{\partial g} 
= g \frac{\partial \swav{S}^{\lbrack 0 \rbrack}}{\partial g}. 
\label{11.178} 
\ee  
Using these relations, eq.(\ref{11.172}) is changed into, 
\be   
\begin{array}{rcl} 
\swav{S}^{\lbrack N+1 \rbrack} &=& 
\swav{S}^{\lbrack 0 \rbrack} 
\lbrack 
\sqrt{Z_1^{\lbrack N \rbrack}}C_{\mu}^{\alpha},  
\sqrt{Z_2^{\lbrack N \rbrack}}\bar\eta_{\alpha}, 
\sqrt{Z_3^{\lbrack N \rbrack}}\eta^{\alpha},    
\sqrt{Z_4^{\lbrack N \rbrack}}K^{\mu}_{\alpha}, 
\sqrt{Z_5^{\lbrack N \rbrack}}L_{\alpha}, 
      Z_g^{\lbrack N \rbrack} g,  
      Z_{\alpha}^{\lbrack N \rbrack} \alpha 
\rbrack  \\ 
&&\\ 
&& - \int {\rm d}^4x \left\lbrack 
(\frac{\alpha_{N+1}(\varepsilon)}{2} 
- \beta_{N+1}(\varepsilon) )\left( C_{\mu}^{\alpha}(x) 
\frac{\delta  \swav{S}^{\lbrack 0 \rbrack} } 
{\delta C_{\mu}^{\alpha}(x) } 
+  L_{\alpha}(x) 
\frac{\delta  \swav{S}^{\lbrack 0 \rbrack} } 
{\delta  L_{\alpha}(x) } \right) \right.  \\ 
&&\\ 
&& \left. 
+ \frac{\beta_{N+1}(\varepsilon)+\gamma_{N+1}(\varepsilon)}{2} 
 \left( \eta^{\alpha}(x) 
\frac{\delta  \swav{S}^{\lbrack 0 \rbrack} } 
{\delta \eta^{\alpha} } 
+ \bar\eta_{\alpha}(x) 
\frac{\delta  \swav{S}^{\lbrack 0 \rbrack} }  
{\delta  \bar\eta_{\alpha}(x) } 
 +  K^{\mu}_{\alpha}(x) 
\frac{\delta  \swav{S}^{\lbrack 0 \rbrack} } 
{\delta K^{\mu}_{\alpha}(x) } 
\right) 
\right\rbrack \\ 
&&\\ 
&&  + \frac{\alpha_{N+1}(\varepsilon)}{2} 
g \frac{\partial  \swav{S}^{\lbrack 0 \rbrack} }{\partial g } 
- 2 (\frac{\alpha_{N+1}(\varepsilon)}{2} 
- \beta_{N+1}(\varepsilon) ) 
\alpha \frac{\partial \swav{S}^{\lbrack 0 \rbrack} } 
{\partial \alpha} + o(\hbar^{N+2}) 
\end{array} 
\label{11.179} 
\ee   
We can see that this relation has just the form of first order functional 
expansion. Using this relation, we can determine the form of  
$\swav{S}^{\lbrack N+1 \rbrack}$. It is 
\be   
\begin{array}{l} 
\swav{S}^{\lbrack N+1 \rbrack} \lbrack 
C_{\mu}^{\alpha}, \bar\eta_{\alpha}, \eta^{\alpha}, 
K^{\mu}_{\alpha}, L_{\alpha}, g, \alpha 
\rbrack   \\ 
\\ 
= \swav{S}^{\lbrack 0 \rbrack} 
\lbrack 
\sqrt{Z_1^{\lbrack N+1 \rbrack}}C_{\mu}^{\alpha}, 
\sqrt{Z_2^{\lbrack N+1 \rbrack}}\bar\eta_{\alpha}, 
\sqrt{Z_3^{\lbrack N+1 \rbrack}}\eta^{\alpha}, 
\sqrt{Z_4^{\lbrack N+1 \rbrack}}K^{\mu}_{\alpha}, 
\sqrt{Z_5^{\lbrack N+1 \rbrack}}L_{\alpha},  
      Z_g^{\lbrack N+1 \rbrack} g,  
      Z_{\alpha}^{\lbrack N+1 \rbrack}  \alpha 
\rbrack, 
\end{array} 
\label{11.180} 
\ee   
where 
\be  
\sqrt{Z_1^{\lbrack N+1 \rbrack}} 
= \sqrt{Z_5^{\lbrack N+1 \rbrack}} 
= \sqrt{Z_{\alpha}^{\lbrack N+1 \rbrack}} 
= 1- \sum_{L=1}^{N+1} 
\left( \frac{\alpha_L (\varepsilon)}{2}  
-\beta_L (\varepsilon) \right), 
\label{11.181} 
\ee  
\be        
\sqrt{Z_2^{\lbrack N+1 \rbrack}} 
= \sqrt{Z_3^{\lbrack N+1 \rbrack}} 
= \sqrt{Z_4^{\lbrack N+1 \rbrack}} 
= 1- \sum_{L=1}^{N+1} 
\frac{\beta_L (\varepsilon)+\gamma_L  (\varepsilon) }{2} , 
\label{11.182}                                      	   
\ee                                                    
\be        
 Z_g^{\lbrack N+1 \rbrack}  
= 1 + \sum_{L=1}^{N+1} 
\frac{\alpha_L (\varepsilon)}{2}.    
\label{11.183}                                      	   
\ee                                                    
Denote 
\be  
\sqrt{Z^{\lbrack N+1 \rbrack}} 
\define \sqrt{Z_1^{\lbrack N+1 \rbrack}} 
= \sqrt{Z_5^{\lbrack N+1 \rbrack}} 
= \sqrt{Z_{\alpha}^{\lbrack N+1 \rbrack}}, 
\label{11.184} 
\ee        
\be 
  \sqrt{\swav{Z}^{\lbrack N+1 \rbrack}}        
\define \sqrt{Z_2^{\lbrack N+1 \rbrack}}   
= \sqrt{Z_3^{\lbrack N+1 \rbrack}} 
= \sqrt{Z_4^{\lbrack N+1 \rbrack}}. 
\label{11.185}                                             
\ee                                
The eq.(\ref{11.180}) is changed into 
\be   
\begin{array}{l} 
\swav{S}^{\lbrack N+1 \rbrack} \lbrack 
C_{\mu}^{\alpha}, \bar\eta_{\alpha}, \eta^{\alpha}, 
K^{\mu}_{\alpha}, L_{\alpha}, g, \alpha 
\rbrack   \\                     
\\ 
= \swav{S}^{\lbrack 0 \rbrack} 
\lbrack 
\sqrt{Z^{\lbrack N+1 \rbrack}}C_{\mu}^{\alpha}, 
\sqrt{\swav{Z}^{\lbrack N+1 \rbrack}}\bar\eta_{\alpha}, 
\sqrt{\swav{Z}^{\lbrack N+1 \rbrack}}\eta^{\alpha}, 
\sqrt{\swav{Z}^{\lbrack N+1 \rbrack}}K^{\mu}_{\alpha}, 
\sqrt{Z^{\lbrack N+1 \rbrack}}L_{\alpha}, 
      Z_g^{\lbrack N+1 \rbrack} g, 
      Z^{\lbrack N+1 \rbrack} \alpha 
\rbrack. 
\end{array} 
\label{11.186} 
\ee   
Using eq.(\ref{11.186}), we can easily prove that 
\be   
\begin{array}{l} 
\bar{\Gamma}^{\lbrack N+1 \rbrack} \lbrack 
C_{\mu}^{\alpha}, \bar\eta_{\alpha}, \eta^{\alpha}, 
K^{\mu}_{\alpha}, L_{\alpha}, g, \alpha 
\rbrack   \\ 
\\ 
= \bar{\Gamma}^{\lbrack 0 \rbrack} 
\left\lbrack 
\sqrt{Z^{\lbrack N+1 \rbrack}}C_{\mu}^{\alpha}, 
\sqrt{\swav{Z}^{\lbrack N+1 \rbrack}}\bar\eta_{\alpha}, 
\sqrt{\swav{Z}^{\lbrack N+1 \rbrack}}\eta^{\alpha}, 
\sqrt{\swav{Z}^{\lbrack N+1 \rbrack}}K^{\mu}_{\alpha}, 
\sqrt{Z^{\lbrack N+1 \rbrack}}L_{\alpha}, 
        Z_g^{\lbrack N+1 \rbrack} g, 
	Z^{\lbrack N+1 \rbrack} \alpha 
\right\rbrack. 
\end{array} 
\label{11.187} 
\ee 
\\

Now, we need to prove that all inductive hypotheses hold at $L=N+1$. The main
inductive hypotheses which is used in the above proof are the following three:
when $L=N$, the following three hypotheses hold,
\begin{enumerate}

\item the lowest divergence of $\bar\Gamma^{\lbrack N \rbrack}$ appears
in the $(N+1)$-loop diagram;

\item $\bar\Gamma^{\lbrack N \rbrack}$ satisfies Ward-Takahashi identities
eqs.(\ref{11.93}-\ref{11.94});

\item after $\hbar^N th$ order renormalization, the action of the system
has the form of eq.(\ref{11.171}).

\end{enumerate}

First, let's see the first hypothesis. According to eq.(\ref{11.169}), after
introducing $(N+1) th$ order counterterm, the $(N+1)$-loop diagram
contribution of $\bar\Gamma^{\lbrack N+1 \rbrack}$ is finite. It means
that the lowest order divergence of $\bar\Gamma^{\lbrack N+1 \rbrack}$
appears in the $(N+2)$-loop diagram. So, the first inductive
hypothesis hold when $L=N+1$.\\

It is known that the
non-renormalized generating functional of regular vertex
\be
\bar{\Gamma}^{\lbrack 0 \rbrack} =
\bar{\Gamma}^{\lbrack 0 \rbrack}
\lbrack C, \bar\eta, \eta, K, L ,g,\alpha \rbrack
\label{11.216}
\ee
satisfies Ward-Takahsshi identities 
eqs.(\ref{11.88}-\ref{11.89}). If we define
\be
\bar{\Gamma} ' =
\bar{\Gamma}^{\lbrack 0 \rbrack}
\lbrack C', \bar\eta', \eta', K', L' ,g',\alpha',
 \rbrack,
\label{11.217}
\ee
then, it must satisfy the following Ward-Takahashi identities
\be
\partial^{\mu}
\frac{\delta \bar{\Gamma}'  }
{\delta K^{\prime\mu}_{\alpha}(x) }
= \frac{\delta \bar{\Gamma} '  }
{\delta \bar\eta'_{\alpha}(x) },
\label{11.218}
\ee
\be
\int {\rm d}^4x \left\lbrace
\frac{\delta \bar{\Gamma} ' }
{\delta K^{\prime\mu}_{\alpha}(x) }
\frac{\delta \bar{\Gamma} ' }
{\delta C_{\mu}^{\prime\alpha}(x) }
+ \frac{\delta \bar{\Gamma} ' }
{\delta L'_{\alpha}(x) }
\frac{\delta \bar{\Gamma} ' }
{\delta \eta^{\prime\alpha}(x) }
\right\rbrace = 0.
\label{11.219}
\ee
Set,
\be
C_{\mu}^{\prime\alpha}
= \sqrt{Z^{\lbrack N+1 \rbrack}} C_{\mu}^{\alpha},
\label{11.222}
\ee
\be
K^{\prime\mu}_{\alpha}
= \sqrt{\swav{Z}^{\lbrack N+1 \rbrack}} K^{\mu}_{\alpha},
\label{11.223}
\ee
\be
L'_{\alpha}
= \sqrt{Z^{\lbrack N+1 \rbrack}} L_{\alpha},
\label{11.224}
\ee
\be
\eta^{\prime\alpha}
= \sqrt{\swav{Z}^{\lbrack N+1 \rbrack}} \eta^{\alpha},
\label{11.225}
\ee
\be
\bar\eta'_{\alpha}
= \sqrt{\swav{Z}^{\lbrack N+1 \rbrack}} \bar\eta_{\alpha},
\label{11.226}
\ee
\be
g' = Z_g^{\lbrack N+1 \rbrack} g,
\label{11.227}
\ee
\be
\alpha' = Z^{\lbrack N+1 \rbrack} \alpha.
\label{11.228}
\ee
In this case, we have
\be
\begin{array}{rcl}
\bar\Gamma' &=& \bar{\Gamma}^{\lbrack 0 \rbrack}
\left\lbrack
\sqrt{Z^{\lbrack N+1 \rbrack}}C_{\mu}^{\alpha},
\sqrt{\swav{Z}^{\lbrack N+1 \rbrack}}\bar\eta_{\alpha},
\sqrt{\swav{Z}^{\lbrack N+1 \rbrack}}\eta^{\alpha},\right.\\
&&\\
&&~~~~~~~\left.
\sqrt{\swav{Z}^{\lbrack N+1 \rbrack}}K^{\mu}_{\alpha},
\sqrt{Z^{\lbrack N+1 \rbrack}}L_{\alpha},
        Z_g^{\lbrack N+1 \rbrack} g,
    Z^{\lbrack N+1 \rbrack} \alpha,
    ,\eta_1,\eta_2
\right\rbrack\\
&&\\
&=& \bar\Gamma^{\lbrack N+1 \rbrack}.
\end{array}
\label{11.231}
\ee
Then eq.(\ref{11.218}) is changed into
\be
\frac{1}{\sqrt{\swav{Z}^{\lbrack N+1 \rbrack}}}
\partial^{\mu}
\frac{\delta \bar{\Gamma}^{\lbrack N+1 \rbrack}  }
{\delta K^{\mu}_{\alpha}(x) }
= \frac{1}{\sqrt{\swav{Z}^{\lbrack N+1 \rbrack}}}
\frac{\delta \bar{\Gamma}^{\lbrack N+1 \rbrack}  }
{\delta \bar\eta_{\alpha}(x) }.
\label{11.232}
\ee
Because $\frac{1}{\sqrt{\swav{Z}^{\lbrack N+1 \rbrack}}}$ does not
vanish, the above equation gives out
\be
\partial^{\mu}
\frac{\delta \bar{\Gamma}^{\lbrack N+1 \rbrack}  }
{\delta K^{\mu}_{\alpha}(x) }
=\frac{\delta \bar{\Gamma}^{\lbrack N+1 \rbrack}  }
{\delta \bar\eta_{\alpha}(x) }.
\label{11.233}
\ee
Eq.(\ref{11.219}) gives out
\be
\int {\rm d}^4x \left\lbrace
\frac{1}{\sqrt{\swav{Z}^{\lbrack N+1 \rbrack}}
\sqrt{Z^{\lbrack N+1 \rbrack}}}
\left\lbrack
\frac{\delta \bar{\Gamma}^{\lbrack N+1 \rbrack} }
{\delta K^{\mu}_{\alpha}(x) }
\frac{\delta \bar{\Gamma}^{\lbrack N+1 \rbrack} }
{\delta C_{\mu}^{\alpha}(x) }
+ \frac{\delta \bar{\Gamma}^{\lbrack N+1 \rbrack} }
{\delta L_{\alpha}(x) }
\frac{\delta \bar{\Gamma}^{\lbrack N+1 \rbrack} }
{\delta \eta^{\alpha}(x) }\right\rbrack
\right\rbrace = 0.
\label{11.234}
\ee
Because $\frac{1}{\sqrt{\swav{Z}^{\lbrack N+1 \rbrack}}
\sqrt{Z^{\lbrack N+1 \rbrack}}}$ does not vanish, 
we can obtain
\be
\int {\rm d}^4x \left\lbrace
\frac{\delta \bar{\Gamma}^{\lbrack N+1 \rbrack} }
{\delta K^{\mu}_{\alpha}(x) }
\frac{\delta \bar{\Gamma}^{\lbrack N+1 \rbrack} }
{\delta C_{\mu}^{\alpha}(x) }
+ \frac{\delta \bar{\Gamma}^{\lbrack N+1 \rbrack} }
{\delta L_{\alpha}(x) }
\frac{\delta \bar{\Gamma}^{\lbrack N+1 \rbrack} }
{\delta \eta^{\alpha}(x) }
 \right\rbrace = 0.
\label{11.235}
\ee
Eq.(\ref{11.233}) and eq.(\ref{11.235}) 
are just the Ward-Takahashi identities
for $L=N+1$. Therefore, the second inductive hypothesis holds when
$L=N+1$. \\

The third inductive hypothesis has already been proved which is shown
in eq.(\ref{11.186}). Therefore, all three inductive hypothesis hold when
$L=N+1$. According to inductive principle, they will hold when $L$ is
an arbitrary non-negative number, especially they hold when $L$
approaches infinity. \\

In above discussions, we have proved that, if we suppose that 
when $L=N$ eq.(\ref{11.171}) holds, then it also holds when $L=N+1$. 
According to inductive principle, we know that 
eq.(\ref{11.186}  -  \ref{11.187}) 
hold for any positive integer $N$, that is 
\be   
\begin{array}{l} 
\swav{S} \lbrack 
C_{\mu}^{\alpha}, \bar\eta_{\alpha}, \eta^{\alpha}, 
K^{\mu}_{\alpha}, L_{\alpha}, g, \alpha 
\rbrack   \\ 
\\ 
= \swav{S}^{\lbrack 0 \rbrack} 
\left\lbrack 
\sqrt{Z}C_{\mu}^{\alpha}, 
\sqrt{\swav{Z}}\bar\eta_{\alpha}, 
\sqrt{\swav{Z}}\eta^{\alpha}, 
\sqrt{\swav{Z}}K^{\mu}_{\alpha}, 
\sqrt{Z}L_{\alpha}, 
      Z_g g, 
      Z \alpha 
\right\rbrack, 
\end{array} 
\label{11.188} 
\ee   
\be   
\begin{array}{l} 
\bar{\Gamma} \lbrack 
C_{\mu}^{\alpha}, \bar\eta_{\alpha}, \eta^{\alpha}, 
K^{\mu}_{\alpha}, L_{\alpha}, g, \alpha 
\rbrack   \\ 
\\ 
= \bar{\Gamma}^{\lbrack 0 \rbrack} 
\left\lbrack 
\sqrt{Z}C_{\mu}^{\alpha}, 
\sqrt{\swav{Z}}\bar\eta_{\alpha}, 
\sqrt{\swav{Z}}\eta^{\alpha}, 
\sqrt{\swav{Z}}K^{\mu}_{\alpha}, 
\sqrt{Z}L_{\alpha}, 
      Z_g  g, 
      Z \alpha 
\right\rbrack, 
\end{array} 
\label{11.189}                                     
\ee   
where 
\be    
\sqrt{Z} \define  \lim_{N \to \infty} 
 \sqrt{Z^{\lbrack N \rbrack}} 
= 1- \sum_{L=1}^{\infty} 
\left( \frac{\alpha_L (\varepsilon)}{2} 
-\beta_L (\varepsilon) \right), 
\label{11.190} 
\ee  
\be  
\sqrt{\swav{Z}} 
\define \lim_{N \to \infty} \sqrt{\swav{Z}^{\lbrack N \rbrack}} 
= 1- \sum_{L=1}^{\infty} 
\frac{\beta_L (\varepsilon)+\gamma_L  (\varepsilon) }{2} , 
\label{11.191}                                            
\ee                                                         
\be 
Z_g \define \lim_{N \to \infty}                                                          
Z_g^{\lbrack N \rbrack}                           
= 1 + \sum_{L=1}^{\infty} 
\frac{\alpha_L (\varepsilon)}{2}. 
\label{11.192}                   
\ee                             
$\swav{S} \lbrack C_{\mu}^{\alpha}, \bar\eta_{\alpha},  
\eta^{\alpha}, K^{\mu}_{\alpha}, L_{\alpha}, g, \alpha \rbrack$ 
and $\bar{\Gamma} \lbrack C_{\mu}^{\alpha}, \bar\eta_{\alpha},  
\eta^{\alpha}, K^{\mu}_{\alpha}, L_{\alpha}, g, \alpha \rbrack$ 
are renormalized action and generating functional of regular vertex.  
The generating functional of regular vertex $\bar{\Gamma}$ contains 
no divergence. All kinds of vertex that generated from $\bar{\Gamma}$  
are finite. From eq.(\ref{11.188}) 
and eq.(\ref{11.189}), we can see that we only  
introduce three known parameters which are $\sqrt{Z}$, $\sqrt{\swav{Z}}$ 
and $Z_g$. Therefore, gravitational gauge theory is a renormalizable 
theory. \\

From eq.(\ref{11.189}) and eqs.(\ref{11.88} - \ref{11.89}), 
we can deduce that the renormalized 
generating functional of regular vertex satisfies Ward-Takahashi  
identities, 
\be  
\partial^{\mu} 
\frac{\delta \bar{\Gamma}  } 
{\delta K^{\mu}_{\alpha}(x) } 
= \frac{\delta \bar{\Gamma}  } 
{\delta \bar\eta_{\alpha}(x) }, 
\label{11.193} 
\ee  
\be  
\int {\rm d}^4x \left\lbrace 
\frac{\delta \bar{\Gamma} } 
{\delta K^{\mu}_{\alpha}(x) } 
\frac{\delta \bar{\Gamma} } 
{\delta C_{\mu}^{\alpha}(x) }    
+ \frac{\delta \bar{\Gamma} } 
{\delta L_{\alpha}(x) } 
\frac{\delta \bar{\Gamma} } 
{\delta \eta^{\alpha}(x) } 
\right\rbrace = 0. 
\label{11.194} 
\ee 
It means that the renormalized theory has the structure of gauge symmetry. 
If we define 
\be  
C_{0 \mu}^{\alpha} = \sqrt{Z} C_{\mu}^{\alpha}, 
\label{11.195} 
\ee  
\be  
\eta_0^{\alpha} = \sqrt{\swav{Z}} \eta^{\alpha},   
\label{11.196} 
\ee 
\be  
\bar\eta_{0 \alpha} = \sqrt{\swav{Z}} \bar\eta_{\alpha},    
\label{11.197} 
\ee 
\be  
K^{\mu}_{0 \alpha} = \sqrt{\swav{Z}} K^{\mu}_{\alpha},   
\label{11.198} 
\ee 
\be 
L_{0 \alpha} = \sqrt{Z} L_{\alpha}, 
\label{11.199} 
\ee 
\be 
g_0 = Z_g g, 
\label{11.200} 
\ee 
\be 
\alpha_0 = Z \alpha. 
\label{11.201} 
\ee 
Therefore, eqs.(\ref{11.188} - \ref{11.189}) are changed into 
\be   
\swav{S} \lbrack 
C_{\mu}^{\alpha}, \bar\eta_{\alpha}, \eta^{\alpha}, 
K^{\mu}_{\alpha}, L_{\alpha}, g, \alpha 
\rbrack    
= \swav{S}^{\lbrack 0 \rbrack} 
\lbrack 
C_{0 \mu}^{\alpha}, \bar\eta_{0 \alpha}, \eta_0^{\alpha}, 
K^{\mu}_{0 \alpha}, L_{0 \alpha}, g_0, \alpha_0 
\rbrack, 
\label{11.202} 
\ee   
\be 
\bar{\Gamma} \lbrack 
C_{\mu}^{\alpha}, \bar\eta_{\alpha}, \eta^{\alpha}, 
K^{\mu}_{\alpha}, L_{\alpha}, g, \alpha 
\rbrack    
= \bar{\Gamma}^{\lbrack 0 \rbrack} 
\lbrack 
C_{0 \mu}^{\alpha}, \bar\eta_{0 \alpha}, \eta_0^{\alpha}, 
K^{\mu}_{0 \alpha}, L_{0 \alpha}, g_0, \alpha_0 
\rbrack.    
\label{11.203} 
\ee   
$C_{0 \mu}^{\alpha}$, $\bar\eta_{0 \alpha}$ and 
$\eta_0^{\alpha}$ are renormalized wave function,  
$K^{\mu}_{0 \alpha}$ and $L_{0 \alpha}$ are renormalized  
external sources, and $g_0$ is the renormalized gravitational  
coupling constant. \\

The action $\swav{S}$ which is given by eq.(\ref{11.202}) is invariant
under the following generalized BRST transformations,
\be
\delta C_{0\mu}^{\alpha}
= -  {\mathbf D}_{0\mu~\beta}^{\alpha} \eta_0^{\beta}
\delta \lambda,
\label{11.252}
\ee
\be
\delta \eta_0^{\alpha} = g_0 \eta_0^{\sigma}
(\partial_{\sigma}\eta_0^{\alpha}) \delta \lambda,
\label{11.253}
\ee
\be
\delta \bar\eta_{0\alpha} = \frac{1}{\alpha_0}
\eta_{\alpha \beta} f_0^{\beta} \delta \lambda,
\label{11.254}
\ee
\be
\delta \eta^{\mu \nu} = 0,
\label{11.255}
\ee
where,
\be
{\mathbf D}^{\alpha}_{0 \mu \beta} =
\delta_{\beta}^{\alpha} \partial_{\mu}
- g_0 \delta^{\alpha}_{\beta} C^{\sigma}_{0 \mu} \partial_{\sigma}
+ g_0 (\partial_{\beta} C^{\alpha}_{0\mu}),
\label{11.258}
\ee
\be
f_0^{\alpha} = \partial^{\mu} C_{0 \mu}^{\alpha}.
\label{11.259}
\ee
Therefore, the normalized action has generalized BRST symmetry, which
means that the normalized theory has the structure of gauge theory.\\

\section{Einstein-like Field Equation with Cosmological Trem}
\setcounter{equation}{0}

In the above chapters, the quantum gravity is formulated
in the traditional framework of quantum field theory, i.e.,
the physical space-time is always flat and the space-time
metric is always selected to be the Minkowski metric. In this
picture, gravity is treated as physical interactions
in flat physical space-time. Our gravitational gauge
transformation does not act on physical space-time
coordinates, but act on
physical fields, so gravitational gauge transformation
does not affect the structure of physical space-time.
This is one picture of gravity, or call it  one
representation of gravity theory. For the sake of simplicity,
we call it physical representation of gravity.\\

There is another representation of gravity which is widely
used in Einstein's general relativity. This representation
is essentially a geometrical representation of gravity.
For gravitational gauge theory, if we treat
$G^{\alpha}_{\mu}$ ( or $G^{-1 \mu}_{\alpha}$ )
as a fundamental physical input of the
theory, we can also set up the geometrical representation
of gravity\cite{c44}. For gravitational gauge theory, the geometrical
nature of gravity essentially originates from the
geometrical nature of the gravitational gauge transformation.
In the geometrical picture of gravity, gravity is not
treated as a kind of physical interactions, but is
treated as the geometry of space-time. So, there
is no physical gravitational interactions in space-time
and space-time is curved if
gravitational field does not vanish. In this case, the
space-time metric is not Minkowski metric, but
$g^{\alpha\beta}$ (or $g_{\alpha\beta}$). In other words,
Minkowski metric is the space-time metric if we discuss
gravity in physical representation while metric tensor
$g^{\alpha\beta}$ ( or $g_{\alpha\beta}$) is space-time
metric if we discuss gravity in geometrical representation.
So, if we use Minkowski metric in our discussion,
it means that we are in physical representation of
gravity; if we use metric tensor
$g^{\alpha\beta}$ (or $g_{\alpha\beta}$) in our discussion,
it means that we are in geometrical representation. \\

In one representation, gravity is treated as physical
interactions, while in another representation, gravity
is treated as geometry of space-time. But we know that
there is only one physics for gravity, so two
representations of gravity must be equivalent. This
equivalence means that the nature of gravity is
physics-geometry duality, i.e., gravity is a kind
of physical interactions which has the characteristics
of geometry; it is also a geometry of space-time
which is essentially a kind of physical interactions.
Now, let's go into the geometrical representation of
gravity and use $g^{\alpha\beta}$ and $g_{\alpha\beta}$
as space-time metric tensors. In this picture, we can
obtain an Einstein-like field equation with
cosmological term. In this chapter, we will first
calculate out the affine connection, curvature rensor,
Recci tensor and curvature scalar from gravitational 
gauge field. Then we will deduce the Einstein-like field 
equation with cosmological constnat. 
\\

In the previous discussions, we have given out the relations between
space-time metric and gravitational gauge fields, which is shown in
eq.(\ref{4.706} - \ref{4.707}). Using these relations, we can
change the lagrangian for scalar field into the following form
\be
{\cal L} = -\frac{1}{2} g^{\alpha \beta} \partial_{\alpha} \phi
\partial_{\beta} \phi  -  \frac{m^2}{2} \phi^2.
\label{20.19}
\ee
In this lagrangian,  we can not see any gravitational gauge
field in it directly. It can be selected as  the lagrangian 
of scalar field in the geometrical representation of gravity. 
In this case, gravity is treated as geometry of space-time. 
Generally speaking, in gravitational gauge theory, we can 
change all lagrangians into the form  which is expressed 
in terms of $G^{\alpha}_{\mu}$. In other words, if we
did not discuss quantum behavior of gravity, from
mathematical point of view, we can use
$G^{\alpha}_{\mu}$ as a fundamental quantity to represent
gravitational field. Plese note that, in gauge theory of
gravity, $G^{\alpha}_{\mu}$ is the quantum counterpart
of Cartan tetrad in Cartan geometry. If we use 
$G^{\alpha}_{\mu}$ as a fundamental quantity of gravity
theory, we can set up the geometrical representation
of gauge theory of gravity, which is quite similar 
to the general relativity in Cartan geometry\cite{c44}.
\\

Eq.(\ref{20.19}) gives out the lagrangian in curved
space-time. But we can make a local transformation
to transform it into flat space-time. 
Making the following coordinate transformation
\be
{\rm d}x^{\mu}  \to {\rm d}x^{\prime \mu}
= \frac{\partial x^{\prime \mu}}{\partial x^{\nu}} {\rm d}x^{\nu},
\label{20.20}
\ee
where $\frac{\partial x^{\prime \mu}}{\partial x^{\nu}}$ is given by,
\be
\frac{\partial x^{\prime \mu}}{\partial x^{\nu}}
= (G^{-1}) ^{\mu}_{\nu}.
\label{20.21}
\ee
It can be proved that
\be
\begin{array}{rcl}
g^{\alpha \beta} \to g^{\prime \alpha \beta}
& = & \frac{\partial x^{\prime \alpha}}{\partial x^{\mu}}
\frac{\partial x^{\prime \beta}}{\partial x^{\nu}} g^{\mu \nu}  \\
& = & \eta^{\alpha \beta}.
\end{array}
\label{20.28}
\ee
Therefore, under this coordinates transformation, the space-time
metric becomes flat, in other words, we go into an inertial
reference system. In this inertial reference system, the Lagrangian
eq.(\ref{20.19}) becomes
\be
{\cal L} = -\frac{1}{2} \eta^{\alpha \beta} \partial_{\alpha} \phi
\partial_{\beta} \phi  -  \frac{m^2}{2} \phi^2.
\label{20.29}
\ee
Eq.(\ref{20.29}) is just the Lagrangian for real 
scalar fields in flat Minkowski space-time. \\

Using space-time metric tensors $g_{\alpha\beta}$
and $g^{\alpha\beta}$, we can calculate the affine
connection and curvature tensor. 
The affine connection $\Gamma^{\lambda}_{\mu \nu}$ is defined by
\be
\Gamma^{\lambda}_{\mu \nu}
= \frac{1}{2} g^{\lambda \sigma}
\left( \frac{\partial g_{\mu \sigma}}{\partial x^{\nu}}
+ \frac{\partial g_{\nu \sigma}}{\partial x^{\mu}}
-\frac{\partial g_{\mu \nu}}{\partial x^{\sigma}} \right).
\label{20.33}
\ee
Using the following relation,
\be
g F^{\lambda}_{\rho \sigma}
= G^{\nu}_{\rho} G^{\mu}_{\sigma}
\lbrack ( G^{-1} \partial_{\mu} G)^{\lambda}_{\nu}
- ( G^{-1} \partial_{\nu} G)^{\lambda}_{\mu} \rbrack,
\label{20.34}
\ee
where $ F^{\lambda}_{\rho \sigma}$ is the component field strength of
gravitational gauge field, we get
\be
\begin{array}{rcl}
\Gamma^{\lambda}_{\mu \nu}
&=& - \frac{1}{2}
\lbrack ( G^{-1} \partial_{\mu} G)^{\lambda}_{\nu}
+ ( G^{-1} \partial_{\nu} G)^{\lambda}_{\mu} \rbrack \\
&& + \frac{1}{2} g \eta^{\alpha_1 \beta_1} \eta_{\alpha \beta}
F^{\rho}_{\mu_1 \beta_1} G^{\lambda}_{\alpha_1} G^{-1 \alpha}_{\rho}
( G^{-1 \beta}_{\nu} G^{-1 \mu_1}_{\mu}
+ G^{-1 \beta}_{\mu} G^{-1 \mu_1}_{\nu} ).
\end{array}
\label{20.35}
\ee
From this expression, we can see that, if there is no gravity in space-time,
that is
\be
C_{\mu}^{\alpha} = 0~~,~~
F^{\lambda}_{\mu  \nu} = 0,
\label{20.36}
\ee
then the affine connection $\Gamma^{\lambda}_{\mu \nu}$ will vanish,
which is what we expect in general relativity.  \\

The curvature tensor $R^{\lambda}_{\mu \nu \kappa}$ is defined by
\be
R^{\lambda}_{\mu \nu \kappa}
\define \partial_{\kappa} \Gamma^{\lambda}_{\mu \nu}
-\partial_{\nu} \Gamma^{\lambda}_{\mu \kappa}
+\Gamma^{\eta}_{\mu \nu} \Gamma^{\lambda}_{\kappa \eta}
- \Gamma^{\eta}_{\mu \kappa} \Gamma^{\lambda}_{\nu \eta},
\label{20.37}
\ee
the Ricci tensor $R_{\mu \kappa}$ is defined by
\be
R_{\mu \kappa} \define
R^{\lambda}_{\mu \lambda \kappa},
\label{20.38}
\ee
and the curvature scalar $R$ is defined by
\be
R \define g^{\mu \kappa} R_{\mu \kappa}.
\label{20.39}
\ee
The explicit expression for Ricci tensor $R_{\mu \kappa}$ is
\be
\begin{array}{rcl}
R_{\mu \kappa} &=&
-(\partial_{\kappa} \partial_{\mu} G\cdot G^{-1}) ^{\alpha}_{\alpha}
+ 2 ( \partial_{\kappa} G \cdot G^{-1}
\cdot \partial_{\mu} G \cdot G^{-1} )^{\alpha}_{\alpha}  \\
&& + \eta^{\rho \sigma} \eta_{\alpha \beta}
( \partial_{\mu} G \cdot G^{-1} )^{\alpha}_{\rho}
( \partial_{\kappa} G \cdot G^{-1} )^{\beta}_{\sigma} \\
&&+ \frac{1}{2} g^{\lambda \nu} \eta_{\alpha \beta}
( G^{-1} \cdot \partial_{\nu} G \cdot G^{-1} \cdot \partial_{\lambda} G
\cdot G^{-1} )^{\alpha}_{\mu} G^{-1 \beta}_{\kappa}  \\
&&  - \frac{1}{2} g^{\lambda \nu} \eta_{\alpha \beta}
( G^{-1} \cdot \partial_{\nu} \partial_{\lambda} G
\cdot G^{-1} )^{\alpha}_{\mu} G^{-1 \beta}_{\kappa} \\
&& + \frac{1}{2} g^{\lambda \nu} \eta_{\alpha \beta}
( G^{-1} \cdot \partial_{\lambda}G \cdot G^{-1} \cdot
\partial_{\nu} G \cdot G^{-1} ) ^{\alpha}_{\mu} G^{-1 \beta}_{\kappa} \\
&&+ \frac{1}{2} g^{\lambda \nu} \eta_{\alpha \beta}
G^{-1 \alpha }_{\mu}
( G^{-1} \cdot \partial_{\nu} G \cdot G^{-1} \cdot \partial_{\lambda} G
\cdot G^{-1} )^{\beta}_{\kappa} \\
&& -\frac{1}{2} g^{\lambda \nu} \eta_{\alpha \beta}
G^{-1 \alpha }_{\mu}
( G^{-1} \cdot \partial_{\nu} \partial_{\lambda}G
\cdot G^{-1} )^{\beta}_{\kappa} \\
&& + \frac{1}{2} g^{\lambda \nu} \eta_{\alpha \beta}
G^{-1 \alpha }_{\mu}
( G^{-1} \cdot \partial_{\lambda} G \cdot G^{-1} \cdot \partial_{\nu} G
\cdot G^{-1} )^{\beta}_{\kappa}  \\
&&+  g^{\lambda \nu} \eta_{\alpha \beta}
( G^{-1} \cdot \partial_{\nu} G \cdot G^{-1} ) ^{\alpha}_{\mu}
( G^{-1} \cdot \partial_{\lambda} G \cdot G^{-1} )^{\beta}_{\kappa} \\
&& -\frac{1}{2} ( G^{-1} \cdot \partial_{\kappa} G \cdot G^{-1}
\cdot \partial_{\lambda} G ) ^{\lambda}_{\mu}
-\frac{1}{2} ( G^{-1} \cdot \partial_{\lambda} G \cdot G^{-1}
\cdot \partial_{\kappa} G ) ^{\lambda}_{\mu}
+ \frac{1}{2} ( G^{-1} \cdot \partial_{\kappa}
\partial_{\lambda} G ) ^{\lambda}_{\mu} \\
&& - \eta^{\rho \sigma} \eta_{\alpha \beta} G^{\lambda}_{\rho}
( G^{-1} \cdot \partial_{\lambda} G \cdot G^{-1} )^{\alpha}_{\mu}
(\partial_{\kappa} G \cdot G^{-1})^{ \beta}_{\sigma} \\
&& - \frac{1}{2} \eta^{\rho \sigma} \eta_{\alpha \beta}
G^{\lambda}_{\rho} G^{-1 \alpha }_{\mu}
 ( \partial_{\kappa} G \cdot G^{-1} \cdot \partial_{\lambda} G
\cdot G^{-1} ) ^{\beta}_{\sigma}  \\
&& + \frac{1}{2}\eta^{\rho \sigma} \eta_{\alpha \beta} G^{\lambda}_{\rho}
G^{-1 \alpha }_{\mu}
( \partial_{\kappa} \partial_{\lambda} G \cdot G^{-1} )^{\beta}_{\sigma}  \\
&& - \frac{1}{2} \eta^{\rho \sigma} \eta_{\alpha \beta}
G^{\lambda}_{\rho} G^{-1 \alpha }_{\mu}
 ( \partial_{\lambda} G \cdot G^{-1} \cdot \partial_{\kappa} G
\cdot G^{-1} ) ^{\beta}_{\sigma}  \\
&&-\frac{1}{2} ( G^{-1} \cdot \partial_{\nu} G \cdot G^{-1}
\cdot \partial_{\mu} G ) ^{\nu}_{\kappa}
-\frac{1}{2} ( G^{-1} \cdot \partial_{\mu} G \cdot G^{-1}
\cdot \partial_{\nu} G ) ^{\nu}_{\kappa}
+ \frac{1}{2} ( G^{-1} \cdot \partial_{\nu}
\partial_{\mu} G ) ^{\nu}_{\kappa} \\
&&- \eta^{\rho \sigma} \eta_{\alpha \beta} G^{\nu}_{\sigma}
( G^{-1} \cdot \partial_{\mu} G \cdot G^{-1} )^{\alpha}_{\kappa}
(\partial_{\nu} G \cdot G^{-1})^{ \beta}_{\rho} \\
&&- \frac{1}{2} \eta^{\rho \sigma} \eta_{\alpha \beta}
G^{\nu}_{\sigma} G^{-1 \alpha }_{\kappa}
 ( \partial_{\nu} G \cdot G^{-1} \cdot \partial_{\mu} G
\cdot G^{-1} ) ^{\beta}_{\rho}  \\
&& + \frac{1}{2}\eta^{\rho \sigma} \eta_{\alpha \beta} G^{\nu}_{\sigma}
G^{-1 \alpha }_{\kappa}
( \partial_{\nu} \partial_{\mu} G \cdot G^{-1} )^{\beta}_{\rho}
- \frac{1}{2} \eta^{\rho \sigma} \eta_{\alpha \beta}
G^{\nu}_{\sigma} G^{-1 \alpha }_{\kappa}
 ( \partial_{\mu} G \cdot G^{-1} \cdot \partial_{\nu} G
\cdot G^{-1} ) ^{\beta}_{\rho}  \\
&& + \frac{1}{2} \eta_{\alpha \beta} \eta^{\alpha_1 \beta_1}
G^{\nu}_{\beta_1} ( \partial_{\nu} G \cdot G^{-1})^{\alpha}_{\alpha_1}
( G^{-1} \cdot \partial_{\mu} G \cdot G^{-1})^{\beta}_{\kappa} \\
&&+ \frac{1}{2} \eta_{\alpha \beta} \eta^{\alpha_1 \beta_1}
G^{\nu}_{\beta_1} ( \partial_{\nu} G \cdot G^{-1})^{\alpha}_{\alpha_1}
( G^{-1} \cdot \partial_{\kappa} G \cdot G^{-1})^{\beta}_{\mu}  \\
&& - \frac{1}{4} \eta_{\alpha \beta} \eta^{\alpha_1 \beta_1}
( \partial_{\kappa} G \cdot G^{-1})^{\alpha}_{\alpha_1}
( \partial_{\mu} G \cdot G^{-1})^{\beta}_{\beta_1} \\
&&- \frac{1}{4} \eta_{\alpha \beta} \eta^{\alpha_1 \beta_1} G^{\nu}_{\beta_1}
( \partial_{\kappa} G \cdot G^{-1})^{\alpha}_{\alpha_1}
( G^{-1} \cdot \partial_{\nu} G \cdot G^{-1})^{\beta}_{\mu}\\
&& - \frac{1}{4} \eta_{\alpha \beta} \eta^{\alpha_1 \beta_1} G^{\lambda}_{\alpha_1}
(G^{-1} \cdot \partial_{\lambda} G \cdot G^{-1})^{\alpha}_{\kappa}
( \partial_{\mu} G \cdot G^{-1})^{\beta}_{\beta_1} \\
&& - \frac{1}{4} \eta_{\alpha \beta} \eta^{\alpha_1 \beta_1}
G^{\lambda}_{\alpha_1}G^{\nu}_{\beta_1}
( G^{-1} \cdot \partial_{\lambda} G \cdot G^{-1})^{\alpha}_{\kappa}
( G^{-1} \cdot \partial_{\nu} G \cdot G^{-1})^{\beta}_{\mu}\\
&& - \frac{g}{2} \eta_{\alpha \beta} \eta^{\alpha_3 \beta_3}
F^{\rho}_{\mu_1 \beta_1} G^{\nu}_{\beta_3} G^{-1 \alpha}_{\rho}
G^{-1 \beta}_{\kappa} G^{-1 \mu_1}_{\mu}
( \partial_{\nu} G \cdot G^{-1} )^{\beta_1}_{\alpha_3}  \\
&& - \frac{g}{2} \eta_{\alpha \beta} \eta^{\alpha_3 \beta_3}
F^{\rho}_{\mu_1 \beta_1} G^{\nu}_{\beta_3} G^{-1 \alpha}_{\rho}
G^{-1 \beta}_{\mu} G^{-1 \mu_1}_{\kappa}
( \partial_{\nu} G \cdot G^{-1} )^{\beta_1}_{\alpha_3} \\
&& - \frac{g}{2} F^{\rho}_{\alpha \beta} G^{-1 \alpha}_{\rho}
( G^{-1} \cdot \partial_{\mu} G \cdot G^{-1})^{\beta}_{\kappa}
- \frac{g}{2} F^{\rho}_{\alpha \beta} G^{-1 \alpha}_{\rho}
( G^{-1} \cdot \partial_{\kappa} G \cdot G^{-1})^{\beta}_{\mu}\\
&& + \frac{g}{4} \eta^{\alpha_3 \mu_1} \eta_{\alpha \beta}
F^{\rho}_{\mu_1 \beta_1} G^{-1 \alpha}_{\rho} G^{-1 \beta}_{\mu}
(\partial_{\kappa}G \cdot G^{-1} )^{\beta_1}_{\alpha_3}
+ \frac{g}{4} F^{\rho}_{\alpha \beta}
(G^{-1} \cdot \partial_{\kappa} G \cdot G^{-1} )^{\beta}_{\rho}
G^{-1 \alpha}_{\mu}  \\
&&+ \frac{g}{4} \eta^{\alpha_3 \mu_1} \eta_{\alpha \beta}
F^{\rho}_{\mu_1 \beta_1} G^{-1 \alpha}_{\rho} G^{-1 \beta}_{\mu}
G^{\lambda}_{\alpha_3}
(G^{-1} \cdot \partial_{\lambda}G \cdot G^{-1} )^{\beta_1}_{\kappa} \\
&&+ \frac{g}{4} F^{\rho}_{\alpha \beta}
(G^{-1} \cdot \partial_{\rho} G \cdot G^{-1} )^{\beta}_{\kappa}
G^{-1 \alpha}_{\mu}
+ \frac{g}{4} F^{\rho}_{\alpha \beta}
(G^{-1} \cdot \partial_{\rho} G \cdot G^{-1} )^{\beta}_{\mu}
G^{-1 \alpha}_{\kappa} \\
&&+ \frac{g}{4} \eta^{\alpha_3 \mu_1} \eta_{\alpha \beta}
F^{\rho}_{\mu_1 \beta_1} G^{-1 \alpha}_{\rho} G^{-1 \beta}_{\kappa}
 (\partial_{\mu}G \cdot G^{-1} )^{\beta_1}_{\alpha_3}
+ \frac{g}{4} F^{\rho}_{\alpha \beta}
(G^{-1} \cdot \partial_{\mu} G \cdot G^{-1} )^{\beta}_{\rho}
G^{-1 \alpha}_{\kappa}  \\
&&+ \frac{g}{4} \eta^{\alpha_3 \mu_1} \eta_{\alpha \beta}
F^{\rho}_{\mu_1 \beta_1} G^{-1 \alpha}_{\rho} G^{-1 \beta}_{\kappa}
G^{\lambda}_{\alpha_3}
(G^{-1} \cdot \partial_{\lambda}G \cdot G^{-1} )^{\beta_1}_{\mu} \\
&& + \frac{g^2}{2} \eta^{\sigma \beta_1} \eta_{\alpha_2 \beta_2}
F^{\rho}_{\alpha \sigma} F^{\rho_2}_{\mu_2 \beta_1}
G^{-1 \alpha}_{\rho} G^{-1 \alpha_2}_{\rho_2}
( G^{-1 \beta_2}_{\mu} G^{-1 \mu_2}_{\kappa}
+ G^{-1 \beta_2}_{\kappa} G^{-1 \mu_2}_{\mu}  )  \\
&& - \frac{g^2}{4} \eta_{\alpha \beta} \eta_{\alpha_2 \beta_2}
\eta^{\alpha_1 \beta_1} \eta^{\mu_1 \sigma_2}
F^{\rho}_{\mu_1 \beta_1} F^{\rho_1}_{\sigma_2 \alpha_1}
G^{-1 \alpha }_{\rho}  G^{-1 \alpha_2}_{\rho_1}
G^{-1 \beta}_{\kappa}  G^{-1 \beta_2}_{\mu} \\
&&- \frac{g^2}{4} \eta_{\alpha \beta} \eta^{\alpha_1 \beta_1}
F^{\rho}_{\mu_1 \beta_1} F^{\rho_1}_{\mu_3 \alpha_1}
G^{-1 \alpha }_{\rho}  G^{-1 \mu_1}_{\rho_1}
G^{-1 \beta}_{\kappa}  G^{-1 \mu_3}_{\mu}\\
&& - \frac{g^2}{4}  \eta_{\alpha_2 \beta_2} \eta^{\alpha_1 \beta_1}
F^{\rho}_{\mu_1 \beta_1} F^{\rho_1}_{\alpha \alpha_1}
G^{-1 \alpha }_{\rho}  G^{-1 \alpha_2}_{\rho_1}
G^{-1 \mu_1}_{\kappa}  G^{-1 \beta_2}_{\mu} \\
&&- \frac{g^2}{4} \eta_{\alpha \beta} \eta^{\alpha_1 \beta_1}
F^{\rho}_{\mu_1 \beta_1} F^{\rho_1}_{\mu_3 \alpha_1}
G^{-1 \alpha }_{\rho}  G^{-1 \beta}_{\rho_1}
G^{-1 \mu_1}_{\kappa}  G^{-1 \mu_3}_{\mu}
\end{array}
\label{20.40}
\ee
The explicit expression for scalar curvature $R$ is
\be
\begin{array}{rcl}
R & = &
4 g^{\mu \kappa} ( \partial_{\mu} G \cdot G^{-1}
\cdot \partial_{\kappa} G \cdot G^{-1} ) ^{\alpha}_{\alpha}
- 2 g^{\mu \kappa} ( \partial_{\mu} \partial_{\kappa}G
 \cdot G^{-1} ) ^{\alpha}_{\alpha} \\
&& + \frac{3}{2} \eta^{\rho \sigma } \eta_{\alpha \beta} g^{\mu \kappa}
( \partial_{\mu} G  \cdot G^{-1} ) ^{\alpha}_{\rho}
( \partial_{\kappa} G  \cdot G^{-1} ) ^{\beta}_{\sigma}
 - 2 \eta^{\alpha \beta} G^{\kappa}_{\beta}
( \partial_{\kappa} G \cdot G^{-1} \cdot
\partial_{\lambda} G )^{\lambda}_{\alpha}\\
&&- 2 \eta^{\alpha \beta} G^{\kappa}_{\beta}
( \partial_{\lambda} G \cdot G^{-1} \cdot
\partial_{\kappa} G )^{\lambda}_{\alpha}
 + 2 \eta^{\alpha \beta} G^{\kappa}_{\beta}
( \partial_{\kappa} \partial_{\lambda} G )^{\lambda}_{\alpha}\\
&& - \frac{5}{2} \eta_{\alpha \beta} \eta^{\rho \sigma} \eta^{\mu \nu}
G^{\kappa}_{\nu} G^{\lambda}_{\rho}
( \partial_{\lambda} G  \cdot G^{-1} ) ^{\alpha}_{\mu}
( \partial_{\kappa} G  \cdot G^{-1} ) ^{\beta}_{\sigma} \\
&& + \eta_{\alpha \beta} \eta^{\alpha_1 \beta_1} \eta^{\mu_1 \nu_1}
G^{\nu}_{\beta_1} G^{\mu}_{\mu_1}
( \partial_{\nu} G  \cdot G^{-1} ) ^{\alpha}_{\alpha_1}
( \partial_{\mu} G  \cdot G^{-1} ) ^{\beta}_{\nu_1 } \\
&&- 2 g \eta^{\alpha \beta} F^{\rho}_{\alpha_1 \beta_1}
G^{\nu}_{\beta} G^{-1 \alpha_1}_{\rho}
( \partial_{\nu} G  \cdot G^{-1} ) ^{\beta_1}_{\alpha} \\
&& + g \eta^{\alpha \mu} F^{\kappa}_{\mu \beta}
 ( \partial_{\kappa} G  \cdot G^{-1} ) ^{\beta}_{\alpha} \\
&& + g \eta^{\alpha \mu} F^{\rho}_{\alpha \beta}
G^{\kappa}_{\mu}  (G^{-1} \cdot
\partial_{\kappa} G  \cdot G^{-1} ) ^{\beta}_{\rho} \\
&& + g^2 \eta^{\beta \beta_1} F^{\rho}_{\alpha \beta}
F^{\rho_1}_{\alpha_1 \beta_1} G^{-1 \alpha}_{\rho}
G ^{-1 \alpha_1}_{\rho_1} \\
&& - \frac{g^2}{2}  \eta_{\alpha \beta} \eta^{\alpha_1 \beta_1}
\eta^{\mu \sigma}  F^{\rho}_{\mu \beta_1}
F^{\rho_1}_{\sigma \alpha_1 } G^{-1 \alpha}_{\rho}
G ^{-1 \beta}_{\rho_1} \\
&&- \frac{g^2}{2} \eta^{\alpha \beta } F^{\rho}_{\mu \beta}
F^{\rho_1}_{\alpha_1 \alpha} G^{-1 \alpha_1}_{\rho}
G ^{-1 \mu}_{\rho_1}
\end{array}
\label{20.41}
\ee
From these expressions, we can see that, if there is no gravity,
that is $C_{\mu}^{\alpha} $ vanishes, then $R_{\mu \nu}$ and
$R$ all vanish. It means that, if there is gravity, the space-time
is flat, which is what we expected in general relativity.
\\

Because scalar curvature $R$ is invariant under general
coordinates transformation, it transforms covariantly under
gravitational gauge transformation. Its behavior under
gravitational gauge transformation is the same as that of
the lagrangian ${\cal L}_0$. Just from the requirement
of gravitational gauge symmetry, the most general lagrangian
which gives an action with gravitational gauge symmetry is
\be
{\cal L}_0 = - \frac{c_1}{4} \eta^{\mu \rho} \eta^{\nu \sigma}
g_{ \alpha \beta}
F_{\mu \nu}^{\alpha} F_{\rho \sigma}^{\beta} + c_2 R,
\label{20.4101}
\ee
where $c_1$ and $c_2$ are two constant parameters. 
But, if we include $R$ in our lagrangian, our field equation 
for gravitational gauge field will become extremely
complicated, for the expression of scalar curvature $R$
in terms of gravitational gauge field $C_{\mu}^{\alpha}$ 
is very complicated. So, we did not consider scalar
curvature $R$ in this model. Another reason that we did not
consider $R$ in this model is that the lagrangian given by
eq.(\ref{4.20}) is the lagrangian expected by gauge field
theory and it is enough to give reasonable results and
predictions on all possible problems, such as classical
limit, classical tests, cosmological constant, 
cosmological model, $\cdots$ etc..
In a meaning, the model discuss in this paper is the
minimum model for quantun gauge theory of gravity. 
\\

Now, let's discuss some transformation properties of these tensors
under general coordinates transformation. Make a special kind of
local coordinates translation,
\be
x^{\mu }  \to x^{\prime \mu}
= x^{\mu} + \epsilon^{\mu} (x').
\label{20.42}
\ee
Under this transformation, the covariant derivative and gravitational
gauge fields transform as
\be
D_{\mu} (x) \to D'_{\mu} (x')
= \ehat (x') D_{\mu} (x') \ehat^{-1} (x') ,
\label{20.43}
\ee
\be
C_{\mu} (x) \to C'_{\mu} (x')
= \ehat (x') C_{\mu} (x') \ehat^{-1} (x') 
+ \frac{i}{g} \ehat (x')
(\frac{\partial}{\partial x^{\prime \mu}} \ehat^{-1} (x')).
\label{20.44}
\ee
It can be proved that
\be
G^{\alpha}_{\mu} (x) \to G^{\prime \alpha}_{\mu} (x')
= \Lambda^{\alpha}_{~\beta} G^{\beta}_{\mu}(x),
\label{20.45}
\ee
where
\be
\Lambda^{\alpha}_{~\beta} =
\frac{\partial x^{\prime \alpha}}{ \partial x^{\beta}}.
\label{20.46}
\ee
We can see that Lorentz index $\mu$ does not take part in transformation.
In fact, all Lorentz indexes do not take part in this kind of transformation.
Therefore,
\be
\eta^{\mu \nu} \to \eta^{\prime \mu \nu} = \eta^{\mu \nu}.
\label{20.47}
\ee
Using all these relations, we can prove that
\be
g^{\alpha \beta}(x)  \to g^{\prime \alpha \beta}(x')
= \Lambda^{\alpha}_{~\alpha_1} \Lambda^{\beta}_{~\beta_1}
g^{\alpha_1 \beta_1}(x),
\label{20.48}
\ee
\be
R_{\alpha \beta \gamma \delta}(x)  \to
R^{\prime}_{ \alpha \beta \gamma \delta}(x')
= \Lambda_{\alpha}^{~\alpha_1} \Lambda_{\beta}^{~\beta_1}
\Lambda_{\gamma}^{~\gamma_1} \Lambda_{\delta}^{~\delta_1}
R_{\alpha_1 \beta_1\gamma_1 \delta_1}(x),
\label{20.49}
\ee
\be
R_{\alpha \beta }(x)  \to
R'_{ \alpha \beta }(x')
= \Lambda_{\alpha}^{~\alpha_1} \Lambda_{\beta}^{~\beta_1}
R_{\alpha_1 \beta_1}(x),
\label{20.50}
\ee
\be
R(x) \to R'(x') = R(x).
\label{20.51}
\ee
These transformation properties are just what we
expected in general relativity. All these quantities has
the same transformation properties as those in general
relativity. \\

Equivalence principle is one of the most important fundamental
principles of general relativity. But, as we have studied in 
previous chapters, the inertial energy-momentum is not 
equivalent to the
gravitational energy-momentum in gravitational gauge theory.
This result is an inevitable result of gauge principle. But all these
differences are caused by gravitational gauge field. In leading term
approximation, the inertial energy-momentum tensor
$T_{i \alpha}^{\mu}$ is the same as the gravitational energy-momentum
tensor $T_{g \alpha}^{\mu}$. Because gravitational coupling
constant $g$ is extremely small and the strength of gravitational
field is also weak, it is hard to detect the difference between
inertial mass and gravitational mass. Using gravitational gauge
field theory, we can calculate the difference of inertial mass
and gravitational mass for different kinds of matter and help us
to test the validity of equivalence principle. This is a fundamental
problem which will help us to understand the nature of gravitational
interactions. In general relativity, equvilence principle is
the foundation of geometrical nature of gravity. But now, the
the foundation of geometrical nature of gravity is no longer
equivalence principle, but geometrical nature of translation
transformation. \\

Define
\be  \label{50.1}
\Lambda \define \frac{1}{2} ( R + 4 g^2 {\cal L}_0).
\ee
Then action given by eq.(\ref{3.22}) will be changed into
\be \label{50.2}
S = S_g + S_M,
\ee
where
\be  \label{50.3}
S_g = \frac{-1 }{16 \pi G} \int {\rm d}^4x
\sqrt{- g} (R - 2 \Lambda),
\ee
\be  \label{50.4}
S_M = \int {\rm d}^4x {\cal L}_M,
\ee
where $G$ is the Newtonian gravitational constant, which
is given by
\be  \label{50.44}
G = \frac{g^2}{4 \pi},
\ee
$R$ is the scalar curvature, $\Lambda$ is the cosmological
term, ${\cal L}_M$ is the lagrangian density for
matter fields. Scalar curvature $R$ can be expressed
by gravitational gauge field $C_{\mu}^{\alpha}$.
We have added the action for matter
fields into eq.(\ref{50.2}) and denoted the action for
pure gravitational gauge field as $S_g$.
 Using the following relations
\be \label{50.5}
\delta \sqrt{-g} = \frac{1}{2}
\sqrt{-g } g^{\mu\nu} \delta g_{\mu\nu},
\ee
\be \label{50.6}
\sqrt{-g } g^{\mu\nu} \delta R_{\mu\nu}
= \partial_{\lambda} W^{\lambda},
\ee
\be \label{50.7}
T_m^{\mu\nu} = \frac{2}{\sqrt{-g }}
\frac{\delta S_M}{\delta g_{\mu\nu}(x)},
\ee
where $T_m^{\mu\nu}$ is the energy-momentum tensor of
matter fields and $W^{\lambda}$ is a  contravariant
vector, we can obtain the Einstein's field equation
with cosmological term $\Lambda$,
\be \label{50.8}
R_{\mu\nu} - \frac{1}{2} g_{\mu\nu} R
+ \Lambda g_{\mu\nu}
= - 8 \pi G T_{\mu\nu},
\ee
where $T_{\mu\nu}$ is the revised energy-momentum tensor,
whose definition is
\be \label{50.9}
T_{\mu\nu} \define T_{m \mu\nu}
- \frac{1}{4 \pi G}
\frac{\delta \Lambda}{\delta g^{\mu\nu}}.
\ee
In eq.(\ref{50.9}), the difinition of
$\frac{\delta \Lambda}{\delta g^{\mu\nu}}$
is not clear, because $\Lambda$ is a function of
$G^{\alpha}_{\mu}$, not a function of
$g^{\mu \nu}$. So, we need to give out an explicite
definition of $\frac{\delta \Lambda}{\delta g^{\mu\nu}}$.
According to eq.(\ref{4.706}), we have
\be \label{50.10}
\frac{\partial g^{\mu\nu}}
{\partial G^{\lambda}_{\alpha}}
= \delta^{\mu}_{\lambda}
\eta^{\alpha\beta} G^{\nu}_{\beta}
+  \delta^{\nu}_{\lambda}
\eta^{\alpha\beta} G^{\mu}_{\beta}.
\ee
Therefore, we have
\be \label{50.11}
\frac{\delta \Lambda}{\delta G^{\lambda}_{\alpha}}
= 2 \eta^{\alpha \beta} G^{\nu}_{\beta}
\frac{\delta \Lambda}{\delta g^{\lambda\nu}}
\ee
It gives out
\be \label{50.12}
\frac{\delta \Lambda}{\delta g^{\mu\nu}}
= \frac{1}{4} \eta_{\alpha \beta}
\left( G_{\nu}^{-1 \beta}
\frac{\delta \Lambda}{\delta G^{\mu}_{\alpha}}
+  G_{\mu}^{-1 \beta}
\frac{\delta \Lambda}{\delta G^{\nu}_{\alpha}}
\right),
\ee
which gives out the explicite expression for
$\frac{\delta \Lambda}{\delta g^{\mu\nu}}$.
Eq.(\ref{50.8}) is quite like the Einstein's field
equation with cosmological constant in form, but it is
not the traditional Einstein field equation, so
we call it the Einstein-like field equation
with cosmological term.
\\

\section{Discussions}
\setcounter{equation}{0}
 
In this paper, a new kind of gauge gravity is 
formulated in the framework of traditional quantum
field theory, where gravity is treated as a kind
of physical interactions and space-time is always
kept falt. This treatment satisfies the fundamental
spirit of traditional quantum field theory, and
go along this way, four different kinds of fundamental
interactions can be unified on the same fundamental
baseline\cite{7}. The most advantage of this approach is that
the renormalizability of the quantum gravity is easy
to be proved. Its transcendental  foundation is gauge
principle. Gravitational gauge interactions is completely 
determined by gauge symmetry. In other words, the 
Lagrangian of the system is completely determined by 
gauge symmetry. Using the langurage of Cartan tetrad,
we set up the geometrical formulation of this new
quantum gauge theory of gravity and to study its
geometrical foundation. So, gravity theory has both
physical picture and geometrical picture, which is 
the reflection of the physics-geometry duality of
gravity. In this chaper, we give some simple 
discussions on some interested problem related to
gauge theory of graivty. The content of this chapter
is not in the main topic of this paper, so our
discussion on them are quite simple, and detailed 
discussion can be found in some related literature. 
\\

\begin{enumerate}
 
\item \lbrack {\bf Schwarzchild-like Solution
	and Classical Tests} \rbrack

It is known that General Relativity is tested by three
main classical tests, which are  perihelion procession,
deflection of light and gravitational red-shift. All these
three tests are related to geodesic curve equation and
schwarzachild solution in general relativity.
If we know geodesic curve equation and space-time
metric, we can calculate perihelion procession,
deflection of light and gravitational red-shift.
In this chapter, we discuss this problem
from the point of view of gauge theory of gravity.
\\

In order to discuss classical tests of gravity, for
the sake of convience, we use the geometrical
representation of gravity. As we have stated above,
in the geometrical representation of gravity,
gravity is not treated as physical interactions
in space-time. In the gemetrical representation
of gravity, we use the same manner which is widely
used in general relativity to discuss classical
tests and to explain the predictions with
observations. In the geometrical representation of
gravity, if there is no other
physical interactions in space time, any mass point
can not feel any physical forces when it moves in
space-time. So, it must move along the curve which
has the least length.
Suppose that a particle is moving from point
A to point B along a curve. Define
\be  \label{16.01}
T_{BA} = \int_A^B
\sqrt{ - g_{\mu\nu} \frac{{\rm d}x^{\mu}}{{\rm d}p}
\frac{{\rm d}x^{\nu}}{{\rm d}p} } {\rm d} p,
\ee
where $p$ is a parameter that describe the orbit.
The real curve that the particle moving along
corresponds to the minimum of $T_{AB}$. The
minimum of $T_{AB}$ gives out the following
geodesic curve equation
\be  \label{16.02}
\frac{{\rm d}^2 x^{\mu}}{{\rm d}p^2}
+ \Gamma^{\mu}_{\nu\lambda}
\frac{{\rm d}x^{\nu}}{{\rm d}p}
\frac{{\rm d}x^{\lambda}}{{\rm d}p} =0,
\ee
where $\Gamma^{\mu}_{\nu\lambda}$ is the affine
connection
\be
\Gamma^{\lambda}_{\mu \nu}
= \frac{1}{2} g^{\lambda \sigma}
\left( \frac{\partial g_{\mu \sigma}}{\partial x^{\nu}}
+ \frac{\partial g_{\nu \sigma}}{\partial x^{\mu}}
-\frac{\partial g_{\mu \nu}}{\partial x^{\sigma}} \right).
\label{16.03}
\ee
Eq.(\ref{16.02}) gives out the curve that a free particle
moves along in curved space-time if we discuss physics
in the geometrical representation of gravity.
\\

Now, we need to calculate a schwarzchild-like solution
in gauge theory of gravity. In chapter 4,
we have obtained a solution of $C_{\mu}^{\alpha}$
for static spherically symmetric gravitational fields
in linear approximation of $gC_{\mu}^{\alpha}$. But
expetimental tests, especially  perihelion procession,
are sensitive to second order of $gC_{\mu}^{\alpha}$.
The best way to do this is to solve the equation of
motion of gravitational gauge field in the second
order approximation of $gC_{\mu}^{\alpha}$. But this
equation of motion is a non-linear second order
partial differential equations. It is rather difficult
to solve them. So, we had to find some other method
to do this.  The perturbation method is used to do
this.  After considering
corrections from gravitational energy of the sun in vacuum
space and gravitational interactive energy between the sun and
the Mercury, the equivalent gravitational gauge field
in the second order approximation is\cite{wu4}
\be  \label{16.1}
g C_{0}^{0} = - \frac{GM}{r}
- \frac{3}{2} \frac{G^2 M^2}{r^2}
+ O\left(\frac{G^3 M^3}{r^3} \right).
\ee
Then using eq.(\ref{4.707}), we can obtain the following
solution
\be  \label{16.2}
\ba{rcl}
d \tau^2 &=& \left\lbrack
1 - \frac{2 GM}{r} + O\left(\frac{G^3 M^3}{r^3} \right)
\right\rbrack d t^2
-  \left\lbrack
1 + \frac{2 GM}{r} + O\left(\frac{G^2 M^2}{r^2} \right)
\right\rbrack d r^2 \\
&&\\
&& - r^2 d \theta^2 - r^2 {\rm sin}^2\theta d \varphi^2,
\ea
\ee
where we have use the following gauge for gravitational
gauge field,
\be  \label{16.3}
C_{\mu}^{\mu} = 0.
\ee
This solution is quite similar to schwarzschild solution,
but it is not schwarzschild solution, so we call it
schwarzschild-like solution.
If we use Eddington-Robertson expansion, we will find that
for the present schwarzschild-like solution\cite{wei},
\be  \label{16.4}
\alpha = \beta = \gamma = 1.
\ee
They have the same values as schwarzschild solution in
general relativity and all three tests are only sensitive
to these three parameters,  so gauge theory of gravity
gives out the same  theoretical predictions as those of
general relativity\cite{wu4}. More
detailed discussions on classical tests can be found
in literature \cite{wu4}. (This result hold for those
models which have other choice
of $\eta_2$ and $J(C)$ which is duscussed in \cite{wu01,wu02}.)\\

\item \lbrack {\bf Gravitational Wave} \rbrack
In gravitational gauge theory, the gravitational gauge field
is represented by $C_{\mu}$.  From the point
of view of quantum field theory, gravitational gauge field
$C_{\mu}$ is a vector field and it obeys dynamics of vector field.
In other words, gravitational wave is vector wave. Suppose that the
gravitational gauge field is very weak in vacuum, then in leading
order approximation, the equation of motion of gravitational wave
is
\be
\partial^{\mu} F_{0 \mu \nu}^{\alpha}= 0 ,
\label{12.1}
\ee
where $F_{0 \mu \nu}^{\alpha}$ is given by eq.(\ref{5.9}). If we set
$g C_{\mu}^{\alpha}$ equals zero, we can obtain eq.(\ref{12.1})
from eq.(\ref{4.41}). Eq.(\ref{12.1}) is very similar to the famous
Maxwell equation in vacuum. Define
\be
F^{\alpha}_{ij}= - \varepsilon_{ijk} B^{\alpha}_{k}
~~,~~
F^{\alpha}_{ 0i} = E^{\alpha}_{i},
\label{12.2}
\ee
then eq.(\ref{12.1}) is changed into
\be
\nabla \cdot \svec{E}^{\alpha}=0,
\label{12.3}
\ee
\be
\frac{\partial}{\partial t} \svec{E}^{\alpha}
- \nabla \times \svec{B}^{\alpha} =0.
\label{12.4}
\ee
From definitions eq.(\ref{12.2}), we can prove that
\be
\nabla \cdot  \svec{B}^{\alpha} =0,
\label{12.5}
\ee
\be
\frac{\partial}{\partial t} \svec{B}^{\alpha}
+ \nabla \times \svec{E}^{\alpha} =0.
\label{12.6}
\ee
If there were no superscript $\alpha$, eqs.(\ref{12.3}-\ref{12.6}) would be
the ordinary Maxwell equations. In ordinary case, the
strength of gravitational field in vacuum is extremely weak,
so the gravitational wave in vacuum is composed of four independent
vector waves. \\

Though gravitational gauge field is a vector field, its
component fields $C_{\mu}^{\alpha}$ have one Lorentz index
$\mu$ and one group index $\alpha$. Both indexes have
the same behavior under Lorentz transformation. According to
the behavior of Lorentz transformation, gravitational field
likes a tensor field. We call it pseudo-tensor field. The
spin of a field is determined according to its behavior under
Lorentz transformation, so the spin of gravitational field
is 2. In conventional quantum field theory, spin-1 field is a vector
field, and vector field is a spin-1 field. In gravitational gauge field,
this correspondence is violated. The reason is that, in gravitational
gauge field theory, the group index contributes to the spin of a field,
while in ordinary gauge field theory, the group index do not
contribute to the spin of a field. In a word, gravitational field
is a spin-2 vector field.   \\

\item \lbrack {\bf Gravitational Magnetic Field} \rbrack
From eq.(\ref{12.3}-\ref{12.6}), we can see that the equations of motion of
gravitational wave in vacuum are quite similar to those of
electromagnetic wave. The phenomenological behavior of gravitational
wave must also be similar to that of electromagnetic wave. In
gravitational gauge theory, $\svec{B}^{\alpha}$ is called the
gravitational magnetic field. It will transmit gravitational magnetic
interactions between two rotating objects. In first order approximation,
the equation of motion of gravitational gauge field is
\be
\partial^{\mu} F_{\mu \nu}^{\alpha}
= - g \eta_{\nu \tau} \eta_2^{\alpha \beta}
T^{\tau}_{g \beta}.
\label{12.7}
\ee
Using eq.(\ref{12.2}) and eq.(\ref{12.7}), we can get the following equations
\be
\nabla \cdot \svec{E}^{\alpha}=
- g \eta_2^{\alpha \beta} T^0_{g \beta},
\label{12.8}
\ee
\be
\frac{\partial}{\partial t} \svec{E}^{\alpha}
- \nabla \times \svec{B}^{\alpha} =
+ g  \eta_2^{\alpha \beta} \svec{T}_{g \beta},
\label{12.9}
\ee
where $ \svec{T}_{g \beta}$ is a simplified notation whose explicit
definition is given by the following relation
\be
( \svec{T}_{g \beta})^i =  \svec{T}^i_{g \beta}.
\label{12.10}
\ee
On the other hand, it is easy to prove that(omit self
interactions of graviton)
\be
\partial_{\mu} F^{\alpha}_{\nu \lambda}
+ \partial_{\nu} F^{\alpha}_{\lambda \mu}
+ \partial_{\lambda} F^{\alpha}_{\mu \nu}
 = 0.
\label{12.11}
\ee
From eq.(\ref{12.11}), we can get
\be
\nabla \cdot  \svec{B}^{\alpha} =0,
\label{12.12}
\ee
\be
\frac{\partial}{\partial t} \svec{B}^{\alpha}
+ \nabla \times \svec{E}^{\alpha} =0.
\label{12.13}
\ee
Eq.(\ref{12.8}) means that energy-momentum density of the system is
the source of gravitational electric fields, eq.(\ref{12.9}) means that
time-varying gravitational electric fields give rise to gravitational
magnetic fields, and eq.(\ref{12.13}) means that time-varying gravitational
magnetic fields give rise to gravitational electric fields.
Suppose that the angular momentum of an rotating object is $J_i$,
then there will be a coupling between angular momentum and
gravitational magnetic fields. The interaction Hamiltionian of this
coupling is proportional to
$(\frac{P_{\alpha}}{m}\!\svec{J} \cdot \svec{B}^{\alpha})$.
The existence of gravitational magnetic fields is important for
cosmology. It is known that almost all galaxies in the universe
rotate. The global rotation of galaxy will give rise to gravitational
magnetic fields in space-time. The existence of gravitational magnetic
fields will affect the moving of stars in (or near) the galaxy.
This influence contributes to the formation of the galaxy and
can explain why almost all galaxies have global large scale
structures. In other words, the gravitational magnetic fields
contribute great to the large scale structure of galaxy and
universe. \\

\item \lbrack {\bf Lorentz Force} \rbrack
There is a force when a particle is moving in a gravitational
magnetic field. In electromagnetic field theory, this force is usually
called Lorentz force. As an example, we discuss gravitational interactions
between gravitational field and Dirac field. Suppose that the gravitational
field is static.
According to eqs.(\ref{6.2}-\ref{6.3}), the interaction Lagrangian is
\be
{\cal L}_I = g J(C) \bar\psi
\gamma^{\mu} \partial_{\alpha} \psi C_{\mu}^{\alpha}.
\label{12.14}
\ee
For Dirac field, the gravitational energy-momentum of Dirac field is
\be
T_{g \alpha}^{\mu} = \bar\psi
\gamma^{\mu} \partial_{\alpha} \psi.
\label{12.15}
\ee
Substitute eq.(\ref{12.15}) into eq.(\ref{12.14}), we get
\be
{\cal L}_I = g J(C) T_{g \alpha}^{\mu} C_{\mu}^{\alpha}.
\label{12.16}
\ee
The interaction Hamiltonian density ${\cal H}_I$ is
\be
{\cal H}_I = - {\cal L}_I  =
- g J(C) T_{g \alpha}^{\mu}(y,\svec{x}) C_{\mu}^{\alpha}(y).
\label{12.17}
\ee
Suppose that the moving particle is a mass point at point $\svec{x} $,
in this case
\be
 T_{g \alpha}^{\mu}(y,\svec{x}) =  T_{g \alpha}^{\mu}
\delta(\svec{y} - \svec{x} ),
\label{12.18}
\ee
where $ T_{g \alpha}^{\mu}$ is independent of space coordinates.
Then, the interaction Hamiltonian $H_I$ is
\be
H_I = \int {\rm d}^3 \svec{y} {\cal H}_I (y)
= - g \int {\rm d}^3 \svec{y}
J(C)  T_{g \alpha}^{\mu}(y,\svec{x}) C_{\mu}^{\alpha}(y).
\label{12.19}
\ee
The gravitational force that acts on the mass point is
\be
f_i = g \int {\rm d}^3y J(C) T^{\mu}_{g \alpha}(y,\svec{x})
F^{\alpha}_{i \mu}
+g \int {\rm d}^3y J(C) T^{\mu}_{g \alpha}(y,\svec{x})
\frac{\partial}{\partial y^{\mu}} C_i^{\alpha}.
\label{12.20}
\ee
For quasi-static system, if we omit higher order contributions, the
second term in the above relation vanish.
For mass point, using the technique of Lorentz covariance analysis,
we can proved that
\be
P_{g \alpha} U^{\mu} = \gamma T_{g \alpha}^{\mu},
\label{12.21}
\ee
where $U^{\mu}$ is velocity, $\gamma$ is the rapidity, and
$P_{g \alpha}$ is the gravitational energy-momentum. According
eq.(\ref{12.18}), $P_{g \alpha}$ is given by
\be
P_{g \alpha} = \int {\rm d}^3 \svec{y}
 T_{g \alpha}^0(y) =  T_{g \alpha}^0.
\label{12.22}
\ee
Using all these relations and eq.(\ref{12.2}), we get
\be
\svec{f} = -g J(C) P_{g \alpha} \svec{E}^{\alpha}
- g J(C) P_{g \alpha} \svec{v} \times \svec{B}^{\alpha}.
\ee
For quasi-static system, the dominant contribution of the above
relation is
\be
\svec{f} = g J(C) M \svec{E}^0
+ g J(C) M \svec{v} \times \svec{B}^0,
\label{12.23}
\ee
where $\svec{v}= \svec{U}/\gamma$ is the velocity of the mass point.
The first term of eq.(\ref{12.23}) is the classical Newton's gravitational
interactions. The second term of eq.(\ref{12.23}) is the Lorentz force. The
direction of this force is perpendicular to the direction of the motion
of the mass point. When
the mass point is at rest or is moving along the direction
of the gravitational magnetic field, this force vanishes.
Lorentz force is important for cosmology, because the rotation
of galaxy will generate gravitational magnetic field and this gravitational
magnetic field will affect the motion of stars and affect the large scale
structure of galaxy. \\

\item \lbrack {\bf Negative Energy} \rbrack
First, let's discuss inertial energy of pure gravitational wave.
Suppose that the gravitational wave is not so strong, so the higher
order contribution is very small. We only consider leading order
contribution here. For pure gravitational field, we have
\be
\frac{\partial {\cal L}_0}{\partial \partial_{\mu} C_{\nu}^{\beta}}
= - \eta^{\mu \rho} \eta^{\nu \sigma} \eta_{2 \beta \gamma}
F_{\rho \sigma}^{\gamma}
+ g \eta^{\lambda \rho} \eta^{\nu \sigma} \eta_{2 \beta \gamma}
C^{\mu}_{\lambda} F^{\gamma}_{\rho \sigma}.
\label{12.24}
\ee
From eq.(\ref{4.32}), we can get the inertial energy-momentum tensor
of gravitational field in the leading order approximation, that is
\be
T^{\mu}_{i \alpha} = J(C)
\lbrack
+ \eta^{\mu \rho} \eta^{\nu \sigma} \eta_{2 \beta \gamma}
F_{\rho \sigma}^{\gamma} \partial_{\alpha} C_{\nu}^{\beta}
+ \delta^{\mu}_{\alpha}{\cal L}_0
\rbrack.
\label{12.25}
\ee
Using eq.(\ref{12.2}), Lagrangian given by eq.(\ref{4.20}) can be changed into
\be
{\cal L}_0 = \frac{1}{2}
\eta_{2 \alpha \beta}
( \svec{E}^{\alpha} \cdot \svec{E}^{\beta}
- \svec{B}^{\alpha} \cdot \svec{B}^{\beta} ).
\label{12.26}
\ee
Space integral of time component of inertial energy-momentum tensor
gives out inertial energy $H_i$ and inertial momentum $\svec{P}_i$.
They are
\be
H_i = \int {\rm d}^3 \svec{x}
J(C) \left\lbrack  \frac{1}{2}
\eta_{2 \alpha \beta}
( \svec{E}^{\alpha} \cdot \svec{E}^{\beta}
+ \svec{B}^{\alpha} \cdot \svec{B}^{\beta} )
\right\rbrack,
\label{12.27}
\ee
\be
\svec{P}_i =   \int {\rm d}^3 \svec{x} J(C)
\eta_{2 \alpha \beta}  \svec{E}^{\alpha} \times \svec{B}^{\beta}.
\label{12.28}
\ee
In order to obtain eq.(\ref{12.27}), eq.(\ref{12.3}) is used.
Let consider the inertial energy-momentum of gravitaional
field $C_{\mu}^0$. Because,
\be
\eta_{2 0 0} = -1,
\label{12.29}
\ee
eq.(\ref{12.27}) gives out
\be
H_i (C^0) = - \frac{1}{2} \int {\rm d}^3 \svec{x}
J(C) ( \svec{E}^0 \cdot \svec{E}^0
+\svec{B}^0 \cdot \svec{B}^0 ).
\label{12.30}
\ee
$H_i(C^0)$ is a negative quantity. It means that the inertial energy
of gravitational field $C_{\mu}^0$ is negative. The gravitational
energy-momentum of pure gravitational gauge field is given by
eq.(\ref{3.25}). In leading order approximation, it is
\be
\begin{array}{rcl}
T_{g \alpha}^{\mu} & =&
\eta^{\mu \rho} \eta^{\nu \sigma} \eta_{2 \beta \gamma}
F_{\rho \sigma}^{\gamma} ( \partial_{\alpha} C_{\nu}^{\beta} )
+ \eta^{\mu}_{1 \alpha} {\cal L}_0 \\
&&\\
&&- \eta^{\lambda \rho} \eta^{\mu \sigma} \eta_{2 \alpha \beta }
\partial_{\nu} ( C_{\lambda}^{\nu} F_{\rho \sigma}^{\beta} )
+ \eta^{\nu \rho} \eta^{\mu \sigma} \eta_{2 \alpha \beta }
\eta^{\lambda}_{1 \tau}
( \partial_{\nu} C_{\lambda}^{\tau}) F_{\rho \sigma}^{\beta}.
\end{array}
\label{12.31}
\ee
After omitting surface terms,
the gravitational energy of the system is
\be
H_g  =  \int {\rm d}^3 \svec{x}
\left\lbrack  \frac{1}{2}  \eta_{2 \alpha \beta }
( \svec{E}^{\alpha} \cdot \svec{E}^{\beta}
+ \svec{B}^{\alpha} \cdot \svec{B}^{\beta} )
 - \eta^{i j } \partial_{0} (C_j^0 E_i^0 )
\right\rbrack.
\label{12.32}
\ee
The gravitational energy  of gravitational field $C_{\mu}^0$ is,
\be
H_g (C^0) = - \frac{1}{2} \int {\rm d}^3 \svec{x}
( \svec{E}^0 \cdot \svec{E}^0
+ \svec{B}^0 \cdot \svec{B}^0
+ 2 \eta^{i j } \partial_{0}  ( C_j^0 E_i^0)) .
\label{12.33}
\ee
$H_g$ is also negative. It means that the gravitational energy of
gravitational field $C_{\mu}^0$ is negative. In other words, gravitational
gauge field $C_{\mu}^0$ has negative gravitational energy and
negative inertial energy. But,
inertial mass is not equivalent to gravitational mass for pure
gravitational gauge field.
\\

\item \lbrack {\bf Repulsive Force} \rbrack
The classical gravitational interactions are attractive interactions.
But in gravitational gauge theory, there are repulsive interactions as
well as attractive interactions. The gravitational force is given by
eq.(\ref{12.23}). The first term corresponds to classical gravitational
force. It is
\be
f_i =  g J(C)  T_{g \alpha}^0 (\partial_i C_0^{\alpha}).
\label{12.59}
\ee
For quasi-static gravitational field, it is changed into
\be
\begin{array}{rcl}
f_i &=& -  g J(C)  P_{g \alpha} E_i^{\alpha}  \\
&&\\
&=& g J(C) ( M_1 E_i^0 - P_{gj} E_i^j   ),
\end{array}
\label{12.60}
\ee
where $M_1$ is the gravitational mass of the mass point which
is moving in gravitational field. Suppose that the gravitational field
is generated by another mass point whose gravitational
energy-momentum is $Q_g^{\alpha}$ and gravitational mass is $M_2$.
For quasi-static gravitational field, we can get
\be
E_i^{\alpha} =
- \frac{g}{4 \pi r^3} Q_g^{\alpha} r_i
\label{12.61}
\ee
Substitute eq.(\ref{12.61}) into eq.(\ref{12.60}), we get
\be
\svec{f} = J(C) \frac{g^2}{4 \pi r^3}
\svec{r} ( - E_{1g} E_{2g} + \svec{P}_g \cdot \svec{Q}_g ),
\label{12.62}
\ee
where $E_{1g}$  and $E_{2g}$ are gravitational energy of two mass point.
From eq.(\ref{12.62}), we can see that, if $\svec{P}_g \cdot \svec{Q}_g$ is
positive, the corresponding gravitational force between two momentum
is repulsive. This repulsive force is important for the stability of
some celestial object. For relativistic system, all mass point moving
at a high speed which is near the speed of light. Then the term
$\svec{P}_g \cdot \svec{Q}_g$ has approximately the same
order of magnitude as
that of $E_{1g} E_{2g}$, therefore, for relativistic systems, the
gravitational attractive force is not so strong as the force when
all mass points are at rest.
\\

\item \lbrack {\bf Equivalence Principle} \rbrack
Equivalence principle is one of the most important foundations of
general relativity, but it is not a logic starting point of
gravitational gauge field theory. The logic starting point of
gravitational gauge field theory is gauge principle. However,
one important inevitable result of gauge principle is that
gravitational mass is not equivalent to inertial mass. The origin
of violation of equivalence principle is gravitational field. If there
were no gravitational field, equivalence principle would strictly
hold. But if gravitational field is strong, equivalence principle
will be strongly violated. For some celestial objects which have
strong gravitational field, such as quasar and black hole, their
gravitational mass will be higher than their inertial mass.
But on earth, the gravitational field is very weak, so the equivalence
principle almost exactly holds. We need to test the validity of
equivalence principle in astrophysics experiments. \\

\end{enumerate}

In this paper, we have discussed a completely new quantum gauge
theory of gravity. Finally, we give a simple summary to the
whole theory.

\begin{enumerate}

\item In leading order approximation, the gravitational gauge field
theory gives out classical Newton's theory of gravity.

\item In first order approximation and for vacuum, the gravitational
gauge field theory gives out Einstein's general theory of relativity.

\item Gravitational gauge field theory is a renormalizable quantum
theory.

\item Combining cosmological principle with the field equation
of gravitational gauge field, we can also set up
a cosmological model with is consistent with recent
observations\cite{wu5}.

\end{enumerate}

\end{document}